\documentclass[twoside,11pt]{article}

\usepackage{jmlr2e}
\usepackage{amsmath}
\usepackage{xcolor} 
\usepackage{multirow}
\usepackage{float}
\newtheorem{Proposition}{Proposition}
\newtheorem{Lemma}{Lemma}
\newtheorem{Theorem}{Theorem}

\newcommand{\R}{\mathbb{R}}
\newcommand{\RR}{\mathbf{R}}
\newcommand{\PP}{{\mathbb{P}}}
\newcommand{\E}{\mathbf{E}}
\newcommand{\xnew}{x_{*}}
\newcommand{\argmin}{\arg\min}

\ShortHeadings{Inference for Case Probability}{Guo, Rakshit, Herman and Chen}
\firstpageno{1}

\begin{document}

\title{Inference for the Case Probability in High-dimensional Logistic Regression}

\author{\name Zijian Guo \email zijguo@stat.rutgers.edu \\
\name Prabrisha Rakshit \email pr412@scarletmail.rutgers.edu\\
        \addr Department of Statistics\\
       Rutgers University\\
       Piscataway, New Jersey, USA\\
       \\
       \name Daniel S. Herman \email daniel.herman2@pennmedicine.upenn.edu \\
       \addr Department of Pathology and Laboratory Medicine\\
       University of Pennsylvania\\
       Philadelphia, Pennsylvania, USA
       \AND
       \name Jinbo Chen \email jinboche@upenn.edu  \\ 
       \addr Department of Pathology and Laboratory Medicine\\
       University of Pennsylvania\\
       Philadelphia, Pennsylvania, USA}

\editor{Francis Bach, David Blei and Bernhard Sch{\"o}lkopf}  
\maketitle
\begin{abstract}
Labeling patients in electronic health records with respect to their statuses of having a disease or condition, i.e. case or control statuses, has increasingly relied on prediction models using high-dimensional variables derived from structured and unstructured electronic health record data. A major hurdle currently is a lack of valid statistical inference methods for the case probability. In this paper, considering high-dimensional sparse logistic regression models for prediction, we propose a novel bias-corrected estimator for the case probability through the development of linearization and variance enhancement techniques. We establish asymptotic normality of the proposed estimator for any loading vector in high dimensions. We construct a confidence interval for the case probability and propose a hypothesis testing procedure for patient case-control labelling.  We demonstrate the proposed method via extensive simulation studies and application to real-world electronic health record data. 
\end{abstract}


\section{Introduction}
Electronic health record (EHR) data provides an unprecedented resource for clinical and translational research. Since EHRs were initially designed to support documentation for medical billing, patients' data are frequently not represented with sufficient precision and nuance for accurate phenotyping. Therefore, heuristic rules and statistical methods are needed to identify patients with a specific health condition. Logistic regression models have been frequently adopted for this ``EHR phenotyping" task \citep{ml1, ml2, ml3, ml4}. These methods commonly require a curated set of patients who are accurately labeled with regard to the presence or absence of a phenotype (e.g. disease or health condition). To obtain such a dataset, medical experts need to retrospectively review EHR charts and/or prospectively evaluate patients to label them. For many phenotypes, the labor and cost of the label assignment processes limit the achievable sample size, which is typically in the range of $50$ to $1,000$. On the other hand, potential predictors in EHRs may include hundreds or thousands of variables derived from billing codes, demographics, disease histories, co-morbid conditions, laboratory test results, prescription codes, and concepts extracted from doctors' notes through methods such as natural language processing. The dimension of these predictors is usually large in comparison to the sample size of the curated dataset \citep{highEHR}. 

One important example of phenotyping goal that would benefit from accurate risk prediction models leveraging large EHR data is primary aldosteronism (PA), the most common identifiable and specifically treatable cause of secondary high blood pressure \citep{PAtreat1, PAtreat2, PAtreat3}. PA is thought, based on epidemiological studies, to affect up to $1\%$ of US adults \citep{PAprev1, PAprev2}, but is diagnosed in many fewer individuals. Endocrine Society Guidelines recommend screening for PA in specific subgroups of hypertension patients, including patients with treatment-resistant high blood pressure or high blood pressure with low blood potassium \citep{PAscreening}. While simple, expert-curated heuristics can be used to identify patients that meet PA screening guidelines, it is of great interest to derive more sensitive and specific prediction models by leveraging the larger set of available potential features in the EHR. One goal of the current paper is to use data extracted from the Penn Medicine EHR and develop preliminary prediction models to help identify patients with hypertension and subsets thereof for which PA screening is recommended by guidelines.

\subsection{Problem Formulation}
We introduce a general statistical problem, which is motivated by EHR phenotyping.
For the $i$-th observation, the outcome $y_i\in \{0,1\}$ indicates whether the interest condition (e.g. PA) is present and $X_{i\cdot}\in \R^{p}$ denotes the observed high-dimensional covariates. Here we assume that $\{y_{i},X_{i\cdot}\}_{1\leq i\leq n}$ are independent and identically distributed and allow the number of covariates $p$ to be larger than the sample size $n$ as often seen in analyzing EHR data. We consider the following high-dimensional logistic regression model, for $1\leq i\leq n$,
\begin{equation}
\PP(y_{i}=1|X_{i\cdot})=h(X_{i\cdot}^{\intercal}\beta) \quad \text{with}\quad h(z)=\exp(z)/[1+\exp(z)]
\label{eq: logistic model}
\end{equation}
where $\beta\in \R^{p}$ denotes the high-dimensional vector of odds ratio parameters. The high-dimensional vector $\beta$ is assumed to be sparse throughout the paper.

The quantity of interest is the case probability $\PP(y_{i}=1|X_{i\cdot}=\xnew) \equiv {h\left(\xnew^{\intercal}\beta\right)}$, which is the conditional probability of $y_i=1$ given $X_{i\cdot}=\xnew \in \R^{p}$.  
The outcome labeling problem in EHR phenotyping is formulated as testing the following null hypothesis on the case probability,
\begin{equation}
H_0: h(\xnew^{\intercal}\beta)<1/2. 
\label{eq: null hypothesis}
\end{equation}
{Here, the threshold $1/2$ can be replaced by other positive numbers in $(0,1)$, which are decided by domain scientists. Throughout the paper, we use the threshold $1/2$ to illustrate the main idea of EHR phenotyping.}

Although the statistical inference problem is motivated from EHR phenotyping, the proposed inference procedure in the high-dimensional logistic model has a broader scope of applications. The linear functional $\xnew^{\intercal}\beta$ itself and the conditional probability of being a case are important quantities in statistics. 
Additionally, the case probability $h(X_{i\cdot}^{\intercal}\beta)$ is the same as the propensity score in causal inference, which is a central quantity for both matching \citep{pearl2000causality,rosenbaum1983central} and double robustness estimators \citep{bang2005doubly,kang2007demystifying}.

\subsection{Our Results and Contribution}
The penalized maximum likelihood estimation methods have been well developed to estimate $\beta\in \R^{p}$ in the high-dimensional logistic model 
\citep{bunea2008honest,bach2010self,buhlmann2011statistics,meier2008group,negahban2009unified,huang2012estimation}.
The penalized estimators enjoy desirable estimation accuracy properties. However, these methods do not lend themselves directly to statistical inference for the case probability mainly because the bias of the penalized estimator dominates the total uncertainty. Our proposed method is built upon the idea of bias correction that has been first developed to aid confidence interval construction for individual regression coefficients in high-dimensional linear regression models \citep{van2014asymptotically,javanmard2014confidence,zhang2014confidence}. This idea has also been extended to making inference for $\beta_j$ for $1\leq j\leq p$ in high-dimensional logistic regression models \citep{van2014asymptotically,ning2017general, ma2018global}. However, there is a lack of methods and theories for inference for the case probability $\PP(y_{i}=1|X_{i\cdot}=\xnew)$, which depends on the high-dimensional loading vector $\xnew \in \R^{p}$ and involves the entire regression vector $\beta \in \R^{p}.$

We propose a novel two-step bias-corrected estimator of the case probability.  In the first step, we estimate $\beta$ by a penalized maximum likelihood  estimator $\widehat{\beta}$ and construct the plug-in estimator $h(\xnew^{\intercal}\widehat{\beta})=\exp(\xnew^{\intercal}\widehat{\beta})/[1+\exp(\xnew^{\intercal}\widehat{\beta})]$. In the second step, we correct the bias of this plug-in estimator. The existing bias correction method \citep{van2014asymptotically} requires an accurate estimator of the high-dimensional vector $[\E \widehat{H}(\beta)]^{-1}\xnew \in \R^{p}$ where $\widehat{H}(\beta)$ denotes the sample Hessian matrix of the negative log-likelihood (see Section \ref{sec: initial estimator} for its definition). However, it is challenging to extend this idea to inference for the case probability since the Hessisan matrix $\E \widehat{H}(\beta)$ is complicated in the logistic model and $\xnew\in \R^{p}$ can be an arbitrary high-dimensional vector (with no sparsity structure). 

We address these challenges through development of linearization and variance enhancement techniques.
The linearization technique is introduced to handle the complex form of the Hessian matrix in the logistic model. Particularly, instead of assigning equal weights,  we conduct a weighted average by reweighing $X_{i\cdot}[y_i - h(\xnew^{\intercal}\widehat{\beta})]$ by ${1}/{{\rm Var}(y_i\mid X_i)}$, which leads to a re-weighted Hessian matrix $n^{-1}\sum_{i=1}^{n} X_{i\cdot} X_{i\cdot}^{\intercal}$. 
We refer to this re-weighting step as ``Linearization" since the re-weighted Hessian matrix  corresponds to the Hessian matrix of the least square loss in the linear model. 
 In addition, to develop an inference procedure for any high-dimensional vector $\xnew$, we introduce an extra constraint in constructing the projection direction for bias correction. The additional constraint is to enhance the variance component of the proposed bias-corrected estimator such that its variance dominates its bias for any high-dimensional loading vector $\xnew$. We refer to the proposed inference method as Linearization with Variance Enhancement, shorthanded as LiVE.

We establish the asymptotic normality of the proposed LiVE estimator for any high-dimensional loading vector $\xnew\in \R^{p}$. We then construct a confidence interval for the case probability and conduct the hypothesis testing \eqref{eq: null hypothesis} related to the outcome labelling. We develop new technical tools to establish the asymptotic normality for the re-weighted estimator; see Section \ref{sec: challenges}.

We conduct a large set of simulation studies to compare the finite-sample performance of the proposed LiVE estimator with the existing state-of-the art methods: the plug-in Lasso estimator, post-selection method, the plug-in \texttt{hdi} \citep{dezeure2015high}, the plug-in \texttt{WLDP} \citep{ma2018global} and generalization of the transformation method \citep{zhu2018linear,tripuraneni} to logistic models. 
 We demonstrate the proposed method using Penn Medicine EHR data to identify patients with hypertension and two subsets thereof that should be screened for PA, per specialty guidelines.

To sum up, the contribution of the current paper is two-fold.
\begin{enumerate}
\item We propose a novel bias-corrected estimator of the case probability and establish its asymptotic normality. To our best knowledge, this is the first inference method for the case probability in high dimensions, which is computationally efficient and statistically valid for any high-dimensional vector $\xnew$.
\item The theoretical justification on establishing the asymptotic normality of the re-weighted estimators is of independent interest and can be used to handle other inference problems in high-dimensional nonlinear models.
\end{enumerate}

{
Our proposed LiVE estimator has been implemented in the R package \texttt{SIHR}, which is available from CRAN. More detailed illustration of the R package \texttt{SIHR} can be found in \cite{rakshit2021sihr}.}
\subsection{Comparison with Existing Literature }
\label{sec: Literature}
We have proposed a two-step bias correction procedure to make inference for $h(\xnew^{\intercal}\beta)$ in the high-dimensional logistic model. Specifically, the linearization and variance enhancement techniques are introduced to ensure that our proposed confidence intervals are valid for any $\xnew\in \R^{p}$ and a broad class of design covariance matrix.
We shall mention other related works and discuss the connections and differences.

Post-selection inference \citep[e.g.]{belloni2013least} is a commonly used method in constructing confidence intervals, where the first step is to conduct model selection and the second step is to run a low-dimensional logistic model with the selected sub-model. However, such a method typically requires the consistency of model selection in the first step. Otherwise, the constructed confidence intervals are not valid as the uncertainty of model selection in the first step is not properly accounted for. It has been observed in Section \ref{sec: sim} that the post-selection method has produced under-covered confidence intervals in finite samples; see  Tables \ref{tab: Setting 1 i} and \ref{tab: decaying coef i} for a detailed comparison.

Inference for a linear combination of regression coefficients in high-dimensional linear model has been investigated in \cite{cai2017confidence, athey2018approximate,zhu2018linear,cai2019individualized}. However, these methods cannot be directly applied to make inference for the case probability in the logistic model. 
Our proposed linearization technique is useful in generalizing the inference methods for linear models to logistic models. The connection established by the linearization is also useful for simplifying the sufficient conditions for estimating the precision matrix or the inverse Hessian matrix. Specifically, the established results in the current paper impose no sparsity condition on the precision matrix or the inverse Hessian matrix, where such a requirement has typically been imposed in theoretical justifications on inference for individual regression coefficients in the logistic regression setting \citep{van2014asymptotically, ning2017general, ma2018global}. More detailed comparisons are provided in Section \ref{sec: comparison}. 

In Section \ref{sec: sim}, we provide detailed numerical comparisons to the inference methods by \citet{van2014asymptotically, ma2018global,zhu2018linear,tripuraneni}. 

\citet{belloni2014high,farrell2015robust, chernozhukov2018double} studied inference for treatment effects in high-dimensional regression models while the current paper focuses on inference for a different quantity, the case probability. 
\citet{sur2019modern,sur2019likelihood} studied inference in high-dimensional logistic regression and focused on the regime where the dimension $p$ is a fraction of the sample size $n$. The current paper considered the regime allowing for the dimension $p$ being much larger than the sample size $n$ with imposing additional sparsity conditions on $\beta$.

Another related work is the iterated re-weighted least squares (IRLS) \citep{fox2015applied}, which is the standard technique used to maximize the likelihood of the logistic model. The weighting is used in IRLS to facilitate the optimization problem. In contrast, the weighting used in the current paper is to facilitate the bias-correction for the statistical inference.

\subsection{Notation}
\label{sec:notation}

For a matrix $X\in \R^{n\times p}$, $X_{i\cdot}$,  $X_{\cdot j}$ and $X_{i,j}$  denote respectively the $i$-th row,  $j$-th column, $(i,j)$ entry of the matrix $X$. $X_{i,-j}$ denotes the sub-row of $X_{i\cdot}$ excluding the $j$-th entry. Let $[p]=\{1,2,\cdots,p\}$. For a subset $J\subset[p]$ and a vector $x\in \R^{p}$, $x_{J}$ is the subvector of $x$ with indices in $J$ and $x_{-J}$ is the subvector with indices in $J^{c}$. For a vector $x\in \R^{p}$, the $\ell_q$ norm of $x$ is defined as $\|x\|_{q}=\left(\sum_{i=1}^{p}|x_i|^q\right)^{\frac{1}{q}}$ for $q > 0$ with $\|x\|_0 $ denoting the cardinality of the support of $x$ and $\|x\|_{\infty}=\max_{1\leq j \leq p}|x_j|$. We use $e_i$ to denote the $i$-th standard basis vector in $\R^p$. We use $\max|X_{i,j}|$ as a shorthand for $\max_{1\leq i\leq n, 1\leq j \leq p}|X_{i, j}|$. For a symmetric matrix $A$, $\lambda_{\min}\left(A\right)$ and $\lambda_{\max}\left(A\right)$  denote respectively the smallest and largest eigenvalues of $A$.  We use $c$ and $C$ to denote generic positive constants that may vary from place to place. For two positive sequences $a_n$ and $b_n$,  $a_n \lesssim b_n$ means $a_n \leq C b_n$ for all $n$ and $a_n \gtrsim b_n $ if $b_n\lesssim  a_n$ and $a_n \asymp b_n $ if $a_n \lesssim b_n$ and $b_n \lesssim a_n$, and $a_n \ll b_n$ if $\limsup_{n\rightarrow\infty} {a_n}/{b_n}=0$.

\section{Methodology}
\label{sec: method}
We describe the proposed  method for the case probability under the high-dimensional logistic model \eqref{eq: logistic model}. In Section \ref{sec: initial estimator}, we review the penalized  maximum likelihood  estimation of $\beta$ and highlight the challenges of inference for the case probability. Then we introduce the linearization technique in Section \ref{sec: method logistic} and the variance enhancement technique in Section \ref{sec: uniform}. In Section \ref{sec: CI construction}, we construct a point estimator and a confidence interval for the case probability and conduct hypothesis testing related to outcome labelling. In Section \ref{sec: comparison}, we compare with the existing estimators \citep{van2014asymptotically,ma2018global,zhu2018linear,tripuraneni,bickel1975one}.

\subsection{Challenges Underlying Inference for the Case Probability}
\label{sec: initial estimator}
The negative log-likelihood function for the data $\{(X_{i\cdot},y_i)\}_{1\leq i\leq n}$ under the logistic regression model \eqref{eq: logistic model} is written as 
$
\ell(\beta)=\sum_{i=1}^{n}\left[\log\left(1+\exp\left(X_{i\cdot}^{\intercal}\beta\right)\right)-y_i\cdot\left(X_{i\cdot}^{\intercal}\beta\right)\right].
$
The penalized log-likelihood estimator $\widehat{\beta}$ is defined as \citep{buhlmann2011statistics},
\begin{equation}
\widehat{\beta}=\argmin_{\beta}\ell(\beta)+\lambda \|\beta\|_1,
\label{eq: penalized estimator}
\end{equation}
with the tuning parameter $\lambda\asymp \sqrt{{\log p}/{n}}$.  It has been shown that $\widehat{\beta}$ satisfies certain nice estimation accuracy and variable selection properties. However, the plug-in estimator $h(\xnew^{\intercal}\widehat{\beta})$ cannot be directly used for confidence interval construction and hypothesis testing, because its bias can be as large as its variance as demonstrated in later simulation studies; see Table \ref{tab: Setting 1,2 ii} in the supplement for the numerical illustration.

Our proposed method is built on the idea of correcting the bias of the plug-in estimator $\xnew^{\intercal}\widehat{\beta}$ and then applying the $h$ function to estimate the case probability. We conduct the bias correction through estimating the error of the plug-in estimator  $\xnew^{\intercal}\widehat{\beta}-\xnew^{\intercal}\beta=\xnew^{\intercal}(\widehat{\beta}-\beta).$  
Before proposing the method, we review the existing bias-correction idea in high-dimensional linear and logistic models \citep{van2014asymptotically,javanmard2014confidence,zhang2014confidence}. 
In particular, a bias-corrected estimator of $\beta_j$ can be constructed as 
\begin{equation}
\widehat{\beta}_j+\widehat{u}^{\intercal}\frac{1}{n}\sum_{i=1}^{n}X_{i\cdot}(y_i-h(X_{i\cdot}^{\intercal}\widehat{\beta}))
\label{eq: generic form}
\end{equation}
where $\widehat{u}\in \R^{p}$ is the projection direction used for correcting the bias of $\widehat{\beta}_j$. Define the error $\epsilon_i=y_i-h(X_{i\cdot}^{\intercal}\beta)$ for $1\leq i \leq n$. We apply the Taylor expansion of the $h$ function and obtain
\begin{equation*}
\begin{aligned}
y_i - h(X_{i\cdot}^{\intercal}\widehat{\beta}) = h(X_{i\cdot}^{\intercal}\beta) - h(X_{i\cdot}^{\intercal}\widehat{\beta}) + \epsilon_i = h^{\prime}(X_{i\cdot}^{\intercal}\widehat{\beta})X_{i\cdot}^{\intercal}(\beta - \widehat{\beta}) + R_i + \epsilon_i
\end{aligned}
\end{equation*}
with the approximation error $R_i = \int_{0}^{1} (1-t)h''(X_{i\cdot}^{\intercal}\widehat{\beta}+tX^{\intercal}_{i\cdot}(\beta-\widehat{\beta})) dt \cdot (X^{\intercal}_{i\cdot}(\widehat{\beta}-\beta))^2$.
Since $h'(x)=h(x)(1-h(x))$ for any $x\in \R,$ we simplify the above expression as \begin{equation}
y_i-h(X_{i\cdot}^{\intercal}\widehat{\beta})=h(X_{i\cdot}^{\intercal}\widehat{\beta})(1-h(X_{i\cdot}^{\intercal}\widehat{\beta}))[X^{\intercal}_{i\cdot}(\beta-\widehat{\beta})+\Delta_i]+\epsilon_i \quad \text{with} \quad \Delta_i= {R_i}/{h'(X_{i\cdot}^{\intercal}\widehat{\beta})}.
\label{eq: key decomposition}
\end{equation}
By multiplying both sides of \eqref{eq: key decomposition} by $X_{i}$ and summing over $i$, we obtain
\begin{equation}
\frac{1}{n}\sum_{i=1}^{n}X_{i\cdot}(y_i-h(X_{i\cdot}^{\intercal}\widehat{\beta}))=\widehat{H}(\widehat{\beta})(\beta-\widehat{\beta})+\frac{1}{n}\sum_{i=1}^{n}\epsilon_i X_{i\cdot}+\frac{1}{n}\sum_{i=1}^{n}h(X_{i\cdot}^{\intercal}\widehat{\beta})(1-h(X_{i\cdot}^{\intercal}\widehat{\beta}))\Delta_i X_{i\cdot},
\label{eq: equal weight}
\end{equation}
where $
\widehat{H}({\beta})=\frac{1}{n}\sum_{i=1}^{n}h(X_{i\cdot}^{\intercal}{\beta})(1-h(X_{i\cdot}^{\intercal}{\beta}))X_{i\cdot}X^{\intercal}_{i\cdot}$ is the Hessian matrix of the negative log-likelihood $\ell(\beta)$.

To conduct the bias-correction, \cite{van2014asymptotically}  construct the projection direction $\widehat{u}\in \R^{p}$ in \eqref{eq: generic form} such that $\widehat{H}(\widehat{\beta}) \widehat{u}\approx e_j$ and hence $$\widehat{u}^{\intercal}\frac{1}{n}\sum_{i=1}^{n}X_{i\cdot}(y_i-h(X_{i\cdot}^{\intercal}\widehat{\beta}))\approx \widehat{u}^{\intercal} \widehat{H}(\widehat{\beta})(\beta-\widehat{\beta})\approx \beta_j-\widehat{\beta}_j.$$ Such an approximation has been shown to be accurate by assuming a sparse $[\E\widehat{H}({\beta})]^{-1}e_j$ \citep{van2014asymptotically}.  
 However, $[\E \widehat{H}({\beta})]^{-1}\xnew$ can be an arbitrarily dense vector and hence it is challenging to accurately estimate $[\E\widehat{H}({\beta})]^{-1}\xnew$ and generalize the bias-correction procedure in \cite{van2014asymptotically}. Specifically, $[\E \widehat{H}({\beta})]^{-1}$ may be dense for the following two reasons: (1) the columns of $[\E \widehat{H}({\beta})]^{-1}$ are dense;
(2) $\xnew$ is a dense vector.

In the following two sections, we develop new techniques, which can effectively correct the bias for an arbitrary loading $\xnew\in \R^{p}$ in the high-dimensional logistic regression.

\subsection{Linearization: Connecting Logistic to Linear}
\label{sec: method logistic}
We introduce a linearization technique to simplify the Hessian matrix.  Instead of averaging with equal weights as in \eqref{eq: equal weight}, we introduce the following re-weighted summation, 
\begin{equation*}
\frac{1}{n}\sum_{i=1}^{n}\underbrace{[h(X_{i\cdot}^{\intercal}\widehat{\beta})(1-h(X_{i\cdot}^{\intercal}\widehat{\beta}))]^{-1}}_{\text{weight for}\; i-th \;\text{observation}}X_{i\cdot}(y_i-h(X_{i\cdot}^{\intercal}\widehat{\beta})).
\end{equation*}
In contrast to \eqref{eq: equal weight}, the above re-weighted summation has the following decomposition:
\begin{equation*}
\frac{1}{n}{\sum_{i=1}^{n}[h(X_{i\cdot}^{\intercal}\widehat{\beta})(1-h(X_{i\cdot}^{\intercal}\widehat{\beta}))]^{-1}\epsilon_iX_{i\cdot}+\widehat{\Sigma}(\beta-\widehat{\beta})+\frac{1}{n}\sum_{i=1}^{n}\Delta_i X_{i\cdot}}, \; \text{with}\;\widehat{\Sigma}=\frac{1}{n}\sum_{i=1}^{n}X_{i\cdot}X^{\intercal}_{i\cdot}.
\label{eq: after linearization}
\end{equation*}
 The main advantage of the re-weighting step is that the second component $\widehat{\Sigma}(\beta-\widehat{\beta})$ on the right hand side is multiplication of the sample covariance matrix $\widehat{\Sigma}$ and the vector difference $\widehat{\beta}-\beta$. In contrast to \eqref{eq: equal weight}, it is sufficient to invert $\widehat{\Sigma}$, instead of the more complicated Hessian matrix $\widehat{H}(\widehat{\beta}).$ Since the main purpose of this re-weighting step is to match the re-weighted Hessian matrix to that of the least square loss in the linear models, we refer to this as the ``{Linearization}" technique. We shall point out that, although {linearization} connects the logistic model to the linear model, it also poses challenges in the theoretical justification of the proposed method. The corresponding technical challenge will be addressed in Section \ref{sec: challenges} with suitable empirical process techniques.

\subsection{Variance Enhancement: Uniform Procedure for \texorpdfstring{$\xnew$}{xnew}}
\label{sec: uniform}
We apply the linearization technique and correct the bias of the plug-in estimator $\xnew^{\intercal}\widehat{\beta}$ as, \begin{equation}
\widehat{\xnew^{\intercal}{\beta}}=\xnew^{\intercal}\widehat{\beta}+\widehat{u}^{\intercal}\frac{1}{n}\sum_{i=1}^{n}[h(X_{i\cdot}^{\intercal}\widehat{\beta})(1-h(X_{i\cdot}^{\intercal}\widehat{\beta}))]^{-1}X_{i\cdot}(y_i-h(X_{i\cdot}^{\intercal}\widehat{\beta})),
\label{eq: correction}
\end{equation}
with $\widehat{u}\in \R^{p}$ denoting a projection direction to be constructed. To see how to construct $\widehat{u}$, we decompose the estimation error  $\widehat{\xnew^{\intercal}{\beta}}-\xnew^{\intercal}\beta$ as 
\begin{equation}
\frac{1}{n}\sum_{i=1}^{n}[h(X_{i\cdot}^{\intercal}\widehat{\beta})(1-h(X_{i\cdot}^{\intercal}\widehat{\beta}))]^{-1}\epsilon_i\widehat{u}^{\intercal}X_{i\cdot}+(\widehat{\Sigma}\widehat{u}-\xnew)^{\intercal}(\beta-\widehat{\beta})+\frac{1}{n}\sum_{i=1}^{n}\Delta_i \widehat{u}^{\intercal} X_{i\cdot},
\label{eq: decomposition}
\end{equation}
and notice that all three terms depend on our constructed projection direction $\widehat{u}\in \R^{p}.$

Motivated by the decomposition in \eqref{eq: decomposition}, we construct $\widehat{u} \in \R^{p}$ as the solution to the following optimization problem, 
\begin{align}
\widehat{u}=\;\argmin_{u\in \RR^{p}} u^{\intercal}\widehat{\Sigma}u \quad
\text{subject to}&\; \|\widehat{\Sigma}u-\xnew\|_{\infty}\leq  \|\xnew\|_2 \lambda_{n} \label{eq: constraint 1} \\
&\;|\xnew^{\intercal}\widehat{\Sigma}u-\|\xnew\|_2^2 |\leq \|\xnew\|_2^2\lambda_{n} \label{eq: constraint 2}\\
&\;\|X u\|_{\infty} \leq \|\xnew\|_2 \tau_n  \label{eq: constraint 3}
\end{align}
where $\lambda_{n}\asymp \left({\log p}/{n}\right)^{1/2}$ and $\tau_n\asymp (\log n)^{1/2}.$ The details on implementing the above algorithm with tuning parameters selection are presented in Section \ref{sec: implementation}.

We now provide some explanations on the construction of $\widehat{u}$ in \eqref{eq: constraint 1} to \eqref{eq: constraint 3} by connecting it to the error decomposition \eqref{eq: decomposition}. The objective function in \eqref{eq: constraint 1} scaled by $1/n$, $u^{\intercal}\widehat{\Sigma}u/n$, is of the same order of magnitude as the variance of the first term in the error decomposition \eqref{eq: decomposition}. The constraints \eqref{eq: constraint 1} and \eqref{eq: constraint 3} are introduced to control the second and third terms in the error decomposition \eqref{eq: decomposition}, respectively. Hence, the objective function, together with the constraints \eqref{eq: constraint 1} and \eqref{eq: constraint 3}, ensure that the error $\widehat{\xnew^{\intercal}{\beta}}-\xnew^{\intercal}\beta$ is controlled to be small. Such an optimization idea has been proposed in the linear model \citep{javanmard2014confidence,zhang2014confidence} and is shown to be effective when $\xnew=e_j$ \citep{javanmard2014confidence,zhang2014confidence}, a sparse $\xnew$ \citep{cai2017confidence} and $\xnew$ with a bounded $\ell_2$ norm \citep{athey2018approximate}. We shall emphasize that such an idea cannot be extended to general loadings $\xnew$ since the variance level of $\frac{1}{n}\sum_{i=1}^{n}[h(X_{i\cdot}^{\intercal}\widehat{\beta})(1-h(X_{i\cdot}^{\intercal}\widehat{\beta}))]^{-1}\epsilon_i\widehat{u}^{\intercal}X_{i\cdot}$ is not guaranteed to dominate the other two bias terms in \eqref{eq: decomposition}, without the additional constraint  \eqref{eq: constraint 2}.  \cite{cai2019individualized} has presented examples where such bias correction method will fail; see Proposition 2 of \cite{cai2019individualized}.

To resolve this, we introduce the additional constraint \eqref{eq: constraint 2} such that the variance component  $\frac{1}{n}\sum_{i=1}^{n}[h(X_{i\cdot}^{\intercal}\widehat{\beta})(1-h(X_{i\cdot}^{\intercal}\widehat{\beta}))]^{-1}\epsilon_i\widehat{u}^{\intercal}X_{i\cdot}$ is the dominating term in the error decomposition \eqref{eq: decomposition}, for any high-dimensional vector $\xnew\in \R^{p}$. In particular, this constraint enhances the variance component in the error decomposition \eqref{eq: decomposition} and hence we refer to the above construction of projection direction $\widehat{u}$ in \eqref{eq: constraint 1} to \eqref{eq: constraint 3} as ``{variance enhancement}".

\begin{remark}\rm
We have shown in Theorem \ref{thm: limiting distribution} that, with a high probability,  $u^{*}=\Sigma^{-1} \xnew$ belongs to  the feasible set  defined by \eqref{eq: constraint 1}, \eqref{eq: constraint 2} and \eqref{eq: constraint 3}. Although  $\widehat{u}$ defined by the optimization problem \eqref{eq: constraint 1} to \eqref{eq: constraint 3} is targeting at $u^{*}=\Sigma^{-1} \xnew$, the asymptotic normality of the proposed LiVE estimator defined in \eqref{eq: correction} does not rely on $\widehat{u}$ to be an accurate estimator of $u^{*}.$ This explains why the proposed bias-corrected estimator does not require any sparsity condition on $\Sigma^{-1}$, $\xnew$ or $\Sigma^{-1}\xnew.$ See Theorem \ref{thm: limiting distribution} and its proof for details.
\end{remark}

\begin{remark} \rm
In the high-dimensional linear model, the variance enhancement idea has been proposed in constructing the bias corrected estimator for $\xnew^{\intercal}\beta$  \citep{cai2019individualized}. However, the method developed for linear models in \cite{cai2019individualized} cannot be directly applied to the inference problem for the case probabilities due to the complexity of the Hession matrix, as highlighted in Section \ref{sec: method logistic}. A valid inference procedure for the case probability depends on both Linearization and Variance Enhancement techniques. \end{remark}

\begin{remark}\rm
 The idea of adding the constraint $\|X u\|_{\infty} \leq \|\xnew\|_2 \tau_n$ was first introduced in \cite{javanmard2014confidence} to establish the asymptotic normality for the non-Gaussian error in the linear model. In our analysis, this additional constraint is not just introduced to establish the asymptotic normality for the  non-Gaussian error $\epsilon_i$, but also facilitates the empirical process proof. The range of values for $\tau_n$ is also different from that in \cite{javanmard2014confidence}, where equation (54) of \cite{javanmard2014confidence} has $\|\xnew\|_2=1$ and $\tau_n \asymp n^{\delta_0}$ with $1/4<\delta_0<1/2$ while $\tau_n$ in our paper is required to satisfy $\left(\log n\right)^{1/2}\lesssim \tau_{n}\ll n^{1/2}$. We have set $\tau_n\asymp \left(\log n\right)^{1/2}$ throughout the rest of the paper.
\end{remark} 
 
\subsection{LiVE: Inference for Case Probabilities}
\label{sec: CI construction}
We propose to estimate $\xnew^{\intercal}\beta$ by $\widehat{\xnew^{\intercal}\beta}$ as defined in \eqref{eq: correction}, with the initial estimator $\widehat{\beta}$ defined in \eqref{eq: penalized estimator} and the projection direction $\widehat{u}$ defined in \eqref{eq: constraint 1} to \eqref{eq: constraint 3}. Subsequently, we estimate the case probability $\PP(y_i=1|X_{i\cdot}=\xnew)$ by 
\begin{equation}
\widehat{\PP}(y_{i}=1|X_{i\cdot}=\xnew)=h(\widehat{\xnew^{\intercal}\beta})
\label{eq: proposed estimator}
\end{equation}

From the above construction, the asymptotic variance of $\widehat{\xnew^{\intercal}{\beta}}$ can be estimated by
\begin{equation*}
\widehat{\rm V}=\widehat{u}^{\intercal}\left[\frac{1}{n^2}\sum_{i=1}^{n}[h(X_{i\cdot}^{\intercal}\widehat{\beta})(1-h(X_{i\cdot}^{\intercal}\widehat{\beta}))]^{-1}X_{i\cdot}X_{i\cdot}^{\intercal}\right]\widehat{u}.
\label{eq: estimated asymptotic variance}
\end{equation*}
We construct the confidence interval for the case probability $\PP(y_i=1|X_{i\cdot}=\xnew)$  as follows:
\begin{equation}
{\rm CI}_{\alpha}(\xnew)=\left[h\left(\widehat{\xnew^{\intercal} \beta}-z_{\alpha/2}\widehat{\rm V}^{1/2}\right),h\left(\widehat{\xnew^{\intercal} \beta}+z_{\alpha/2}\widehat{\rm V}^{1/2}\right)\right],
\label{eq: CI}
\end{equation}
where $z_{\alpha/2}$ is the upper $\alpha/2$-quantile of the standard normal distribution. We conduct the following hypothesis testing related to outcome labeling \eqref{eq: null hypothesis}
\begin{equation}
{\phi}_{\alpha}(\xnew)=\mathbf{1}\left(\widehat{\xnew^{\intercal} \beta}-z_{\alpha}\widehat{\rm V}^{1/2}\geq 0\right).
\label{eq: testing}
\end{equation}
Here, the testing procedure \eqref{eq: testing} will label the observation as a case if $\widehat{\xnew^{\intercal}\beta}$ is above $z_{\alpha}\widehat{\rm V}^{1/2}$; as a control, otherwise. {If the goal is to test the null hypothesis $H_0: h(\xnew^{\intercal}\beta)<c_*$ for $c_*\in (0,1),$ we generalize \eqref{eq: testing} to 
$
{\phi}^{c_*}_{\alpha}(\xnew)=\mathbf{1}\left(\widehat{\xnew^{\intercal} \beta}-z_{\alpha}\widehat{\rm V}^{1/2}\geq h^{-1}(c_*)\right),$
where $h^{-1}$ is the inverse function of $h$ defined in \eqref{eq: logistic model}.}

\subsection{Comparison to Other Estimators}
\label{sec: comparison}

In the following, we discuss the difference between the proposed LiVE method and related methods. A detailed numerical comparison with the methods in \citet{van2014asymptotically,ma2018global,zhu2018linear,tripuraneni} is provided in Section \ref{sec: sim}.

The main distinction is that the existing literature focused on single regression coefficients, instead of the case probability. We shall use $\widetilde{\beta}_j$ to denote the existing coordinate-wise debiased estimator of $\beta_j$ for $1\leq j\leq p$ \citep{van2014asymptotically,ma2018global}. 
The computation cost of the plug-in estimator $\xnew^{\intercal}\widetilde{\beta}$ is much higher than our proposed method, as the proposed method targets at $\xnew^{\intercal}\beta$ directly and requires construction of one projection direction as in \eqref{eq: constraint 1} to \eqref{eq: constraint 3}. In contrast, the plug-in debiased estimator $\xnew^{\intercal}\widetilde{\beta}$ \citep{van2014asymptotically,ma2018global} requires construction of $p$ projection directions. See Tables \ref{tab: Setting 1 i} and \ref{tab: decaying coef i} for a detailed comparison of computation times. We also mention that there exist technical difficulties of controlling the bias of the plug-in debiased estimator; see Section \ref{sec: technical diff} in the supplement.

The re-weighting idea has been proposed in \cite{ma2018global} for inference for the single regression coefficient $\beta_j$ in the high-dimensional logistic model. However, it is not straightforward to extend it to make inference for a general linear combination of regression coefficients. A brief summary of the method can be found in Section \ref{sec: Ma review} in the supplement. Moreover, the analysis in \cite{ma2018global} requires sample splitting, where half of the data was used for constructing an initial estimator of the regression coefficient vector and the other half was used for bias correction. But our empirical process results in Section \ref{sec: challenges} can carry out the analysis for an arbitrary combination of the regression vector and bypass the sample splitting related to the re-weighting step.

The transformation methods have been proposed in \cite{zhu2018linear,tripuraneni} for high-dimensional linear models. We now extend this method to the logistic model.
 With $H_{*}=x_{*} x_{*}^{\intercal} /\left\|x_{*}\right\|_{2}^{2},$ we have 
   $$
    X_{i\cdot}^{\intercal}\beta = X_{i\cdot}^{\intercal}H_{*}\beta + X_{i\cdot}^{\intercal}\left({\rm I} - H_{*}\right)\beta=\frac{X_{i\cdot}^{\intercal} x_{*}}{\left\|x_{*}\right\|_{2}^{2}}\left(x_{*}^{\intercal} \beta\right)+\left(X_{i\cdot}^{\intercal} U\right)\left(U^{\intercal} \beta\right)
    $$
    where ${\rm I}-H_{*}=U U^{\intercal}$ and $U^{\intercal} U={\rm I}_{p-1}$. We construct $U\in \R^{p\times (p-1)}$ with its columns denoting the eigenvectors of ${\rm I} - H_*$ corresponding to its non-zero eigenvalues. This defines an equivalent logistic regression model
    \begin{equation}
    \mathbb{P}(y_i = 1|\check{X}_{i\cdot}) = h(\check{X}_{i\cdot}^{\intercal}\eta) \quad i = 1,2, \cdots, n
    \label{eq: transformed model}
    \end{equation}
    with $\eta_{1}=x_{*}^{\intercal} \beta$, $\eta_{-1} = U^{\intercal}\beta$ and $\check{X}_{i,1}=X_{i,\cdot}^{\intercal} x_{*} /\left\|x_{*}\right\|_{2}^{2}, \check{X}_{i,-1}= X_{i,\cdot}^{\intercal}U.$ 
    We apply our proposed method to the newly defined logistic regression model and obtain the bias-corrected estimator $\widehat{\eta}_1$ and its variance estimator $\widehat{\mathrm{V}}_{U}.$
The CI and the testing procedure related to $\xnew^{\intercal}\beta$ are given by
\begin{equation}
{\rm CI}_{\alpha}(\xnew)=\left[h(\widehat{\eta}_1 - z_{\alpha/2}\widehat{\rm V}_U^{1/2}), h(\widehat{\eta}_1 + z_{\alpha/2}\widehat{\rm V}_U^{1/2})\right], \; {\phi}_{\alpha}(\xnew)=\mathbf{1}\left(\widehat{\eta}_1-z_{\alpha}\widehat{V}_U^{1/2}\geq 0 \right).
\label{eq: transformation}
\end{equation}
 In Section \ref{sec: U Method}, we provide a comparison between \eqref{eq: transformation} and our proposed method. We observe that the transformation method suffers from a larger bias and does not achieve the desired coverage when $\xnew$ is relatively dense; see Table \ref{tab: U_method p=1000} for details.

Furthermore, it can be challenging to  analyze the transformed model \eqref{eq: transformed model} since $\eta$ is not necessarily sparse (even if $\beta$ is sparse). To guarantee a sparse $\eta$, certain special structures (e.g. sparsity) need to be imposed on the loading $\xnew.$ More detailed discussions on \cite{tripuraneni} are provided in Section \ref{sec: tripuraneni} in the supplement.

Our proposed debiased estimator is closely related to the one-step estimator in \cite{bickel1975one}. Our bias correction step corresponds to the following estimating equation \begin{equation}\E\psi(y_i, X_{i\cdot},\beta)=0 \quad \text{with}\quad \psi(y_i, X_{i\cdot}, \beta)=\frac{X_{i\cdot}(y_i-h(X_{i\cdot}^{\intercal}\beta))}{h(X_{i\cdot}^{\intercal}\beta)(1-h(X_{i\cdot}^{\intercal}\beta))},
\label{eq: est equation}
\end{equation}
which is a weighted version of the estimating equation $\E X_{i\cdot}(y_i-h(X_{i\cdot}^{\intercal}\beta))=0.$ We shall point out that the weight in \eqref{eq: est equation} depends on both the data and the unknown parameter $\beta.$

In comparison to \cite{bickel1975one}, we propose our bias-correction step by taking a Taylor expansion of $X_{i\cdot}(y_i-h(X_{i\cdot}^{\intercal}\beta))$ instead of $\psi(y_i, X_{i\cdot},\beta);$ see equation \eqref{eq: key decomposition}. Note that the derivative of $\psi(y_i, X_{i\cdot},\beta)$ with respect to $\beta$ has a complicated form due to the fact that the weight in \eqref{eq: est equation} also involves $\beta$. Hence, it is not straightforward to express our weighted bias-corrected as a one-step estimator since the weight in the estimating equation \eqref{eq: est equation} also depends on $\beta.$

\section{Theoretical Justification}
\label{sec: theory}

\subsection{Model Conditions and Initial Estimators}
\label{sec: penalized logistic rate}
We introduce the following modeling assumptions to facilitate theoretical analysis. 
\begin{enumerate}
\item[(A1)]  The rows $\{X_{i\cdot}\}_{1\leq i\leq n} $ are i.i.d. $p$-dimensional {Sub-gaussian} random vectors with $\Sigma=\E (X_{i\cdot} X_{i\cdot}^{\intercal})$ where $\Sigma$ satisfies $c_0\leq \lambda_{\min}\left(\Sigma\right) \leq \lambda_{\max}\left(\Sigma\right) \leq C_0$ for some positive constants $C_0\geq c_0>0$; The high-dimensional vector $\beta$ is assumed to be of $k$ non-zero entries.  
\item[(A2)]  With probability larger than $1-p^{-c}$, $\min\{h(X_{i\cdot}^{\intercal}{\beta}),1-h(X_{i\cdot}^{\intercal}{\beta})\}\geq c_{\min}$
for $1\leq i\leq n$ and some small positive constant $c_{\min}\in(0,1)$.
\end{enumerate}
Condition {\rm (A1)} imposes the tail condition for the high-dimensional covariates $X_{i\cdot}$ and assumes that the population second order moment matrix is invertible. Condition ${\rm (A2)}$ is imposed such that the case probability is uniformly bounded away from $0$ and $1$ by a small positive constant $c_{\min}$. Condition (A2) requires $X_{i\cdot}^{\intercal}\beta$ to be bounded for all $1\leq i\leq n$ with a high probability. Such a condition has been commonly made in 
in analyzing high-dimensional logistic models \citep{athey2018approximate,van2014asymptotically,ma2018global, ning2017general}. For example, see condition (iv) of Theorem 3.3 of \cite{van2014asymptotically} and the overlap assumption (Assumption 6) in \cite{athey2018approximate}. {{As a remark, Condition (A2) may be stringent for certain applications; we test the robustness of our proposed method to the violation of (A2) in Section \ref{sec: A2}.}}

The following proposition states the theoretical properties of  the penalized  estimator $\widehat{\beta}$ in \eqref{eq: penalized estimator}, which have been established in
\cite{negahban2009unified,huang2012estimation}.

\begin{Proposition}
Suppose that Conditions $\rm (A1)$ and $\rm (A2)$ hold and there exists a positive constant $c>0$ such that  $\max_{i,j}\left|X_{ij}\right| k\lambda_0 \leq c$ with $\lambda_0=\left\|\frac{1}{n}\sum_{i=1}^{n}\epsilon_i X_i\right\|_{\infty}$. For any positive constant $\delta_0>0$ and the proposed estimator $\widehat{\beta}$ in \eqref{eq: penalized estimator} with $\lambda=(1+\delta_0)\lambda_0$, with probability greater than $1-p^{-c}-\exp(-cn),$
\begin{equation}
\|\widehat{\beta}_{S^{c}}-\beta_{S^{c}}\|_1 \leq (2/\delta_0+1)\|\widehat{\beta}_{S}-\beta_{S}\|_1 \quad \text{and}\quad \|\widehat{\beta}-\beta\|_1\leq C k \lambda_0
 \label{eq: est property}
\end{equation}
where $S$ denotes the support of $\beta$ and $C>0$ is a positive constant.
\label{prop: lasso convergence}
\end{Proposition}
We will choose $\lambda_0$ at the scale of $\left(\log p/n\right)^{1/2}$ and then Proposition \ref{prop: lasso convergence} shows that the initial estimator $\widehat{\beta}$ satisfies the following property: 
\begin{enumerate}
\item[(B)] With probability greater than $1-p^{-c}-\exp(-cn)$ for some constant $c>0$, 
\begin{equation*}
\|\widehat{\beta}-\beta\|_1\leq C k \left({\log p}/{n}\right)^{1/2} \quad \text{and} \quad \|\widehat{\beta}_{S^{c}}-\beta_{S^{c}}\|_1 \leq C_0\|\widehat{\beta}_{S}-\beta_{S}\|_1
\end{equation*}
where $S$ denotes the support of $\beta$ and $C>0$ and $C_0>0$ are positive constants.
\end{enumerate}
The asymptotic normality established in next subsection will hold for any initial estimator satisfying condition ${\rm (B)}$, including our initial estimator defined in \eqref{eq: penalized estimator}. 

\subsection{Asymptotic Normality and Statistical Inference}
\label{sec: limiting distribution}
We now establish the limiting distribution for the proposed point estimator $\widehat{\xnew^{\intercal}{\beta}}$. 
\begin{Theorem}
Suppose that Conditions {\rm (A1)} and {\rm (A2)} hold, $\tau_{n}\asymp (\log n)^{1/2}$ defined in \eqref{eq: constraint 3} satisfies $\tau_{n}{k \log p}/{\sqrt{n}}\rightarrow 0$. Then for any initial estimator $\widehat{\beta}$ satisfying condition {\rm (B)} and any constant $0<\alpha<1$, 
$$\PP\left[{{\rm V}^{-1/2}}\left(\widehat{\xnew^{\intercal}\beta}-\xnew^{\intercal}\beta\right)\geq z_{\alpha}\right]\rightarrow \alpha$$
where 
\begin{equation}
{\rm V}=\widehat{u}^{\intercal}\left[\frac{1}{n^2}\sum_{i=1}^{n}[h(X_{i\cdot}^{\intercal}{\beta})(1-h(X_{i\cdot}^{\intercal}{\beta}))]^{-1}X_{i\cdot}X_{i\cdot}^{\intercal}\right]\widehat{u}.
\label{eq: asymptotic variance}
\end{equation}
With probability greater than $1-p^{-c}-\exp(-cn)$,
\begin{equation}
c_0{\|\xnew\|_2}/{n^{1/2}}\leq {\rm V}^{1/2}\leq C_0{\|\xnew\|_2}/{n^{1/2}},
\label{eq: dominating variance}
\end{equation}
for some positive constants $c,c_0,C_0>0$.
\label{thm: limiting distribution}
\end{Theorem}

Theorem \ref{thm: limiting distribution} can be used to justify the validity of the proposed confidence interval.

\begin{Proposition}
Under the same conditions as in Theorem \ref{thm: limiting distribution}, the confidence interval ${\rm CI}_{\alpha}(\xnew)$ proposed in \eqref{eq: CI} satisfies
$
\liminf_{n\rightarrow \infty}\PP\left[{\PP}(y_{i}=1|X_{i\cdot}=\xnew)\in {\rm CI}_{\alpha}(\xnew)\right]\geq 1-\alpha, 
$
and 
$$
\limsup_{n\rightarrow \infty}\PP\left({\mathbf L}({\rm CI}_{\alpha}(\xnew))\geq (1+\delta) \left(\rho^2 \rm V\right)^{1/2}\right)=0,
$$
where ${\mathbf L}({\rm CI}_{\alpha}(\xnew))$ denotes the length of the confidence interval ${\rm CI}_{\alpha}(\xnew)$,  $\delta>0$ is any positive constant, ${\rm V}$ is defined in \eqref{eq: asymptotic variance} and $\rho=h(\xnew^{\intercal}\beta)(1-h(\xnew^{\intercal}\beta)).$
\label{prop: coverage}
\end{Proposition}

A few remarks are in order for Theorem \ref{thm: limiting distribution} and Proposition \ref{prop: coverage}. Firstly, the asymptotic normality in Theorem \ref{thm: limiting distribution} is established without imposing any condition on the high-dimensional vector $\xnew\in \R^{p}.$ The variance enhancement construction of the projection direction $\widehat{u}$ in \eqref{eq: constraint 1} to \eqref{eq: constraint 3} is crucial for establishing such a uniform result over any $\xnew\in \R^{p}.$ Specifically, with the additional constraint \eqref{eq: constraint 2}, we can establish the lower bound of the asymptotic variance in \eqref{eq: dominating variance}, which guarantees that the variance component of \eqref{eq: decomposition} dominates the remaining bias.

Secondly, to establish the asymptotic normality result, we do not impose any sparsity condition on the precision matrix $\Sigma^{-1}$. This has weakened sparsity conditions imposed on the inverse of the Hessian matrix $\E \left(h(X_{i\cdot}^{\intercal} \beta)\left(1-h(X_{i\cdot}^{\intercal} \beta)\right) X_{i\cdot} X_{i\cdot}^{\intercal}\right)$ or the precision matrix $\Sigma^{-1}$   \citep{van2014asymptotically, ning2017general,ma2018global}. 
Thirdly, with $\tau_n \asymp \left(\log n\right)^{1/2}$, the required sparsity condition on $\beta$ is $k \ll {n^{1/2}}/{[\log p \left(\log n\right)^{1/2}]}$. Such sparsity conditions are imposed for confidence interval construction for both high-dimensional linear models and logistic models \citep{javanmard2014confidence,zhang2014confidence,van2014asymptotically, ning2017general,ma2018global}. Regarding confidence interval construction for $\beta_j$ in high-dimensional linear models, \cite{cai2017confidence} have shown that the ultra-sparse condition $k\ll {n^{1/2}}/{\log p}$ is necessary and sufficient for constructing a confidence interval of length $1/\sqrt{n}.$ {{Recently, \cite{cai2021logistic} extended this result to inference for single regression coefficients in the high-dimensional logistic regression.}}

Theorem \ref{thm: limiting distribution} also justifies the validity of the proposed testing procedure.
To study the testing procedure, we introduce the following parameter space for $\theta=(\beta,\Sigma)$, 
$$\Theta(k)=\left\{\theta=(\beta,\Sigma): \|\beta\|_0\leq k, c_0\leq \lambda_{\min}(\Sigma)\leq \lambda_{\max}(\Sigma)\leq C_0\right\}$$ for some positive constants $C_0\geq c_0>0$. We consider the following null parameter space
$
\mathcal{H}_0=\left\{\theta=(\beta,\Sigma)\in \Theta(k): \xnew^{\intercal}\beta \leq 0\right\}
$
and the local alternative parameter space 
\begin{equation*}
\mathcal{H}_1(\mu)=\left\{\theta=(\beta,\Sigma)\in \Theta(k): \xnew^{\intercal}\beta ={\mu}/{n^{1/2}}\right\}, \quad \text{for some}\; \mu>0.
\label{eq: alter space}
\end{equation*}
\vspace{-5mm}
\begin{Proposition}
Under the same conditions as in Theorem \ref{thm: limiting distribution}, for each $\theta\in \mathcal{H}_0$, the proposed testing procedure ${\phi}_{\alpha}(\xnew)$ in \eqref{eq: testing} satisfies
$
\limsup_{n\rightarrow \infty}\PP_{\theta}\left[{\phi}_{\alpha}(\xnew)=1\right]\leq \alpha.
$
For $\theta\in \mathcal{H}_1(\mu)$, we have 
\begin{equation}
\limsup_{n\rightarrow \infty}\left|\PP_{\theta}\left[{\phi}_{\alpha}(\xnew)=1\right]-[1-\Phi^{-1}(z_{\alpha}-{\mu}/{\left(n {\rm V}\right)^{1/2}})]\right|=0,
\label{eq: power}
\end{equation}
where $\Phi^{-1}$ is the inverse of the cumulative function of standard normal distribution.
\label{prop: testing}
\end{Proposition}
The proposed hypothesis testing procedure is shown to have a well-controlled type I error rate. The asymptotic power expression in \eqref{eq: power} holds for any $\mu.$ Since \eqref{eq: dominating variance} implies $c_0\|\xnew\|_2\leq \left(n {\rm V}\right)^{1/2}\leq C_0\|\xnew\|_2,$ the power of the proposed test in \eqref{eq: power} is nontrivial if $\mu\geq C \|\xnew\|_2$ holds for a large positive constant $C$. If $\mu/\|\xnew\|_2\rightarrow \infty$ or equivalently ${n^{1/2}} \xnew^{\intercal}\beta /{\|\xnew\|_2} \rightarrow \infty$, then the power will be 1 in the asymptotic sense.
It has also been observed in Section \ref{sec: sim} that the finite sample performance of the proposed procedure depends on the sample size $n$ and the $\ell_2$ norm $\|\xnew\|_2$; see Tables \ref{tab: Setting 1 i} and \ref{tab: decaying coef i} for details.

\subsection{Analysis Related to Reweighting in Linearization}
\label{sec: challenges}
In the following, we provide more insights on how to establish the asymptotic normality and summarize technical tools for analyzing the re-weighted estimator obtained by the linearization procedure. 
Regarding the decomposition \eqref{eq: decomposition}, the first term captures the stochastic error due to the model error $\epsilon_i$, the second term is a bias component arising from estimating $\Sigma^{-1}\xnew$, and the third term appears due to the nonlinearity of the logistic regression model.  The following proposition controls the second and third terms.
\begin{Proposition} Suppose that Conditions {\rm (A1)} and ${\rm (A2)}$ hold. For any estimator $\widehat{\beta}$ satisfying Condition {\rm (B)}, with probability larger than $1-p^{-c}-\exp(-cn)$ for some positive constant $c>0$,
	\begin{equation}
	n^{1/2}\left|(\widehat{\Sigma}\widehat{u}-\xnew)^{\intercal}(\widehat{\beta}-\beta)\right|\leq n^{1/2}\|\xnew\|_2\lambda_{n}\|\widehat{\beta}-\beta\|_1\lesssim \|\xnew\|_2 {k \log p}\cdot{n^{-1/2}},
	\label{eq: error bound 1}
	\end{equation}
	and 
	\begin{equation}
	n^{1/2}|\widehat{u}^{\intercal}\frac{1}{n}\sum_{i=1}^{n}X_{i\cdot}\Delta_i|\leq  \tau_{n} \|\xnew\|_2 {k \log p}\cdot{n^{-1/2}}
	\label{eq: error bound 3}
	\end{equation}
	\label{prop: decomposition}
\end{Proposition}
Together with \eqref{eq: decomposition}, it remains to establish the asymptotic normality of the following term,
\begin{equation}
\widehat{u}^{\intercal}\frac{1}{n}\sum_{i=1}^{n} [h(X_{i\cdot}^{\intercal}\widehat{{\beta}})(1-h(X_{i\cdot}^{\intercal}\widehat{\beta}))]^{-1}X_{i\cdot}\epsilon_i.
\label{eq: reweight-term}
\end{equation}
Because of the dependence between the weight $[h(X_{i\cdot}^{\intercal}\widehat{{\beta}})(1-h(X_{i\cdot}^{\intercal}\widehat{\beta}))]^{-1}$ and the model error $\epsilon_i$, it is challenging to establish the asymptotic normality of this re-weighted summation \eqref{eq: reweight-term}. 

We decouple the correlation between $\widehat{\beta}$ and $\epsilon_i$ through the following expression,
\begin{equation}
\begin{aligned}
&\widehat{u}^{\intercal}\frac{1}{n}\sum_{i=1}^{n} [h(X_{i\cdot}^{\intercal}\widehat{{\beta}})(1-h(X_{i\cdot}^{\intercal}\widehat{\beta}))]^{-1}X_{i\cdot}\epsilon_i=\widehat{u}^{\intercal}\frac{1}{n}\sum_{i=1}^{n} [h(X_{i\cdot}^{\intercal}{\beta})(1-h(X_{i\cdot}^{\intercal}{\beta}))]^{-1}X_{i\cdot}\epsilon_i\\
&+{\widehat{u}}^{\intercal}\frac{1}{{n}}\sum_{i=1}^{n}\left([h(X_{i\cdot}^{\intercal}\widehat{\beta})(1-h(X_{i\cdot}^{\intercal}\widehat{\beta}))]^{-1}-[h(X_{i\cdot}^{\intercal}{\beta})(1-h(X_{i\cdot}^{\intercal}{\beta}))]^{-1}\right)X_{i\cdot}\epsilon_i.
\end{aligned}
\label{eq: technical decomposition}
\end{equation}
The first component on the right hand side of the above summation is not involved with the estimator $\widehat{\beta}$, so that the standard probability argument can be applied to establish the asymptotic normality. The second component on the right hand side of \eqref{eq: technical decomposition} captures the error incurred on estimating $\beta$ by $\widehat{\beta}$. We now provide a sharp control of this error term by suitable empirical process theory. 

\begin{Lemma} Suppose that Conditions {\rm (A1)} and ${\rm (A2)}$ hold and the initial estimator $\widehat{\beta}$ satisfies Condition {\rm (B)}, then with probability greater than $1-p^{-c}-\exp(-cn)-{1}/{t_0}$,
{\small
\begin{equation}
\begin{aligned}
\left|\widehat{u}^{\intercal}\frac{1}{n^{1/2}}\sum_{i=1}^{n}\left([h(X_{i\cdot}^{\intercal}\widehat{\beta})(1-h(X_{i\cdot}^{\intercal}\widehat{\beta}))]^{-1}-[h(X_{i\cdot}^{\intercal}{\beta})(1-h(X_{i\cdot}^{\intercal}{\beta}))]^{-1}\right)X_{i\cdot}\epsilon_i\right| \leq C t_0 \tau_n \|\xnew\|_2\frac{k \log p}{n^{1/2}} 
\end{aligned}
\label{eq: error bound 2}
\end{equation}
}
where $\tau_{n}$ is defined in \eqref{eq: constraint 3}, $t_0>1$ is a large positive constant and $c>0$ and $C>0$ are positive constants.
\label{lem: bias control}
\end{Lemma}

The main step in establishing the above lemma is to apply a contraction principle for i.i.d. symmetric random variables taking values $\{-1,1,0\}$. See Lemma \ref{lem: contraction} for the precise statement. This extends the existing results on contraction principles for i.i.d Rademacher random variables \citep{koltchinskii2011oracle}.  This lemma and the related analysis are particularly useful for carefully characterizing the approximation error in \eqref{eq: error bound 2} and can be of independent interest in establishing asymptotic normality of other re-weighted estimators in high dimensions. The proof of Lemma \ref{lem: bias control} is presented in Section \ref{sec: proof}.
\begin{remark}
In comparison, in case of the linear model or the logistic model without re-weighting \citep{van2014asymptotically,javanmard2014confidence,zhang2014confidence}, such a challenge does not exist since the corresponding term is of the form $\widehat{u}^{\intercal}\frac{1}{n}\sum_{i=1}^{n} X_{i\cdot}\epsilon_i$ and the direction $\widehat{u}$, defined in  \citet{van2014asymptotically,javanmard2014confidence,zhang2014confidence}, is either directly independent of $\epsilon_i$ or can be replaced by $u^{*}=\Sigma^{-1}\xnew$ (by assuming $\Sigma^{-1}\xnew$ to be sparse). 
\end{remark}

\section{Numerical Studies}
\label{sec: sim}
\subsection{Algorithm Implementation and Method Comparision}
\label{sec: implementation}
We provide details on how to implement the LiVE estimator defined in \eqref{eq: correction}. 
The initial estimator $\widehat{\beta}$ defined in \eqref{eq: penalized estimator} is computed using the R-package \texttt{cv.glmnet} \citep{glmnet} with the tuning parameter $\lambda$ chosen by cross-validation. To compute the projection direction $\widehat{u}\in \R^{p},$ we implement the following constrained optimization, 
\begin{equation}
\begin{aligned}
\widehat{u}=\;\argmin_{u\in \RR^{p}} u^{\intercal}\widehat{\Sigma}u \quad
\text{subject to}&\; \|\widehat{\Sigma}u-\xnew\|_{\infty}\leq  \|\xnew\|_2 \lambda_{n}, \quad |\xnew^{\intercal}\widehat{\Sigma}u-\|\xnew\|_2^2 |\leq \|\xnew\|_2^2\lambda_{n}.\\
\end{aligned}
\label{eq: implementation}
\end{equation}

This construction does not include the  constraint \eqref{eq: constraint 3}, which is mainly imposed to facilitating the theoretical proof. We have conducted an additional check in simulations and observed that our constructed $\widehat{u}$ in \eqref{eq: implementation} satisfies $\|X\widehat{u}\|_{\infty}\leq C\sqrt{\log n}\|x_*\|_2$; see Section \ref{sec: tau} in the supplementary material for details. 

We solve the dual problem of \eqref{eq: implementation},
\begin{equation}
\widehat{v}=\arg \min _{v \in \mathbb{R}^{p+1}} \frac{1}{4} v^{\intercal} H^{\intercal} \widehat{\Sigma} H v+b^{\intercal} H v+\lambda_n\|v\|_{1} \text { with } H=\left[b, \mathbb{I}_{p \times p}\right], b=\frac{1}{\left\|\xnew\right\|_{2}}\xnew
\label{eq: dual}
\end{equation}
and then solve the primal problem \eqref{eq: implementation} as $\widehat{u}=-\left(\widehat{v}_{-1}+\widehat{v}_{1} b\right) / 2 .$ We refer to Proposition 2 in \cite{cai2019individualized} for the
the detailed derivation of the dual problem \eqref{eq: dual}.
In this dual problem, when $\widehat{\Sigma}$ is singular and the tuning parameter $\lambda_n>0$ gets sufficiently close to $0$, the dual problem cannot be solved as the minimum value converges to negative infinity. Hence, we choose the smallest $\lambda_n>0$ such that the dual problem has a finite minimum value. The tuning parameter $\lambda_n$ selected in this manner is at the scale of $\sqrt{\log p/n}$. We investigate the ratio $\lambda_n/\sqrt{\log p/n}$ in Section \ref{sec: mu and lambda} in the supplement.

We compare our proposed LiVE estimator with the following state-of-the-art methods.
\begin{itemize}
\vspace{-1.5mm}
\item {\bf Plug-in Lasso.} Estimate $\xnew^{\intercal}{\beta}$ by $\xnew^{\intercal}\widehat{\beta}$ with $\widehat{\beta}$ denoting the penalized estimator in \eqref{eq: penalized estimator}.
\vspace{-7mm}
\item {\bf Post-selection method.}
First select important predictors through penalized logistic regression estimator $\widehat{\beta}$ in \eqref{eq: penalized estimator} and then fit a standard logistic regression with the selected predictors. The post-selection estimator $\widehat{\beta}_{\rm PL}\in \mathbf{R}^{p}$ is used to estimate $\xnew^{\intercal}\beta$ by $\xnew^{\intercal}\widehat{\beta}_{\rm PL}$. The variance of this post-selection estimator $\xnew^{\intercal}\widehat{\beta}_{\rm PL}$ can be obtained by the inference results in the classical low-dimensional logistic regression, denoted by $\widehat{\mathrm{V}}_{\rm PL}$.
\vspace{-1.5mm}
\item {\bf Plug-in \texttt{hdi} \citep{dezeure2015high}.}
The R package \texttt{hdi} is implemented to obtain the coordinate debiased Lasso estimator  $\widehat{\beta}_{\texttt{hdi}} \in \mathbf{R}^{p}$ and the plug-in estimator $\xnew^{\intercal}\widehat{\beta}_{\texttt{hdi}}$ is used to estimate $\xnew^{\intercal}\beta$, with the variance estimator as $\widehat{\mathrm{V}}_{\texttt{hdi}}$.
\vspace{-1.5mm}
\item {\bf Plug-in  \texttt{WLDP} \citep{ma2018global}.}
We compute the debiased lasso estimator $\widehat{\beta}_{\texttt{WLDP}}\in \mathbf{R}^{p}$ by the Weighted LDP algorithm in Table 1 of \cite{ma2018global}.
 The plug-in estimator of $\xnew^{\intercal}\beta$ and the associated variance are given by $\xnew^{\intercal}\widehat{\beta}_{\texttt{WLDP}}$ and $\widehat{\mathrm{V}}_{\texttt{WLDP}}$ respectively. 
\vspace{-6mm}
 \item {\bf Generalization of Transformation Method \citep{zhu2018linear} in \eqref{eq: transformation}.}  
\end{itemize}
\vspace{-1.5mm}

We compare the above estimators with the proposed LiVE estimator in \eqref{eq: correction} in terms of Root Mean Square Error (RMSE), standard error and bias. 
Since the plug-in Lasso estimator is not useful for CI construction due to its large bias, we compare with Post-selection method, plug-in \texttt{hdi}, \texttt{WLDP} and the transformation method, from the perspectives of CI construction and  hypothesis testing \eqref{eq: null hypothesis}. Recall that our proposed CI and testing procedure for \eqref{eq: null hypothesis} are implemented as in \eqref{eq: CI} and \eqref{eq: testing}, respectively.
The inference procedures based on post-selection method, plug-in \texttt{hdi} and plug-in weighted \texttt{WLDP} are defined as,
\begin{equation*}
{\rm CI}_{\alpha}(\xnew)=\left[h(\xnew^{\intercal} \widetilde{\beta}-z_{\alpha/2}\widetilde{\rm V}^{1/2}),h(\xnew^{\intercal} \widetilde{\beta}+z_{\alpha/2}\widetilde{\rm V}^{1/2})\right],
\; {\phi}_{\alpha}(\xnew)=\mathbf{1}\left(\xnew^{\intercal} \widetilde{\beta}-z_{\alpha}\widetilde{\rm V}^{1/2}\geq 0\right),
\end{equation*}
with replacing $(\widetilde{\beta},\widetilde{\rm V})$ by $(\widehat{\beta}_{\rm PL},\widehat{\mathrm{V}}_{\rm PL})$, $(\widehat{\beta}_{\texttt{hdi}},\widehat{\mathrm{V}}_{\texttt{hdi}})$ and $(\widehat{\beta}_{\texttt{WLDP}},\widehat{\mathrm{V}}_{\texttt{WLDP}})$ respectively. 

Throughout the simulation, we set $X_{i,1}=1$ to represent the intercept and generate the covariates $\{X_{i,-1}\}_{1\leq i \leq n}$ from the multivariate normal distribution with zero mean and covariance matrix $\Sigma$.  Conditioning on $X_{i\cdot}$, the binary outcome is generated by 
$
y_{i} \sim {\rm Bernoulli} \left({h\left(X_{i\cdot}^{\intercal}\beta\right)}\right),$ for $1\leq i\leq n$. We generate the following loadings $x_*$.

\begin{itemize}
\vspace{-1.5mm}
\item {\bf Loading 1}: We set $x_{{\rm basis},1}=1$ and generate $x_{{\rm basis},-1}\in \R^{(p-1)}$ following $N(0,\widetilde{\Sigma})$ with $\widetilde{\Sigma} = \{q \cdot 0.5^{1+|j-l|}\}_{1\leq j,l\leq (p-1)}$ for some $q>0$; for $r\geq 0,$ generate $x_*$ as
\begin{equation}
x_{*,j}=\begin{cases} 
x_{{\rm basis},j} &  \text{for} \; 1\leq j\leq 11\\
r \cdot x_{{\rm basis},j} &  \text{for}\; 12\leq j \leq p\\
\end{cases}.
\label{eq: shrink def}
\end{equation}
\vspace{-6mm}
\item {\bf Loading 2}: $x_{{\rm basis},1}$ is set as $1$ and $x_{{\rm basis},-1}\in \R^{(p-1)}$ is generated as following $N(0,\widetilde{\Sigma})$ with $\widetilde{\Sigma} = \{q \cdot (-0.75)^{|j-l|}/2\}_{1\leq j,l\leq (p-1)}$ for some $q>0$; generate $x_*$ using \eqref{eq: shrink def}.
\end{itemize}
\vspace{-1.5mm}
All simulation results are averaged over $500$ replications. The loadings are only generated once and kept the same across all 500 replications. 

\subsection{Varying sample size \texorpdfstring{$n$}{n} and loading norm \texorpdfstring{$\|\xnew\|_2$}{xnorm}}
\label{sec: varying n and norm}
We investigate the  performance of our method across different sample sizes $n$ and loading norms $\|\xnew\|_2.$ 
We set $p = 501$, $\Sigma = \{0.5^{1+|j-l|}\}_{1 \leq j \leq l \leq (p-1)}$ and vary $n \in \{200,400,600\}$. We carry out the simulations for both {Loading 1} and {Loading 2} with $q = 1$ and $r\in \{1,1/25\}.$ 
We generate the exactly sparse regression vector $\beta$ as \begin{enumerate}
    \item[(S1)] $\beta_1=0$, $\beta_{j}={(j-1)}/{20}$ for $2\leq j\leq 11$ and $\beta_{j}=0$ for $12\leq j\leq p$. 
\end{enumerate}
Here, the scale parameter $r$ in \eqref{eq: shrink def} controls the magnitude of the noise variables in $\xnew$. As $r$ decreases, $\|\xnew\|_2$ decreases but the case probability $h(\xnew^{\intercal}\beta)$ remains the same for all choices of $r$ since only the values of $x_{*,j}$ for $1\leq j\leq 11$ affect $\xnew^{\intercal}\beta$. 
Since the R package \texttt{hdi} and the \texttt{WLDP} algorithm only report the debiased estimators together with their variance estimators for the regression coefficients excluding the intercept, the intercept $\beta_1$ is set as $0$ to have a fair comparison. In Section \ref{sec: exact with intercept} in the supplement, we conduct additional simulation studies for models with a non-zero intercept.

In Table \ref{tab: Setting 1 i}, we compare the proposed LiVE method with post-selection, \texttt{hdi} and \texttt{WLDP}, in terms of CI construction and hypothesis testing for the setting $(\rm S1)$ with $x_*$ generated as Loading 1. The CIs constructed by LiVE and \texttt{hdi} have coverage over different scenarios and the lengths are reduced when a larger sample is used to construct the CI. \texttt{WLDP} overcovers and the post-selection method undercovers. 

\begin{table}[htp!]
\centering
\scalebox{0.7}{
\begin{tabular}{|rrr|r|rrrr|rrrr|rrrr|rrrr|}
\hline
\multicolumn{20}{|c|}{{\bf Setting (S1), Loading 1 with $q = 1$}} \\
\hline
\multicolumn{3}{|c|}{}&&\multicolumn{4}{c|}{LiVE} &\multicolumn{4}{c|}{Post Selection} &\multicolumn{4}{c|}{\texttt{hdi}}&\multicolumn{4}{c|}{\texttt{WLDP}} \\
\hline
$\|\xnew\|_{2}$ &{\rm r} & Prob &$n$& Cov & ERR& Len & t & Cov & ERR & Len & t & Cov & ERR & Len & t & Cov & ERR & Len & t \\
 \hline
 \multirow{3}{*}{16.1} & \multirow{3}{*}{1} & \multirow{3}{*}{0.732} &200& 0.98 & 0.05 & 0.88 & 5 & 0.68 & 0.54 & 0.42 & 1 & 0.97 & 0.06 & 0.93 & 370 & 1.00 & 0.00 & 1.00 & 34 \\   
 & & & 400& 0.97 & 0.10 & 0.81 & 14 & 0.71 & 0.57 & 0.38 & 2 & 0.96 & 0.10 & 0.87 & 751 & 1.00 & 0.00 & 1.00 & 56 \\ 
 & & & 600& 0.95 & 0.13 & 0.74 & 23 & 0.70 & 0.68 & 0.32 & 6 & 0.94 & 0.10 & 0.83 & 3212 & 1.00 & 0.00 & 1.00 & 118 \\ 
 \hline
 \multirow{3}{*}{1.90} & \multirow{3}{*}{$\frac{1}{25}$} & \multirow{3}{*}{0.732} &200& 0.96 & 0.62 & 0.34 & 5 & 0.80 & 0.77 & 0.31 & 1 & 0.92 & 0.86 & 0.31 & 371 & 1.00 & 0.36 & 0.58 & 34\\ 
  & & &400& 0.94 & 0.92 & 0.23 & 14 & 0.83 & 0.93 & 0.24 & 2 & 0.92 & 0.96 & 0.23 & 751 & 1.00 & 0.45 & 0.53 & 54\\ 
  & & &600& 0.95 & 0.95 & 0.19 & 22 & 0.82 & 0.95 & 0.20 & 5 & 0.95 & 0.97 & 0.19 & 3211 & 1.00 & 0.47 & 0.50 & 118 \\ 
 \hline
\end{tabular}
}
\caption{{\bf Varying $n$ and $\|\xnew\|_2$.} ``r" and ``Prob" represent the shrinkage parameter and Case Probability respectively. The columns indexed with ``Cov" and ``Len" represent the empirical coverage and length of the CIs; the column indexed with ``ERR" represents the empirical rejection rate of the test; ``t" represents the averaged computation time (in seconds). The columns under ``LiVE" ,``Post Selection", ``\texttt{hdi}" and ``\texttt{WLDP}" correspond to the proposed estimator, the post selection estimator, the plug-in debiased estimator using \texttt{hdi} and \texttt{WLDP}, respectively.}
\label{tab: Setting 1 i}
\end{table}

Regarding the testing procedure, we report the empirical rejection rate (ERR), which is defined as the proportion of null hypothesis in \eqref{eq: null hypothesis} being rejected out of the $500$ replications. Under the null hypothesis, ERR is an empirical measure of the type I error; under the alternative hypothesis, ERR is an empirical measure of the power.  For Loading 1 (alternative hypothesis), the empirical power increases with sample sizes, for all methods.  For the case that $\|\xnew\|_2$ is relatively small, the proposed LiVE method has a power above $0.90$ when the sample size reaches $400$. For settings with a large $\|\xnew\|_2$, the power is not as high mainly due to the high variance of the bias-corrected estimator. This is consistent with the theoretical results established in Proposition \ref{prop: testing}. 

We have investigated the computational efficiency of all methods and reported the averaged time of implementing each method under the column indexed with ``t" (the units are seconds). The proposed LiVE method is computationally efficient and can be finished within $25$ seconds on average. The \texttt{hdi} algorithm provides valid CIs but requires around an hour to achieve the same goal for $n=600$ and $p=501$. The main reason is that the \texttt{hdi} is not designed for inference for case probabilities and requires the implementation of $p$ high-dimensional penalization algorithms for bias-correction.

The inference results for Loading 2 are similar and reported in Table \ref{tab: Setting 2 i} in the supplement. We report Root Mean Squared Error (RMSE), bias and standard deviation of the proposed LiVE estimator, plug-in Lasso, post-selection, \texttt{hdi} and \texttt{WLDP} in Table \ref{tab: Setting 1,2 ii} in the supplementary material. It is observed that the plug-in Lasso estimator cannot be used for confidence interval construction as its bias component is a dominant component of the RMSE and the uncertainty of the bias component is hard to quantify.

Post selection inference methods can produce incorrect inference due to the fact that the model selection uncertainty  is not quantified. The post-selection method can select either a larger model or a smaller model compared to the true one. In Table \ref{tab: Setting 1 i}, post selection undercovers since post-selection tends to select a relatively large set of variables and this results in a perfect separation in the re-fitting step. In Section \ref{sec: naive} in the supplementary material, we show another setting where the post-selection method selects a smaller model and leads to a substantial omitted variable bias.

In practical settings, the regression vector $\beta$ might not be exactly sparse but can have some large regression coefficients and most others are small but not exactly zero. To simulate these practical settings, we consider the following generation of $\beta$: \begin{enumerate}
 \item[(S2)] $\beta_1= 0$ and $\beta_j= (j-1)^{-\rm decay} \; \text{for}\; 2\leq j\leq p,$ with ${\rm decay} \in \{1, 2\}.$ 
\end{enumerate}      

We illustrate the method comparison using Loading 1.
The inference results are reported in Table \ref{tab: decaying coef i} for  ${\rm decay} = 1$. {The results for ${\rm decay} = 2$ are similar to those for ${\rm decay} = 1$ and summarized in Table \ref{tab: decaying coef i decay2} in the supplement.} The estimation results are reported in Table \ref{tab: decaying coef ii} in the supplement.

\begin{table}[htp!]
\centering
\scalebox{0.7}{
\begin{tabular}{|rrr|r|rrrr|rrrr|rrrr|rrrr|}
  \hline
 \multicolumn{20}{|c|}{{\bf Setting (S2) with decay=1, Loading 1 with $q = 1$}} \\
  \hline
\multicolumn{3}{|c|}{}&&\multicolumn{4}{c|}{LiVE} &\multicolumn{4}{c|}{Post Selection} &\multicolumn{4}{c|}{\texttt{hdi}}&\multicolumn{4}{c|}{\texttt{WLDP}} \\
   \hline
$\|\xnew\|_{2}$ &{\rm r}& Prob &$n$& Cov & ERR& Len & t & Cov & ERR & Len & t & Cov & ERR & Len & t & Cov & ERR & Len & t\\ 
 \hline
\multirow{3}{*}{16.1} & \multirow{3}{*}{1} & \multirow{3}{*}{0.645} &200& 0.96 & 0.05 & 0.93 & 5 & 0.58 & 0.26 & 0.45 & 1 & 0.96 & 0.06 & 0.93 & 370 & 1.00 & 0.00 & 1.00 & 34 \\ 
  & & & 400& 0.96 & 0.04 & 0.85 & 14 & 0.60 & 0.31 & 0.41 & 2 & 0.97 & 0.07 & 0.90 & 751 & 1.00 & 0.00 & 1.00 & 56 \\ 
  & & & 600& 0.97 & 0.05 & 0.80 & 23 & 0.62 & 0.37 & 0.37 & 6 & 0.96 & 0.07 & 0.86 & 3212 & 1.00 & 0.00 & 1.00 & 118 \\ 
 \hline
\multirow{3}{*}{1.09} & \multirow{3}{*}{$\frac{1}{25}$} & \multirow{3}{*}{0.523} &200& 0.96 & 0.06 & 0.40 & 5 & 0.69 & 0.13 & 0.31 & 1 & 0.96 & 0.05 & 0.39 & 371 & 1.00 & 0.00 & 0.75 & 34\\ 
  & & &400& 0.96 & 0.11 & 0.28 & 14 & 0.58 & 0.16 & 0.24 & 2 & 0.94 & 0.11 & 0.28 & 751 & 1.00 & 0.00 & 0.68 & 54\\ 
  & & &600& 0.97 & 0.07 & 0.24 & 22 & 0.71 & 0.09 & 0.21 & 5 & 0.96 & 0.04 & 0.24 & 3211 & 1.00 & 0.00 & 0.65 & 118 \\ 
 \hline
\end{tabular}
}
\caption{{\bf Varying $n$ and $\|x_*\|_2$.}  ``r" and``Prob" represent the shrinkage parameter and Case Probability respectively. The columns indexed with ``Cov" and ``Len" represent the empirical coverage and length of the CIs; the column indexed with ``ERR" represents the empirical rejection rate of the test; ``t" represents the averaged computation time (in seconds). The columns under ``LiVE" ,``Post Selection", ``\texttt{hdi}" and ``\texttt{WLDP}" correspond to the proposed estimator, the post selection estimator, the plug-in debiased estimator using \texttt{hdi} and \texttt{WLDP}  respectively.}
\label{tab: decaying coef i}
\end{table}

Note that as the regression coefficient is decaying, the shrinking parameter $r$ in \eqref{eq: shrink def} plays a role in determining the case probability. The main observations are consistent with those in Table \ref{tab: Setting 1 i}: only the proposed LiVE method and \texttt{hdi} have proper coverage across different scenarios while the CI by post selection undercovers and the CI by \texttt{WLDP} overcovers. The proposed method is computationally more efficient than \texttt{hdi}: for $n=600$, the average computation time for the proposed algorithm is $23$ seconds while \texttt{hdi} with a similar performance requires more than $3200$ seconds.

For $\text{\rm decay}=1$ and $r=1$, the case probability $(0.645)$ is above $0.5$; the proposed LiVE method and \texttt{hdi} achieve the correct coverage level but the testing procedures have low powers. This matches with Proposition \ref{prop: testing}, that is, the power of the proposed testing procedure tends to be low for the observation $\xnew$ with very large $\|\xnew\|_{2}$. For $\text{\rm decay}=1$ and $r={1}/{25}$, the case probability is $0.523$ and this represents an alternative in the indistinguishable region and the power of the proposed testing procedure is low as expected.

\subsection{Comparison with the Transformation Method}
\label{sec: U Method}
We compare LiVE with the Transformation method \citep{zhu2018linear,tripuraneni} based on settings $(\rm S3)$ and $(\rm S4)$, which are variations of setting $(\rm S1).$
\begin{enumerate}
    \item[(S3)] $p = 501$; $\beta_1 = 0$, $\beta_{j}={(j-1)}/{10}$ for $2\leq j\leq 11$ and $\beta_{j}=0$ for $12\leq j\leq 501$.
    \item[(S4)] $p = 1001$; $\beta_1 = 0$; $\beta_{j}={(j-1)}/{20}$ for $2 \leq j \leq 11$ but  $j \neq 3,4,6$; $\beta_j = 1$ for $j = 3, 4, 6$ and $\beta_j = 0$ for $12 \leq j \leq 1001$.
\end{enumerate}
Set $\Sigma = \{0.5^{1+|j-l|}\}_{1 \leq j \leq l \leq (p-1)}$. 
Table \ref{tab: U_method p=1000} summarizes  the results for $(\rm S3)$ and $(\rm S4)$ with {Loading 1} with $q = 1$ and $r \in \{1, 1/2, 1/5, 1/25\}$. We vary $n \in \{200, 400, 600\}.$ 

\begin{table}[htp!]
\centering
\scalebox{0.8}{
\begin{tabular}{|rrr|r|rrrrrr|rrrrrr|}
\hline
\multicolumn{16}{|c|}{{\bf Setting $(\rm S3)$, Loading 1 with $q = 1$}} \\
\hline
\multicolumn{3}{|c|}{}&&\multicolumn{6}{c|}{LiVE} &\multicolumn{6}{c|}{Transformation Method}\\
\hline
$\|x_*\|_{2}$ &{\rm r}& Prob &$n$& Cov & ERR & Len & RMSE & Bias & SE & Cov & ERR & Len & RMSE & Bias & SE \\ 
\hline
\multirow{3}{*}{16.1} & \multirow{3}{*}{1} & \multirow{3}{*}{0.881} &200& 0.99 & 0.08 & 0.92 & 0.28 & -0.12 & 0.25 & 0.84 & 0.06 & 0.82 & 0.34 & -0.26 & 0.22  \\ 
& & & 400& 1.00 & 0.12 & 0.87 & 0.23 & -0.08 & 0.22 & 0.98 & 0.13 & 0.82 & 0.28 & -0.18 & 0.22 \\ 
& & & 600& 0.98 & 0.17 & 0.81 & 0.19 & -0.05 & 0.18 & 0.96 & 0.14 & 0.82 & 0.26 & -0.16 & 0.21  \\ 
\hline
\multirow{3}{*}{8.18} & \multirow{3}{*}{$\frac{1}{2}$} & \multirow{3}{*}{0.881} &200& 0.99 & 0.22 & 0.72 & 0.16 & -0.06 & 0.14 & 0.65 & 0.11 & 0.63 & 0.27 & -0.24 & 0.14 \\
& & &400& 0.98 & 0.38 & 0.62 & 0.13 & -0.04 & 0.12 & 0.90 & 0.20 & 0.64 & 0.23 & -0.17 & 0.15 \\ 
& & &600& 0.97 & 0.58 & 0.51 & 0.10 & -0.03 & 0.10 & 0.92 & 0.28 & 0.60 & 0.20 & -0.14 & 0.14 \\ 
\hline
\multirow{3}{*}{3.66} & \multirow{3}{*}{$\frac{1}{5}$} & \multirow{3}{*}{0.881} &200& 0.94 & 0.78 & 0.40 & 0.10 & -0.06 & 0.09 & 0.60 & 0.54 & 0.41 & 0.19 & -0.15 & 0.12 \\
& & &400& 0.97 & 0.95 & 0.32 & 0.07 & -0.03 & 0.06 & 0.74 & 0.85 & 0.32 & 0.12 & -0.08 & 0.09 \\ 
& & &600& 0.95 & 0.98 & 0.25 & 0.06 & -0.02 & 0.06 & 0.84 & 0.94 & 0.26 & 0.09 & -0.06 & 0.07 \\ 
\hline
\multirow{3}{*}{1.90} & \multirow{3}{*}{$\frac{1}{25}$} & \multirow{3}{*}{0.881} &200& 0.88 & 0.96 & 0.26 & 0.09 & -0.05 & 0.07 & 0.92 & 0.99 & 0.20 & 0.06 & -0.03 & 0.06 \\
& & &400& 0.92 & 0.99 & 0.21 & 0.06 & -0.03 & 0.05 & 0.94 & 0.99 & 0.16 & 0.04 & -0.01 & 0.04 \\ 
& & &600& 0.92 & 0.99 & 0.17 & 0.05 & -0.02 & 0.04 & 0.96 & 1.00 & 0.14 & 0.03 & 0.00 & 0.03 \\ 
\hline
\hline
\multicolumn{16}{|c|}{{\bf Setting $(\rm S4)$, Loading 1 with $q = 1$}} \\
\hline
\multicolumn{3}{|c|}{}&&\multicolumn{6}{c|}{LiVE} &\multicolumn{6}{c|}{Transformation Method}\\
\hline
$\|x_*\|_{2}$ &{\rm r}& Prob &$n$& Cov & ERR & Len & RMSE & Bias & SE & Cov & ERR & Len & RMSE & Bias & SE \\ 
\hline
\multirow{3}{*}{22.2} & \multirow{3}{*}{1} & \multirow{3}{*}{0.814} &200& 0.99 & 0.02 & 0.97 & 0.33 & -0.11 & 0.31 & 0.00 & 0.35 & 0.04 & 0.31 & -0.30 & 0.08  \\ 
& & & 400& 0.99 & 0.04 & 0.94 & 0.30 & -0.09 & 0.29 & 0.04 & 0.10 & 0.31 & 0.28 & -0.27 & 0.09 \\ 
& & & 600& 0.98 & 0.08 & 0.91 & 0.26 & -0.06 & 0.25 & 0.87 & 0.08 & 0.69 & 0.26 & -0.19 & 0.18  \\ 
\hline
\multirow{3}{*}{11.2} & \multirow{3}{*}{$\frac{1}{2}$} & \multirow{3}{*}{0.814} &200& 0.99 & 0.10 & 0.85 & 0.20 & -0.05 & 0.20 & 0.06 & 0.17 & 0.29 & 0.27 & -0.25 & 0.10  \\ 
& & & 400& 0.99 & 0.17 & 0.75 & 0.17 & -0.03 & 0.17 & 0.77 & 0.12 & 0.68 & 0.21 & -0.13 & 0.16 \\ 
& & & 600& 0.99 & 0.22 & 0.71 & 0.15 & -0.03 & 0.15 & 0.87 & 0.20 & 0.64 & 0.19 & -0.11 & 0.15  \\ 
\hline
\multirow{3}{*}{4.90} & \multirow{3}{*}{$\frac{1}{5}$} & \multirow{3}{*}{0.814} &200& 0.96 & 0.50 & 0.52 & 0.10 & -0.02 & 0.10 & 0.89 & 0.50 & 0.42 & 0.14 & -0.08 & 0.11  \\ 
& & & 400& 0.97 & 0.75 & 0.41 & 0.08 & -0.01 & 0.08 & 0.90 & 0.73 & 0.35 & 0.11 & -0.05 & 0.10 \\ 
& & & 600& 0.97 & 0.76 & 0.39 & 0.08 & -0.02 & 0.08 & 0.91 & 0.85 & 0.30 & 0.09 & -0.03 & 0.09  \\ 
\hline
\multirow{3}{*}{2.29} & \multirow{3}{*}{$\frac{1}{25}$} & \multirow{3}{*}{0.814} &200& 0.96 & 0.91 & 0.30 & 0.08 & -0.02 & 0.07 & 0.89 & 0.98 & 0.22 & 0.07 & 0.02 & 0.07 \\ 
& & &400& 0.94 & 0.98 & 0.24 & 0.06 & -0.02 & 0.06 & 0.88 & 0.99 & 0.19 & 0.06 & 0.02 & 0.06 \\ 
& & &600& 0.93 & 0.98 & 0.23 & 0.06 & -0.02 & 0.06 & 0.90 & 1.00 & 0.17 & 0.05 & 0.02 & 0.04 \\ 
\hline
\end{tabular}
}
\caption{\textbf{Comparison with Transformation Method.} ``r" and ``Prob" represent the shrinkage parameter and Case Probability respectively. The columns indexed with ``Cov" and ``Len" represent the empirical coverage and length of the CIs; the column indexed with ``ERR" represents the empirical rejection rate of the test; The columns indexed with ``RMSE", ``Bias" and ``SE" represent the RMSE, bias and standard error, respectively. The columns under ``LiVE" and ``Transformation Method" correspond to LiVE and the transformation method, respectively.}
\label{tab: U_method p=1000}
\end{table}

We observe that the performance of our proposed LiVE estimator and the transformation method are similar for a sparse loading (e.g. $r=1/25$) while the performance of these two methods can be quite different for a dense loading (e.g. $r=1,1/2,1/5$). Specifically, for a dense loading, the bias of the transformation method is typically larger than that of our proposed LiVE estimator; our proposed confidence intervals in general have coverage while the transformation method undercovers. For $(\rm S3)$, when $r=1/5$, our proposed confidence intervals have better coverage than those by the transformation method while their lengths are similar; with a relatively dense loading (e.g. $r=1$ or $r=1/2$) and a smaller sample size ($n = 200$), the transformation method undercovers while the confidence intervals by the LiVE method provide coverage at the expense of longer lengths. For $(\rm S4)$, the observation is similar to that for $(\rm S3).$ 

\subsection{Increasing dimension \texorpdfstring{$p$}{p} and coefficient magnitudes} 
\label{sec: larger p, larger beta}
We vary $p$ across $\{1001, 2001, 5001\}$ and generate $\beta$ as a mixture of large and small signals. 

\vspace{-3mm}
\begin{enumerate}
    \item[(S5)]  $\beta_1 = 0$; $\beta_{j}={(j-1)}/{20}$ for $7 \leq j \leq 11$; $\beta_j = 1$ for $j = 2, 3, 4$; $\beta_j = -1$ for $j = 5,6$ and $\beta_j = 0$ for $12 \leq j \leq p$ 
    \vspace{-1.5mm}
    \item[(S6)] $\beta_1 = 0$; $\beta_j={(j-1)}^{- \rm decay}$ for $7 \leq j \leq p$; $\beta_j = 1$ for $j = 2, 3, 4$ and $\beta_j = -1$ for $j = 5,6$ with $\rm decay \in \{1,2\}$
\end{enumerate}    
Set $\Sigma = \{0.5^{1+|j-l|}/2\}_{1 \leq j \leq l \leq (p-1)}$. Settings $(\rm S5)$ and $(\rm S6)$ are variations of Settings $(\rm S1)$ and $(\rm S2),$ respectively. 
The results for Setting $(\rm S5)$ with respect to {Loading 1} with $q = 1/2$ and $r = 1/5$ are summarized in Table \ref{tab: ES p1000}. We vary $n$ across $\{400,600,1000\}.$ In Section \ref{sec: larger p, larger beta supp} in the supplement, we report the results for Setting (S5) with respect to {Loading 2} with $q = 1/2$ and $r = 1/5$ in Table \ref{tab: ES p1000 Loading 2} and the results for $(\rm S6)$ in Tables \ref{tab: AS decay 1 p large} and \ref{tab: AS decay 2 p large}. {The results are similar to that reported in Table \ref{tab: ES p1000}.} 

\begin{table}[ht]
    \centering
    \scalebox{0.8}{
    \begin{tabular}{|rrrr|r|rrrrrr|}
    \hline
    \multicolumn{11}{|c|}{{\bf Setting $(\rm S5)$, Loading 1 with $q = 1/2$}} \\
    \hline
  $p$&  $\|x_*\|_{2}$ &{\rm r}& Prob &$n$& Cov & ERR & Len & RMSE & Bias & SE \\ 
    \hline
 \multirow{3}{*}{1001}   & \multirow{3}{*}{3.21} & \multirow{3}{*}{$\frac{1}{5}$} & \multirow{3}{*}{0.263} &400& 0.91 & 0.00 & 0.42 & 0.13 & 0.07 & 0.11 \\ 
 &   & & &600& 0.94 & 0.00 & 0.38 & 0.11 & 0.04 & 0.10 \\ 
 &   & & &1000& 0.93 & 0.00 & 0.30 & 0.08 & 0.04 & 0.07 \\
    \hline
    \multirow{3}{*}{2001}   &   \multirow{3}{*}{4.60} & \multirow{3}{*}{$\frac{1}{5}$} & \multirow{3}{*}{0.531} &400& 0.96 & 0.12 & 0.55 & 0.15 & 0.07 & 0.14 \\ 
  &  & & &600& 0.95 & 0.12 & 0.49 & 0.13 & 0.05 & 0.12 \\ 
  &  & & &1000& 0.97 & 0.15 & 0.42 & 0.11 & 0.03 & 0.11 \\
    \hline
\multirow{3}{*}{5001}   &    \multirow{3}{*}{7.07} & \multirow{3}{*}{$\frac{1}{5}$} & \multirow{3}{*}{0.385} &400& 0.97 & 0.00 & 0.72 & 0.19 & 0.02 & 0.19 \\
   & & & &600& 0.98 & 0.00 & 0.65 & 0.15 & 0.02 & 0.15 \\ 
   & & & &1000& 0.97 & 0.00 & 0.57 & 0.13 & 0.01 & 0.13 \\
    \hline
    \end{tabular}
    }
    \caption{\textbf{Inference properties of {\bf LiVE} with increasing $p$ and coefficient magnitudes.} ``r" and ``Prob" represent the shrinkage parameter and Case Probability respectively. The columns indexed with ``Cov" and ``Len" represent the empirical coverage and length of the CIs; the column indexed with ``ERR" represents the empirical rejection rate of the test; The columns indexed with ``RMSE", ``Bias" and ``SE" represent the RMSE, bias and standard error, respectively.} 
    \label{tab: ES p1000}
    \end{table}
    
\noindent The observations are persistent with those in Section $4.2$. The proposed LiVE method has coverage across different scenarios. In Table \ref{tab: ES p1000}, when $p \in \{1001, 5001\}$ the case probabilities ($< 0.5$) correspond to the null hypothesis and the testing procedure has type I error controlled. However when $p = 2001$, the case probability ($0.531$) corresponds to an alternative in the indistinguishable region and consequently the testing procedure does not have power.
    
\subsection{Varying sparsity of \texorpdfstring{$\beta$}{beta}}
\label{sec: sparsity}
We test the sensitivity of our method to the sparsity assumption $k \lesssim {\sqrt{n}}/{\log p}$ by varying the sparsity over a range of values. To mimic the configuration of real data in Section \ref{sec: real}, we set $n = 318$ and $p = 199$. Set $\Sigma = \{0.5^{1+|j-l|}\}_{1 \leq j \leq l \leq (p-1)}$ and generate $\beta$ as
\begin{enumerate}
    \item[(S7)] $\beta_1=0$, $\beta_{j}={(j-1)}/{d}$ for $2\leq j\leq l + 1$ and $\beta_{j}=0$ for $l + 2\leq j\leq 199$. 
\end{enumerate}
We vary $d \in \{5, 10, 20\}$ and $l \in \{5, 10, 12, 15, 20\}.$ With $d$ increasing or $l$ decreasing, the effective sparsity level decreases. Table \ref{tab: sparsity} summarizes the results for the setting (S7) with $x_*$ generated as {Loading 1} with $q = 1/2$ and $r = 1/5$. For $d=5$, our proposed LiVE method is more reliable for $l=5$ or $10.$ Even though the coverage is guaranteed for a larger $l$, the proposed confidence intervals overcover. When $d$ is increased to $10$, the method works well for $l\leq 12$; for $l=15, 20$, the confidence intervals are conservative. For $d=20,$  our proposed method is reliable for $l$ being as large as $20.$ In general, our proposed confidence intervals are reliable for the relatively sparse signals; for the setting with a dense signal, the proposed confidence intervals are conservative and hence less informative.

\begin{table}[ht]
\centering
\scalebox{0.8}{
\begin{tabular}{|rr|r|r|rrrrrr|}
\hline
\multicolumn{10}{|c|}{{\bf Setting $(\rm S7)$, Loading 1 with $q = 1/2$ and $r=1/5$}} \\
\hline
$\|x_*\|_{2}$ & $d$ & $l$ & Prob & Cov & ERR & Len & RMSE & Bias & SE \\ 
\hline
\multirow{5}{*}{2.22}  & \multirow{5}{*}{5}&5& 0.627 & 0.94 & 0.35 & 0.32 & 0.08 & -0.01 & 0.08\\ 
& & 10 &0.939 & 0.95 & 0.38 & 0.66 & 0.08 & -0.06 & 0.05 \\ 
&  & 12 &0.940 & 0.99 & 0.07 & 0.89 & 0.09 & -0.07 & 0.06 \\
&  & 15 &0.951 & 0.99 & 0.01 & 0.98 & 0.13 & -0.11 & 0.07 \\
&  & 20 &0.891 & 0.99 & 0.00 & 0.99 & 0.23 & -0.20 & 0.11 \\
\hline
\multirow{5}{*}{2.22} & \multirow{5}{*}{10}&5& 0.564 & 0.93 & 0.23 & 0.29 & 0.08 & 0.00 & 0.08\\ 
& &  10 &0.798 & 0.94 & 0.88 & 0.32 & 0.08 & -0.04 & 0.07 \\ 
& &  12 &0.798 & 0.95 & 0.62 & 0.43 & 0.08 & -0.04 & 0.08 \\
& &  15 &0.815 & 0.97 & 0.20 & 0.67 & 0.10 & -0.06 & 0.08 \\
& &  20 &0.741 & 0.99 & 0.01 & 0.93 & 0.14 & -0.10 & 0.10 \\
\hline
\multirow{5}{*}{2.22} &  \multirow{5}{*}{20}&5& 0.532 & 0.94 & 0.14 & 0.29 & 0.08 & 0.00 & 0.08\\ 
& &  10 &0.665 & 0.92 & 0.62 & 0.28 & 0.08 & -0.01 & 0.08 \\ 
& &  12 &0.665 & 0.94 & 0.60 & 0.30 & 0.08 & -0.01 & 0.08 \\
& &  15 &0.677 & 0.95 & 0.49 & 0.35 & 0.08 & -0.01 & 0.08 \\
& &  20 &0.629 & 0.97 & 0.08 & 0.53 & 0.10 & -0.04 & 0.10  \\
\hline
\end{tabular}
}
\caption{\textbf{Varying sparsity of $\beta$.}  The columns indexed with ``Cov" and ``Len" represent the empirical coverage and length of the constructed CIs respectively; the column indexed with ``ERR" represents the empirical rejection rate of the testing procedure; The columns indexed with ``RMSE", ``Bias" and ``SE" represent the RMSE, bias and standard error, respectively.} 
\label{tab: sparsity}
\end{table}

\subsection{Robustness to Violation of {\rm (A2)}}
\label{sec: A2}    
We now test the robustness of our proposed method to the violation of Condition ($\rm A2$). 
We generate $\beta$ as
\begin{enumerate}
    \item[(S8)] $\beta_1=0$, $\beta_{j}=1$ for $2\leq j\leq 11$ and $\beta_{j}=0$ for $12\leq j\leq 501$. 
\end{enumerate}
    \vspace{-1mm}
    
We construct different covariance matrices $\Sigma$ such that a certain proportion of the conditional case probability $\left\{h\left(X_{i}^{\top} \beta\right)\right\}_{i=1}^{n}$ are near 0 or 1. 

\begin{figure}[htp!]
\centering
\includegraphics[scale=0.6]{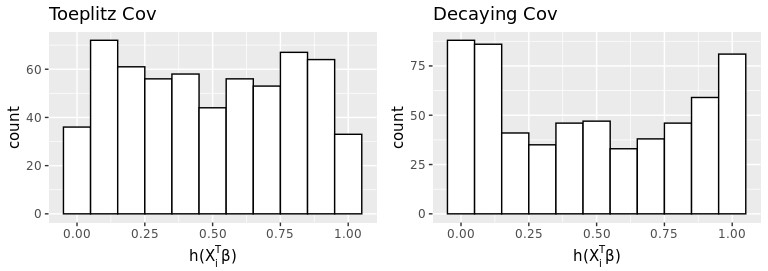}
\caption{Histogram of $\left\{h(X_{i \cdot}^{\intercal}\beta)\right\}_{i=1}^{n}$ for a sample of $n=600$ observations with respect to  setting $(\rm S8)$ with Toeplitz Covariance (left) and Decaying Covariance (right).} 
\label{fig:histogram setting2}
\end{figure}

\begin{enumerate}
    \item[(i)] \textbf{Toeplitz Covariance :} $\Sigma\in \R^{500\times 500}$ is constructed as a block diagonal matrix. Each block, $\Sigma_0$ is a matrix of dimension $50 \times 50$ constructed as:
    $(\Sigma_{0})_{i,i} = 0.5$ for  $1\leq i\leq 50$ and $(\Sigma_{0})_{i,j} = \frac{0.03}{2} (1- {|i-j|}/{49})$ for $1\leq i\neq j\leq 50.$
       \item[(ii)] \textbf{Decaying Covariance :} $\Sigma = \{0.5^{1+|j-l}|\}_{1\leq j \leq l \leq 500}$.\\
\end{enumerate}    
In Figure \ref{fig:histogram setting2}, we plot the case probability for $n = 600$. The left panel of Figure \ref{fig:histogram setting2} corresponds to the setting with the Toeplitz covariance matrix, where $37$ out of $600$ conditional case probabilities lie below $0.1$ while $35$ lie above $0.9$; The right panel of  Figure \ref{fig:histogram setting2} corresponds to the setting with the decaying covariance matrix, where $79$ out of $600$ conditional case probabilities lie below $0.1$ while $77$ lie above $0.9$, which suggests stronger violation of assumption {\rm (A2)}. Due to the deeper U-shape on the right of Figure \ref{fig:histogram setting2}, Condition {\rm (A2)} is more violated for the setting with a decaying covariance matrix.  The inference results are summarized in Table \ref{tab: A2 p=500 Setting 2}.

\begin{table}[htp!]
\centering
\scalebox{0.8}{
\begin{tabular}{|rrr|r|rrrrrr|}
\hline
\multicolumn{10}{|c|}{{\bf Setting $(\rm S8)$, Loading 2 with $q = 1$}} \\
\hline
\multicolumn{10}{|c|}{{\bf Toeplitz Covariance }} \\
\hline
$\|x_*\|_{2}$ &{\rm r}& Prob &$n$& Cov & ERR & Len & RMSE & Bias & SE \\ 
\hline
\multirow{4}{*}{6.21} & \multirow{4}{*}{$\frac{1}{5}$} & \multirow{4}{*}{0.200} &200& 0.97 & 0.00 & 0.75 & 0.28 & 0.09 & 0.27\\ 
& & &400& 0.96 & 0.00 & 0.61 & 0.21 & 0.06 & 0.20\\ 
& & &600& 0.93 & 0.00 & 0.54 & 0.19 & 0.04 & 0.18\\
\hline
\hline
\multicolumn{10}{|c|}{{\bf Decaying Covariance }} \\
\hline
$\|x_*\|_{2}$ &{\rm r}& Prob &$n$& Cov & ERR & Len & RMSE & Bias & SE \\ 
\hline
\multirow{4}{*}{6.21} & \multirow{4}{*}{$\frac{1}{5}$} & \multirow{4}{*}{0.200} &200& 0.98 & 0.00 & 0.95 & 0.30 & 0.11 & 0.28\\ 
& & &400& 0.99 & 0.00 & 0.90 & 0.26 & 0.10 & 0.24\\ 
& & &600& 0.98 & 0.00 & 0.86 & 0.19 & 0.06 & 0.18\\ 
\hline
\end{tabular}
}
\caption{\textbf{Robustness to violation of $(\rm A2)$.} ``r" and``Prob" represent the shrinkage parameter and Case Probability respectively. The columns indexed with ``Cov" and ``Len" represent the empirical coverage and length of the CIs; the column indexed with ``ERR" represents the empirical rejection rate of the test; The columns indexed with ``RMSE", ``Bias" and ``SE" represent the RMSE, bias and standard error, respectively. }
\label{tab: A2 p=500 Setting 2}
\end{table}

In table \ref{tab: A2 p=500 Setting 2}, we observe that the stronger violation of (A2) results in our constructed CIs overcovering. The wider CIs are expected since the weights, $[h(X_{i\cdot}^{\intercal}\beta)(1-h(X_{i\cdot}^{\intercal}\beta))]^{-1}$, can be quite large when a large proportion of $\{h(X_{i\cdot}^{\intercal}\beta)\}_{i=1}^{n}$ are close to 0 or 1.  To summarize, the less U-shaped the histogram of the conditional case probabilities is, the better is the inference produced by the LiVE method.

We plot the histogram of the conditional case probability $\{h(X_{i\cdot}^{\intercal}\beta)\}_{i=1}^{n}$ and Condition (A2) is strongly violated if a large proportion of $\{h(X_{i\cdot}^{\intercal}\beta)\}_{i=1}^{n}$ concentrate around $0$ or $1$. We have plotted the histogram for simulation settings (S1), (S2), (S5) and (S6) in Figure \ref{fig:histogram s1s2s5s6} in the supplement. 

\section{Real Data Analysis}
\label{sec: real}

We applied the proposed methods to develop preliminary models for predicting three related disease conditions, hypertension, hypertension that appears to be resistant to standard treatment (henceforth ``R-hypertension"), and hypertension with unexplained low blood potassium (henceforth ``LP-hypertension"). The data were extracted from the Penn Medicine clinical data repository, including demographics, laboratory results, medication prescriptions, vital signs, and encounter meta information. 
The analysis cohort consisted of $348$ patients who were at least $18$ years old, had at least $5$ office visits over at least three distinct years between $2007$ and $2017$, and at least $2$ office visits were at one of the $37$ primary care practices. Patient charts were reviewed by a dedicated physician to determine each of the three outcome statuses, and unclear cases were secondarily reviewed by an additional expert clinician. The prevalence of the three outcome variables were $39.4\%$, $8.1\%$, and $4.6\%$, respectively. Longitudinal EHR variables, which had varied values over multiple observations, were summarized by minimum, maximum, mean, median, standard deviation, and/or skewness, and these summary statistics were used as predictors after appropriate normalization. Highly right-skewed variables were log-transformed. We included $198$ predictors in the final analyses, after removing those with missing values.  

\begin{figure}[htp!]
\centering
  \includegraphics[scale=0.65]{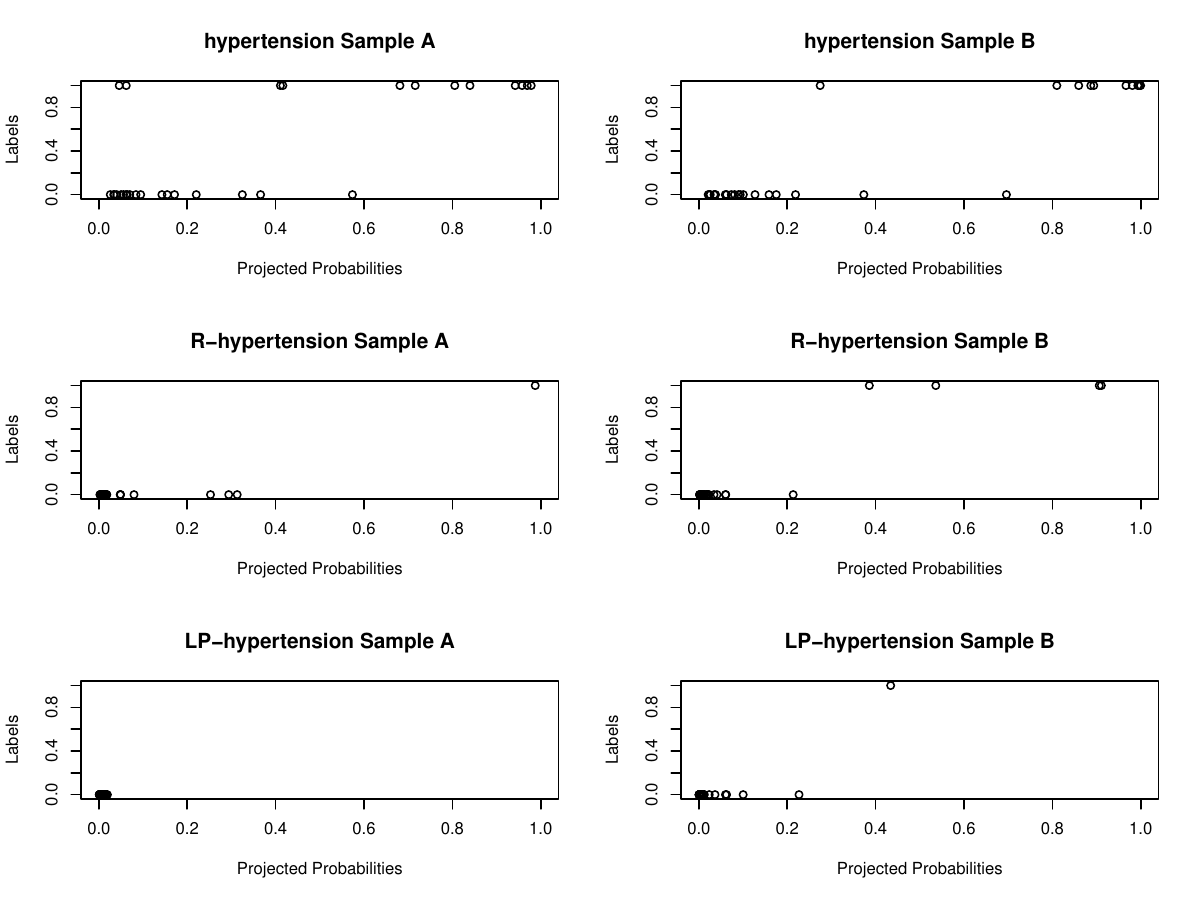}
  \caption{Performance for predicting three phenotypes in two random sub-samples.}
  \label{fig: pred}
\end{figure}

In our analysis, we randomly sampled $30$ patients as the test sample, then their predictor vectors were treated as $\xnew$. A prediction model for each outcome variable was developed using the remaining $318$ patients and then applied to the test sample to obtain bias-corrected estimates of the case probabilities using our method. The left and right columns in Figure \ref{fig: pred} present results on two independent test samples, where the three rows within each column correspond to the three outcome variables. In each panel, the $x$-axis represents the predicted probability generated by our method, and the $y$-axis represents the true outcome status ($1$ or $0$). {In all six panels, the predicted probabilities by the LiVE method for true cases tended to be high and for true controls tended to be low. This illustrates that the LiVE estimator in \eqref{eq: proposed estimator} is predictive for the true outcome status.}

\begin{figure}[htp!]
\centering
  \includegraphics[scale=0.65]{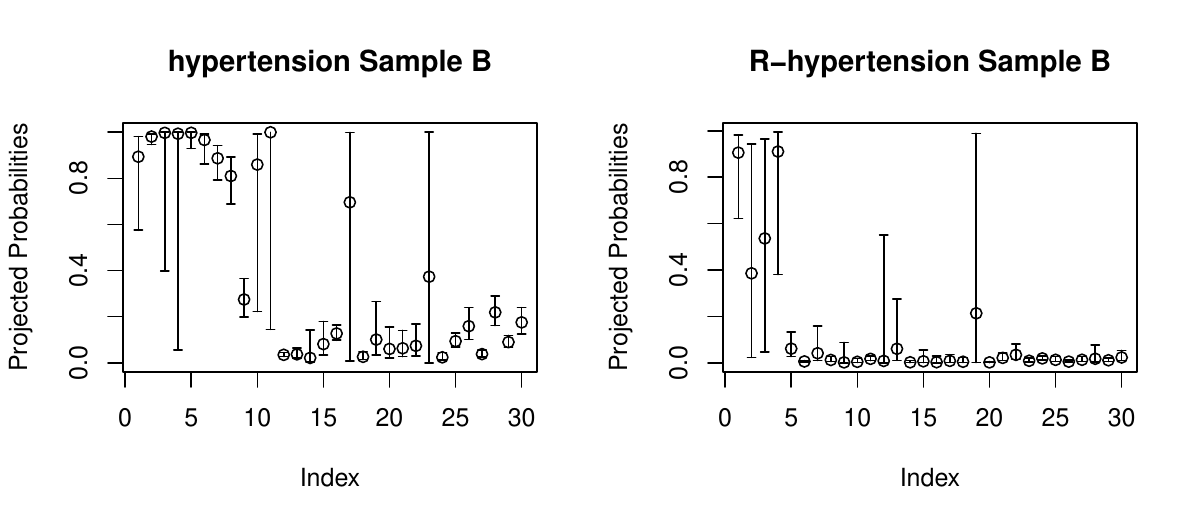}
  \caption{Confidence interval construction: on the left panel, indexes 1 to 11 correspond to observations with hypertension; indices 12 to 30 correspond to those without hypertension. On the right panel, indices 1 to 4 correspond to observations with R-hypertension; indices 5 to 30 correspond to those without R-hypertension.}
  \label{fig: CI}
\end{figure}

{Figure \ref{fig: CI} presented confidence intervals constructed using our method for the case probabilities shown in the top two panels in the right column in Figure \ref{fig: pred}, corresponding to prediction of hypertension and resistant hypertension.} 
The length of the constructed confidence intervals appeared to vary since each patient in the test sample had different observed predictors $\xnew$. This observation is consistent with the established theory in Theorem \ref{thm: limiting distribution}, which states that the length of confidence interval depends on $\|\xnew\|_2$. More interestingly, the constructed confidence intervals appeared to be informative of the outcome statuses for the majority of the test patients. For hypertension, $80\%$ of the confidence intervals lied either above or below $50\%$; For R-hypertension, $83\%$ of the confidence intervals lie either above or below $50\%$. 

{ We further divide the 30 randomly sampled observations into two subgroups by their true status and then investigate the performance of constructed confidence intervals for the subgroup of observations being cases and the other subgroup of observations being controls. On the left panel of Figure \ref{fig: CI}, the observations with indexes between 1 and 11 correspond to cases (observations with hypertension) while the remaining $19$ observations correspond to observations without hypertension. Out of the 11 observations with hypertension, six constructed CIs are predictive with the whole interval above $0.5$, one is misleading as the interval is below $0.5$ and the remaining four are not predictive as the CIs come cross $0.5$; Out of the 19 patients without hypertension, 17 constructed CIs are below $0.5$ and hence predictive but the remaining two are not. On the right hand side of Figure \ref{fig: CI}, the observations with indexes between 1 and 4 correspond to observations with R-hypertension while the remaining $26$ observations correspond to the observations without R-hypertension. Out of the four observations with R-hypertension, only one constructed CI is predictive and the other three are not; out of the 26 observations without R-hypertension, 24 are predictive and the other two are not.} {Overall, the constructed CIs are predictive for the outcome for $77\%$ (hypertension),  $83\%$(R-hypertension), and $77\%$ (LP-hypertension) of subjects, where a constructed CI is predictive if either the constructed CI lies above $0.5$ for the true case or below $0.5$ for the true control. This demonstrated the practical usefulness of the developed models for evaluating the outcome status of patients, the labor-intensive chart review may be avoided for the majority of patients. }

Additional results corresponding to the remaining four panels are presented in Figure \ref{fig: CI comp} in the supplementary materials. The observation is similar to that in Figure \ref{fig: CI}.

\acks{The research of Z. Guo  was supported in part by NSF DMS 1811857, 2015373 and NIH R01GM140463-01,   R56-HL-138306-01. The research of P. Rakshit was supported in part by NSF DMS 1811857 and NIH R01GM140463-01. The research of D. Herman was supported in part by the University of Pennsylvania Department of Pathology and Laboratory Medicine and a Penn Center for Precision Medicine Accelerator Fund Award. The research of J. Chen was supported in part by NIH R56-HL138306, R01-HL138306 and R01GM140463-01. We acknowledge one reviewer for suggesting the comparison with the Transformation Method.
We would like to acknowledge Dr. Qiyang Han for the helpful discussion on contraction principles and Mr. Rong Ma for sharing the \texttt{WLDP} code; We would like to acknowledge the efforts of Xiruo Ding MS and Imran Ajmal MBBS, who were essential to the real data analysis presented. Mr. Ding extracted, wrangled, and engineered the EHR data. Dr. Ajmal performed the chart review for the three clinical phenotypes studied.}

\section{Proof}
\label{sec: proof}
We provide the proof of Theorem \ref{thm: limiting distribution} in Section \ref{sec: thm 1} and that of Lemma \ref{lem: bias control} in Section \ref{sec: proof lemma1}. The remaining proofs are postponed to Section \ref{sec: add proof} in the supplementary material.

We introduce the following events
{\small
$$
\mathcal{A}_1=\left\{\max_{1\leq i\leq n,\; 1\leq j\leq p}\left|X_{ij}\right|\leq C\sqrt{\log n+\log p}\right\}, \;\;
\mathcal{A}_2=\left\{\min_{\|\eta\|_2=1, \|\eta_{S^{c}}\|_1\leq C \|\eta_{S}\|_1}\frac{1}{n}\sum_{i=1}^{n}\left(X_{i\cdot}^{\intercal}\eta\right)^2\geq c\lambda_{\min}\left(\Sigma\right)\right\}$$
$$
\mathcal{A}_3=\left\{\min_{1\leq i\leq n}\frac{\exp\left(X_{i\cdot}^{\intercal}{\beta}\right)}{\left[1+\exp\left(X_{i\cdot}^{\intercal}{\beta}\right)\right]^2}\geq c^2_{\min}\right\}, \;\;
\mathcal{A}_4=\left\{\lambda_0=\left\|\frac{1}{n}\sum_{i=1}^{n}\epsilon_i X_i\right\|_{\infty}\leq C\sqrt{\frac{\log p}{n}}\right\}$$
$$\mathcal{A}_5=\left\{\|\widehat{\beta}-\beta\|_2\leq C \sqrt{\frac{k \log p}{n}}\right\},\;\;
\mathcal{A}_6=\left\{\|(\widehat{\beta}-\beta)_{S^{c}}\|_1\leq C_0 \|(\widehat{\beta}-\beta)_{S}\|_1\right\}$$}
\noindent where $S$ denotes the support of the high-dimensional vector $\beta$.
The following lemma \ref{lem: high prob a} controls the probability of these defined events and the proof is omitted as it is similar to Lemma 4 in \cite{cai2017confidence}. 
\begin{Lemma} Suppose Conditions {\rm (A1)} and {\rm (A2)} hold, then 
$
\PP\left(\cap_{i=1}^{4}\mathcal{A}_i\right)\geq 1-\exp(-cn)-p^{-c}
$
and on the event $\cap_{i=1}^{4}\mathcal{A}_i$, the events $\mathcal{A}_5$ and $\mathcal{A}_6$ hold.
\label{lem: high prob a}
\end{Lemma}
The following Lemma is about the Taylor expansion of logit function and the corresponding proof is presented in Section \ref{sec: logit bound} in the supplementary material.
\begin{Lemma} For $h(x)=\frac{\exp(x)}{1+\exp(x)}$, we have 
\begin{equation}
\left(h'(a)\right)^{-1}\left(h(x)-h(a)\right)=(x-a)+\int_{0}^{1} (1-t)(x-a)^2\frac{h''(a+t(x-a))}{h'(a)} dt.
\label{eq: Taylor expansion}
\end{equation}
where 
$
h'(x)=\frac{\exp(x)}{(1+\exp(x))^2}$ and $h''(x)=\frac{2\exp(2x)}{(1+\exp(x))^3}. 
$
We further have
\begin{equation}
\exp\left(-\left|x-a\right|\right) \leq \frac{h'(x)}{h'(a)}\leq \exp\left(\left|x-a\right|\right) \quad \text{and}\quad \left|\frac{h'(x)}{h'(a)}-1\right|\leq \exp\left(\left|x-a\right|\right)
\label{eq: inequality 1}
\end{equation}
and 
\begin{equation}
\left|\int_{0}^{1} (1-t)(x-a)^2\frac{h''(a+t(x-a))}{h'(a)} dt\right|\leq \exp(|x-a|)(x-a)^2
\label{eq: taylor error bound}
\end{equation}
\label{lem: taylor expansion of logit}
\end{Lemma}

\subsection{{Proof of Theorem \ref{thm: limiting distribution}}}
\label{sec: thm 1}
\underline{Proof of \eqref{eq: dominating variance}.} On the event $\mathcal{A}_3$, we have
$
\widehat{u}^{\intercal}\left[\frac{1}{n^{2}}\sum_{i=1}^{n}X_{i\cdot}X_{i\cdot}^{\intercal}\right]\widehat{u} \leq {\rm V} \leq \frac{1}{c_{\min}^{2}} \widehat{u}^{\intercal}\left[\frac{1}{n^{2}}\sum_{i=1}^{n}X_{i\cdot}X_{i\cdot}^{\intercal}\right]\widehat{u}.$
To control the upper bound part $\sqrt{\rm V}\leq \frac{C_0\|\xnew\|_2}{n}$, we define the following events
\begin{equation}
\begin{aligned}
\mathcal{B}_1&=\left\{\left \|\widehat{\Sigma}\Sigma^{-1}\xnew-\xnew\right\|_{\infty}\leq  \|\xnew\|_2 \lambda_{n}\right\}; \quad
\mathcal{B}_2=\left\{\left\|\xnew^{\intercal}\widehat{\Sigma}\Sigma^{-1}\xnew-\|\xnew\|_2^2 \right|\leq \|\xnew\|_2^2\lambda_{n} \right\} \\
\mathcal{B}_3&=\left\{\|X \Sigma^{-1}\xnew\|_{\infty} \leq \|\xnew\|_2 \tau_n \right\}
\end{aligned}
\end{equation}
The following lemma controls the probability of $\cap_{i=1}^{3}\mathcal{B}_i$ and its proof is presented in section \ref{sec: proof of lem 4}. 
\begin{Lemma} Suppose Condition {\rm (A1)} holds and  $\lambda_{n}\asymp \sqrt{{\log p}/{n}}$ and $\tau_n\lesssim n^{\delta}$ for $0<\delta<\frac{1}{2}$, then 
\begin{equation}
\PP\left(\cap_{i=1}^{3}\mathcal{B}_i\right)\geq 1-n^{-c}-p^{-c}.
\label{eq: high prob b}
\end{equation}
\label{lem: high prob b}
\end{Lemma}
On the event $\cap_{i=1}^{3}\mathcal{B}_i$, then $u=\Sigma^{-1}\xnew$ satisfies the constraints \eqref{eq: constraint 1}, \eqref{eq: constraint 2} and \eqref{eq: constraint 3}. As a consequence, the feasible set is non-empty on the event $\cap_{i=1}^{3}\mathcal{B}_i$ and we further obtain an upper bound for the minimum value, that is, 
$
{\rm V}\leq {\xnew^{\intercal}\Sigma^{-1}\widehat{\Sigma}\Sigma^{-1}\xnew}/{n}.
$

The proof of the lower bound part $\sqrt{\rm V}\geq \frac{c_0\|\xnew\|_2}{n}$ is facilitated by the optimization constraint \eqref{eq: constraint 1}. We define a proof-facilitating optimization problem,
\begin{equation}
\begin{aligned}
\widetilde{u}=\;\argmin_{u\in \R^{p}} u^{\intercal}\left(\frac{1}{n}\sum_{i=1}^{n} X_{i\cdot} X^{\intercal}_{i\cdot}\right)u \quad \text{subject to} \quad |\xnew^{\intercal}\widehat{\Sigma}u-\|\xnew\|_2^2 |\leq \|\xnew\|_2^2\lambda_{n} \end{aligned}
\label{eq: proof-facilitating}
\end{equation}
Note that $\widehat{u}$ satisfies the feasible set of \eqref{eq: proof-facilitating} and hence
\begin{equation}
\begin{aligned}
&\widehat{u}^{\intercal}\left(\frac{1}{n}\sum_{i=1}^{n} X_{i\cdot} X^{\intercal}_{i\cdot}\right)\widehat{u}\geq \widetilde{u}^{\intercal}\left(\frac{1}{n}\sum_{i=1}^{n} X_{i\cdot} X^{\intercal}_{i\cdot}\right)\widetilde{u}\\
&\geq \widetilde{u}^{\intercal}\left(\frac{1}{n}\sum_{i=1}^{n} X_{i\cdot} X^{\intercal}_{i\cdot}\right)\widetilde{u}+t\left((1-\lambda_{n})\|\xnew\|_2^2-\xnew^{\intercal}\widehat{\Sigma}\widetilde{u}\right) \; \text{for any}\; t\geq 0,
\end{aligned}
\label{eq: proof-facilitating inequality 1}
\end{equation}
where the last inequality follows from the constraint of \eqref{eq: proof-facilitating}. For a given $t\geq 0$, we have 
\begin{equation}
\begin{aligned}
&\widetilde{u}^{\intercal}\left(\frac{1}{n}\sum_{i=1}^{n} X_{i\cdot} X^{\intercal}_{i\cdot}\right)\widetilde{u}+t\left((1-\lambda_{n})\|\xnew\|_2^2-\xnew^{\intercal}\widehat{\Sigma}\widetilde{u}\right) \\
&\geq \min_{u\in \R^{p}}{u}^{\intercal}\left(\frac{1}{n}\sum_{i=1}^{n} X_{i\cdot} X^{\intercal}_{i\cdot}\right){u}+t\left((1-\lambda_{n})\|\xnew\|_2^2-\xnew^{\intercal}\widehat{\Sigma}{u}\right).
\end{aligned}
\label{eq: proof-facilitating inequality 2}
\end{equation}
By solving the minimization problem of the right hand side of \eqref{eq: proof-facilitating inequality 2}, we have the minimizer $u^{*}$ satisfies $\widehat{\Sigma}u^{*}=\frac{t}{2} \widehat{\Sigma}\xnew$ and hence the minimized value of the right hand side of \eqref{eq: proof-facilitating inequality 2} is 
$-\frac{t^2}{4}\xnew^{\intercal}\widehat{\Sigma}\xnew+t(1-\lambda_{n})\|\xnew\|_2^2.$
Combined with \eqref{eq: proof-facilitating inequality 1} and \eqref{eq: proof-facilitating inequality 2}, we have 
\begin{equation}
\widehat{u}^{\intercal}\left(\frac{1}{n}\sum_{i=1}^{n} X_{i\cdot} X^{\intercal}_{i\cdot}\right)\widehat{u}\geq \max_{t\geq 0}\left[-\frac{t^2}{4}\xnew^{\intercal}\widehat{\Sigma}\xnew+t(1-\lambda_{n})\|\xnew\|_2^2 \right].
\label{eq: proof-facilitating inequality 3}
\end{equation}
For $t^{*}=2\frac{(1-\lambda_{n})\|\xnew\|_2^2}{\xnew^{\intercal}\widehat{\Sigma}\xnew}>0$, the minimum of the right hand side of \eqref{eq: proof-facilitating inequality 3} is achieved and hence establish 
\begin{equation}
\widehat{u}^{\intercal}\left(\frac{1}{n}\sum_{i=1}^{n} X_{i\cdot} X^{\intercal}_{i\cdot}\right)\widehat{u}\geq \frac{(1-\lambda_{n})^2\|\xnew\|_2^4}{\xnew^{\intercal}\widehat{\Sigma}\xnew}.
\end{equation}
Then $\PP\left[{{\rm V}^{-1/2}}\left(\widehat{\xnew^{\intercal}\beta}-\xnew^{\intercal}\beta\right)\geq z_{\alpha}\right]\rightarrow \alpha$
follows from the decomposition \eqref{eq: decomposition}, the variance control in \eqref{eq: dominating variance}, Lemma \ref{lem: bias control} and Proposition  \ref{prop: decomposition} and the following lemma.
\begin{Lemma} Suppose that Conditions {\rm (A1)} and ${\rm (A2)}$ hold and $\tau_{n}$ defined in \eqref{eq: constraint 3} satisfies $\left(\log n\right)^{1/2}\lesssim \tau_{n}\ll n^{1/2}$, then
$
\frac{1}{V^{1/2}}\widehat{u}^{\intercal}\frac{1}{n}\sum_{i=1}^{n} [h(X_{i\cdot}^{\intercal}{\beta})(1-h(X_{i\cdot}^{\intercal}{\beta}))]^{-1}X_{i\cdot}\epsilon_i\rightarrow N(0,1)
$
where ${\rm V}$ is defined in \eqref{eq: asymptotic variance}.
\label{lem: dominating variance}
\end{Lemma}

\subsection{Proof of Lemma \ref{lem: bias control}}
\label{sec: proof lemma1}

To start the proof, we recall that $h(z)=\frac{\exp(z)}{1+\exp(z)}$ and define the functions $g_i$ for $1\leq i\leq n$ 
\begin{equation*}
g_i(t_i)=\left(\left(\frac{\exp\left(X_{i\cdot}^{\intercal}{\beta}+t_i\right)}{\left(1+\exp\left(X_{i\cdot}^{\intercal}{\beta}+t_i\right)\right)^2}\right)^{-1}-\left(\frac{\exp\left(X_{i\cdot}^{\intercal}{\beta}\right)}{\left(1+\exp\left(X_{i\cdot}^{\intercal}{\beta}\right)\right)^2}\right)^{-1}\right)\widehat{u}^{\intercal}X_{i\cdot}, \end{equation*}
and the space for $\delta\in \R^{p}$ as
 \begin{equation}
 \mathcal{C}=\left\{\delta: \|\delta_{S^{c}}\|_1\leq  c\|\delta_{S}\|_1,\; \|\delta\|_2\leq C^{*} \sqrt{\frac{k \log p}{n}}\right\}.
 \label{eq: restricted space}
 \end{equation}
 for some positive constants $c>0$ and $C^{*}>0$.
We further define 
\begin{equation}
\mathcal{T}=\left\{t=(t_1,\cdots,t_n): t_i=X_{i\cdot}^{\intercal}\delta \;\;\text{where}\;\; \delta \in \mathcal{C}\right\},
\label{eq: mapped restricted space}
\end{equation}
We can rewrite the main component of the left hand side of \eqref{eq: error bound 2} as
\begin{equation}
\begin{aligned}
&\left|\widehat{u}^{\intercal}\frac{1}{\sqrt{n}}\sum_{i=1}^{n}\left(\left(\frac{\exp(X_{i\cdot}^{\intercal}\widehat{\beta})}{\left(1+\exp(X_{i\cdot}^{\intercal}\widehat{\beta})\right)^2}\right)^{-1}-\left(\frac{\exp\left(X_{i\cdot}^{\intercal}{\beta}\right)}{\left(1+\exp\left(X_{i\cdot}^{\intercal}{\beta}\right)\right)^2}\right)^{-1}\right)X_{i\cdot}\epsilon_i\right|\cdot {\bf 1}_{\mathcal{A}_1\cap\mathcal{A}_{3}\cap\mathcal{A}_5\cap \mathcal{A}_6} \\
&\leq \sup_{\delta\in \mathcal{C}}\left|\frac{1}{\sqrt{n}}\sum_{i=1}^{n} g_i(X_{i\cdot}^{\intercal}\delta)\cdot {\bf 1}_{\mathcal{A}_1\cap\mathcal{A}_{3}}\cdot \epsilon_i\right|=\sup_{t\in \mathcal{T}}\left|\frac{1}{\sqrt{n}}\sum_{i=1}^{{n}} g_i(t_i)\cdot {\bf 1}_{\mathcal{A}_1\cap\mathcal{A}_{3}}\cdot \epsilon_i\right|
\end{aligned}
\label{eq: bound over sup a}
\end{equation}
where $\mathcal{C}$ is defined in \eqref{eq: restricted space} and $\mathcal{T}$ is defined in \eqref{eq: mapped restricted space}.  In the following, we control the last part of \eqref{eq: bound over sup a} via applying the symmetrization technique \cite{van2006empirical} stated in Lemma \ref{lem: symmetrization} and the contraction principle in Lemma \ref{lem: contraction}. The proofs of Lemma \ref{lem: symmetrization} and Lemma \ref{lem: contraction} are presented in Sections \ref{sec: proof sym} and \ref{sec: proof contraction} in the supplementary materials, respectively.
\begin{Lemma}
Suppose that $y_i'$ is an independent copy of $y_i$ and $\epsilon'_i$ is defined as $y'_i-\E(y'_i\mid X_i)$. For all convex nondecreasing functions $\Phi: \R_{+}\rightarrow \R_{+}$, then 
\begin{equation}
\E \Phi\left(\sup_{t\in \mathcal{T}}\left|\sum_{i=1}^{n} g_i(t_i) \epsilon_i\right|\right)\leq \E \Phi\left(\sup_{t\in \mathcal{T}}\left|\sum_{i=1}^{n} g_i(t_i) \xi_i\right|\right),
\label{eq: symmetrization}
\end{equation}
where 
$
\xi_i=\epsilon_i-\epsilon'_i=y_i-y'_i.
$
\label{lem: symmetrization}
\end{Lemma}
The following lemma is a modification of Theorem 2.2 in \cite{koltchinskii2011oracle}, where the result in \cite{koltchinskii2011oracle} was only developed for i.i.d Rademacher variables $\xi_i$. The following lemma is more general in the sense that $\xi_1, \xi_2,\cdots, \xi_{n}$ are only required to be independent and satisfy the probability distribution \eqref{eq: distribution}. The following lemma can also be derived by extending the proof of Theorem $4.12$ in \cite{contractionreview}. To be self-contained, we give a proof of Lemma \ref{lem: contraction} in the supplementary section \ref{sec: proof contraction}.
\begin{Lemma} Let $t=(t_1,\cdots,t_n)\in\mathcal{T}\subset R^{n}$ and let $\phi_i: \R\rightarrow\R, i=1,\cdots,n$ be functions such that $\phi_i(0)=0$ and 
$
\left|\phi_i(u)-\phi_i(v)\right|\leq |u-v|, u,v\in \R.
$
For all convex nondecreasing functions $\Phi: \R_{+}\rightarrow \R_{+}$, then 
\begin{equation}
\E \Phi\left(\frac{1}{2}\sup_{t\in \mathcal{T}}\left|\sum_{i=1}^{n}\phi_i(t_i)\xi_i\right|\right)\leq \E \Phi\left(\sup_{t\in \mathcal{T}}\left|\sum_{i=1}^{n}t_i\xi_i\right|\right),
\label{eq: contraction}
\end{equation}
where $\left\{\xi_i\right\}_{1\leq i\leq n}$ are independent random variables with the probability density function 
\begin{equation}
\PP\left(\xi_i=1\right)=\PP\left(\xi_i=-1\right)\in [0, \frac{1}{2}], \; \PP\left(\xi_i=0\right)=1-2\PP\left(\xi_i=1\right).
\label{eq: distribution}
\end{equation}
\label{lem: contraction}
\end{Lemma}

We will apply Lemmas \ref{lem: symmetrization} and \ref{lem: contraction} and  control $\sup_{t\in \mathcal{T}}\left|\frac{1}{\sqrt{n}}\sum_{i=1}^{{n}} g_i(t_i)\cdot {\bf 1}_{\mathcal{A}_1\cap \mathcal{A}_3}\cdot \epsilon_i\right|$ in \eqref{eq: bound over sup a}.
 For $t,s\in \mathcal{T} \subset \R^{n}$, then there exist $\delta^{t},\delta^{s}\in \mathcal{C}\subset \R^{p}$ such that 
$
t_i-s_i=X_{i\cdot}^{\intercal}\left(\delta^{t}-\delta^{s}\right)$ and $t_i=X_{i\cdot}^{\intercal}\delta^{t}$ for $1\leq i\leq n.$
Hence on the event $\mathcal{A}_1,$
\begin{equation}
\max\left\{\max_{1\leq i\leq n}|t_i-s_i|,\max_{1\leq i\leq n}|t_i|\right\}\leq C k \sqrt{\frac{\log p}{n}} \sqrt{\log p+\log n}\leq 1. 
\label{eq: difference control}
\end{equation}
where the last inequality follows as long as $\sqrt{n} \geq k \log p \left(1+\sqrt{\frac{\log n}{\log p}}\right)$ \\
\\
Define $q(x)=\left(\frac{\exp(x)}{(1+\exp(x))^2}\right)^{-1}$
and then 
\begin{equation}
\begin{aligned}
g_i(s_i)-g_{i}(t_i)
=\left(\frac{q\left(X_{i\cdot}^{\intercal}{\beta}+s_i\right)}{q\left(X_{i\cdot}^{\intercal}{\beta}+t_i\right)}-1\right)\frac{q\left(X_{i\cdot}^{\intercal}{\beta}+t_i\right)}{q\left(X_{i\cdot}^{\intercal}{\beta}\right)} q\left(X_{i\cdot}^{\intercal}{\beta}\right)\widehat{u}^{\intercal}X_{i\cdot}.
\end{aligned}
\label{eq: re-express}
\end{equation}
By \eqref{eq: inequality 1}, we have 
\begin{equation}
\left|\frac{q\left(X_{i\cdot}^{\intercal}{\beta}+s_i\right)}{q\left(X_{i\cdot}^{\intercal}{\beta}+t_i\right)}-1\right|\leq \left|\exp(\left|s_i-t_i\right|)-1\right|\leq e|s_i-t_i|,
\label{eq: contra func}
\end{equation}
where the last inequality holds as long as $\left|s_i-t_i\right|$ is sufficiently small, as verified in \eqref{eq: difference control}. Similarly, we establish that $\frac{q\left(X_{i\cdot}^{\intercal}{\beta}+t_i\right)}{q\left(X_{i\cdot}^{\intercal}{\beta}\right)} \leq e.$
Combined with  \eqref{eq: re-express} and \eqref{eq: contra func}, we obtain
\begin{equation}
\left|g_i(s_i)-g_{i}(t_i)\right|\leq \frac{1}{c_{\min}^2} e^{2}\left|s_i-t_i\right| \left|\widehat{u}^{\intercal}X_{i\cdot}\right|\leq  \frac{1}{c_{\min}^2} e^{2}\left|s_i-t_i\right|\|\xnew\|_2 \tau_n,
\label{eq: generic lip}
\end{equation}
where the last inequality follows from the constraint \eqref{eq: constraint 3}. By applying \eqref{eq: generic lip}, we have 
\begin{equation}
\frac{1}{L_{n}}\left|g_i(t_i)-g_{i}(s_i)\right|\cdot {\bf 1}_{\mathcal{A}_1\cap\mathcal{A}_3}\leq  \left|t_i-s_i\right| \;\; \text{where} \;\; L_n=\frac{e^2}{c_{\min}^2} \|\xnew\|_2 \tau_n.
\end{equation}
Define 
$
\phi_{i}(t_i)=\frac{1}{L_{n}}g_{i}(t_i)\cdot {\bf 1}_{\mathcal{A}_1}.$ We then apply \eqref{eq: symmetrization} and \eqref{eq: contraction} with $\Phi(x)=x$ and obtain
\begin{equation*}
\E_{\xi\mid X} \sup_{t\in \mathcal{T}}\left|\frac{1}{n}\sum_{i=1}^{n} \phi_i(t_i)\cdot {\bf 1}_{\mathcal{A}_1\cap\mathcal{A}_3} \xi_i\right|\leq 2\E_{\xi\mid X} \sup_{\delta\in \mathcal{C}} \left|\frac{1}{n}\sum_{i=1}^{n}\delta^{\intercal}X_{i\cdot}\xi_i\right|
\label{eq: contraction app1}
\end{equation*}
and hence 
$
\E\sup_{t\in \mathcal{T}}\left|\frac{1}{n}\sum_{i=1}^{n} \phi_i(t_i)\cdot {\bf 1}_{\mathcal{A}_1\cap\mathcal{A}_3} \xi_i\right|\leq 2\E \sup_{\delta\in \mathcal{C}} \left|\frac{1}{n}\sum_{i=1}^{n}\delta^{\intercal}X_{i\cdot}\xi_i\right|.$
Note that
\begin{equation*}
\E \sup_{\delta\in \mathcal{C}} \left|\frac{1}{n}\sum_{i=1}^{n}\delta^{\intercal}X_{i\cdot}\xi_i\right|\leq \sup_{\delta\in \mathcal{C}}\|\delta\|_1 \E \left\|\frac{1}{n}\sum_{i=1}^{n}X_{i\cdot}\xi_i\right\|_{\infty}\leq \sup_{\delta\in \mathcal{C}}\|\delta\|_1  \sqrt{\frac{2 \log p}{n}} \|X_{i\cdot}\|_{\psi_2},
\end{equation*}
where the last inequality follows from the fact that $\frac{1}{\sqrt{n}}\sum_{i=1}^{n}X_{i\cdot}\xi_i$ is sub-gaussian random variable with sub-gaussian norm $\|X_{i\cdot}\|_{\psi_2}$.
Combined with  $\sup_{\delta\in \mathcal{C}}\|\delta\|_1\leq C k\sqrt{\frac{\log p}{n}}$, we establish  $\E \sup_{\delta\in \mathcal{C}} \left|\frac{1}{n}\sum_{i=1}^{n}\delta^{\intercal}X_{i\cdot}\xi_i\right|\leq C \frac{k \log p}{n}\|X_{i\cdot}\|_{\psi_2}$ and $
\E\sup_{t\in \mathcal{T}}\left|\frac{1}{n}\sum_{i=1}^{n} \phi_i(t_i)\cdot {\bf 1}_{\mathcal{A}_1} \xi_i\right|\leq C \frac{k \log p}{n}\|X_{i\cdot}\|_{\psi_2}.
$
By Chebyshev's inequality,  
$$
\PP\left(\sup_{t\in \mathcal{T}}\left|\frac{1}{n}\sum_{i=1}^{n} g_i(t_i)\cdot {\bf 1}_{\mathcal{A}_1} \xi_i\right|\geq  C t \|\xnew\|_2 \tau_n \frac{k \log p}{n}\|X_{i\cdot}\|_{\psi_2}
\right)\leq \frac{1}{t}.
$$
By \eqref{eq: bound over sup a}, we establish that  \eqref{eq: error bound 2} holds with probability larger than $1-(\frac{1}{t}+p^{-c}+\exp(-cn)).$

\vskip 0.2in
\bibliography{HDRef}

\newpage

\appendix
\counterwithin{figure}{section}
\counterwithin{table}{section}

\vspace*{-10pt}

\section{Additional Discussion}
\subsection{Technical Difficulty of the Plug-in Debiased Estimator}
\label{sec: technical diff}
There exists technical difficulties to establish the asymptotic normality of the plug-in estimators $\xnew^{\intercal}\widetilde{\beta}$ with $\widetilde{\beta}\in \R^{p}$ denoting any coordinate-wise bias-corrected estimator proposed in \cite{van2014asymptotically, ning2017general,ma2018global}. To see this, we can apply the results in  \cite{van2014asymptotically, ning2017general,ma2018global} to show that for $1\leq j\leq p$, $$\widetilde{\beta}_j=\beta_j+M(\widetilde{\beta}_j)+{\rm Bias}(\widetilde{\beta}_j)$$ where $\sqrt{n}M(\widetilde{\beta}_j)$ is asymptotically normal and ${\rm Bias}(\widetilde{\beta}_j)$ is a small bias component. Then we have the following error decomposition 
$$\xnew^{\intercal}\widetilde{\beta}-\xnew^{\intercal}{\beta}=\sum_{j=1}^{p}x_{*,j}M(\widetilde{\beta}_j)+\sum_{j=1}^{p}x_{*,j}{\rm Bias}(\widetilde{\beta}_j).$$ The component  $\sqrt{n}\sum_{j=1}^{p}x_{*,j}M(\widetilde{\beta}_j)$ is asymptotically normal with its standard error of the order $\|\xnew\|_2$ and the bias $\sum_{j=1}^{p}x_{*,j}{\rm Bias}(\widetilde{\beta}_j)$ is upper bounded by $\|\xnew\|_1 k \log p/n$, with a high probability. If $\|\xnew\|_1$ is much larger than $\|\xnew\|_2$, the upper bound for the bias $\sum_{j=1}^{p}x_{*,j}{\rm Bias}(\widetilde{\beta}_j)$ is not necessarily dominated by the standard error of $\sum_{j=1}^{p}x_{*,j}M(\widetilde{\beta}_j)$, even if $k\ll \sqrt{n}/\log p$. 

We shall point out that, the upper bound for the bias depends on $\|\xnew\|_1$ instead of $\|\xnew\|_2$ mainly because the coordinate-wise inference results constrained the bias ${\rm Bias}(\widetilde{\beta}_j)$ separately instead of directly constraining $\sum_{j=1}^{p}x_{*,j}{\rm Bias}(\widetilde{\beta}_j)$ as a total. This makes it challenging to establish asymptotic normality of the plug-in estimators for any high-dimensional loading $\xnew$.

\subsection{A Brief Review of \texorpdfstring{\cite{ma2018global}}{Ma2018global}}
\label{sec: Ma review}
The bias corrected estimator proposed in  \cite{ma2018global} is 
\begin{equation}
\widetilde{\beta}_{j}=\widehat{\beta}_{j}+\frac{\sum_{i=1}^{n} v_{i j}\left(y_{i}-h\left(\widehat{\beta}^{\intercal} X_{i\cdot}\right)\right)}{\sum_{i=1}^{n} v_{i j} h^{\prime}\left(\widehat{\beta}^{\intercal} X_{i\cdot}\right) X_{i j}}, \quad j=1, \ldots, p    
\label{eq: Ma correction}
\end{equation}
where $\widehat{\beta}$ is the penalized logistic estimator of $\beta$ and $v_j = (v_{j1},\cdots,v_{jn})^{\intercal}$ is the score vector to be constructed. 
Let $X_{j}$ and $X_{-j}$ denote the $j$-th column of $X$ and the submatrix of $X$ excluding the $j$-th column, respectively. \cite{ma2018global} construct the score $v_j$ as follows, 
$$
    v_{j}=\widehat{W}^{-1}(X_{j}-X_{-j} \widehat{\gamma}) $$
where $$
    \widehat{\gamma}=\underset{b}{\arg \min }\left\{\frac{\left\|X_{j}-X_{-j} b\right\|_{2}^{2}}{2 n}+\lambda\|b\|_{1}\right\} , \quad \widehat{W} = \operatorname{diag}\left(h^{\prime}(\widehat{\beta}^{\intercal}X_{1\cdot}),\cdots,h^{\prime}(\widehat{\beta}^{\intercal}X_{n\cdot})\right).$$
Then the estimator in \eqref{eq: Ma correction} can be written as 
\begin{equation}
\widetilde{\beta}_{j}=\widehat{\beta}_{j}+\frac{\sum_{i=1}^{n} [h(X_{i\cdot}^{\intercal}\widehat{\beta})(1-h(X_{i\cdot}^{\intercal}\widehat{\beta}))]^{-1} (X_{i,j}-X_{i,-j}^{\intercal}\widehat{\gamma}) \left(y_{i}-h\left(\widehat{\beta}^{\intercal} X_{i\cdot}\right)\right)}{\sum_{i=1}^{n} (X_{i,j}-X_{i,-j}^{\intercal}\widehat{\gamma}) X_{i j}}. 
\label{eq: Ma correction simplified}
\end{equation}

The results in \cite{ma2018global} are about inference for $\beta_j$ instead of an arbitrary linear combination $\xnew^{\intercal}\beta.$ 
This bias-corrected estimator in \eqref{eq: Ma correction} is shown to be effective under a sparsity condition on $\Sigma^{-1}e_j$ where $e_j$ is the $j-$th Euclidean basis. However, it is not straightforward to extend this to deal with arbitrary $x_*$ and non-sparse $\Sigma^{-1}.$ 

\subsection{Additional discussion about \texorpdfstring{\cite{tripuraneni}}{tripuraneni}}
\label{sec: tripuraneni}
The focus of the paper by \cite{tripuraneni} is on  estimation of linear functional or the related prediction problem in high-dimensional linear models. However, the high-dimensional estimation and confidence interval construction can be very different for a dense loading $\xnew.$ With respect to the method proposed in Section 3.1 in \citet{tripuraneni}, this difference has been established in \citet{cai2019individualized} in the high-dimensional linear model. 

Firstly, Proposition $3$ in \cite{cai2019individualized} established that if the loading $x_*$ is of certain dense structure, then the projection direction introduced in Section $3.1$ of \cite{tripuraneni} is zero and hence the ``bias-corrected" estimator is reduced to the plug-in estimator. We believe that this fact is also true in case of high-dimensional logistic regression. We have further shown that the plug-in estimator has a large bias and is not suitable for confidence interval construction.

Secondly, the confidence interval construction in Proposition 4 of \cite{tripuraneni} requires the sparsity of $\beta$, which may be hard to obtain in practical applications. 

\section{Additional Proofs}
\label{sec: add proof}

\subsection{{Proof of Proposition \ref{prop: decomposition}}}

\noindent \underline{Proof of \eqref{eq: error bound 1}} The first inequality of \eqref{eq: error bound 1} follows from Holder's inequality and the second inequality follows from Condition {\rm (B)}.\\
\noindent \underline{Proof of \eqref{eq: error bound 3}} By Cauchy inequality, we have 
\begin{equation}
\sqrt{n}\left|\widehat{u}^{\intercal}\frac{1}{n}\sum_{i=1}^{n}X_{i\cdot}\Delta_i\right|\leq \max_{1\leq i\leq n}\left|\widehat{u}^{\intercal}X_{i\cdot}\right|\frac{1}{\sqrt{n}}\sum_{i=1}^{n} \left|\Delta_i\right|\leq \tau_{n} \|\xnew\|_2 \frac{1}{\sqrt{n}}\sum_{i=1}^{n} \left|\Delta_i\right|
\label{eq: diminishing error b}
\end{equation}
By Lemma \ref{lem: taylor expansion of logit}, we have
$
\left|\Delta_i\right|\leq \exp\left(|X^{\intercal}_{i\cdot}(\widehat{\beta}-\beta)|\right)\left(X^{\intercal}_{i\cdot}(\widehat{\beta}-\beta)\right)^2.
$
On the event $\mathcal{A}=\cap_{i=1}^{6}\mathcal{A}_{i}$, we have 
\begin{equation}
\begin{aligned}
&\sum_{i=1}^{n}\left|\Delta_i\right| \leq \sum_{i=1}^{n}\exp\left(|X^{\intercal}_{i\cdot}(\widehat{\beta}-\beta)|\right)\left(X^{\intercal}_{i\cdot}(\widehat{\beta}-\beta)\right)^2\\
&\leq \exp\left(\max|X_{ij}|\cdot\|\widehat{\beta}-\beta\|_1\right)\sum_{i=1}^{n}\left(X^{\intercal}_{i\cdot}(\widehat{\beta}-\beta)\right)^2\leq C \sum_{i=1}^{n}\left(X^{\intercal}_{i\cdot}(\widehat{\beta}-\beta)\right)^2.
\end{aligned}
\label{eq: inter bound 1}
\end{equation}
where the second inequality follows from Holder inequality and the last inequality follows from the fact that $\sqrt{n}\gg k\log p\left(1+\sqrt{\frac{\log n}{\log p}}\right).$
On the event $\mathcal{A}$, we have 
\begin{equation}
\frac{1}{n}\sum_{i=1}^{n}\left(X^{\intercal}_{i\cdot}(\widehat{\beta}-\beta)\right)^2\leq C \|\widehat{\beta}-\beta\|_2^2\leq C\frac{k \log p}{n}.
\end{equation}
Together with \eqref{eq: diminishing error b} and \eqref{eq: inter bound 1}, we establish that, on the event $\mathcal{A}$,
\begin{equation}
\sqrt{n}\left|\widehat{u}^{\intercal}\frac{1}{n}\sum_{i=1}^{n}X_{i\cdot}\Delta_i\right| \leq C \tau_{n} \|\xnew\|_2 \frac{k \log p}{\sqrt{n}}.
\end{equation}
\subsection{Proof of Lemma \ref{lem: high prob b}}
\label{sec: proof of lem 4}
 Define $u^{*}=\Sigma^{-1}x_*$. Let $D \in \mathbb{R}^{n}$ be defined as $D := \frac{\widehat{\Sigma}u^{*}-x_*}{\|u^{*}\|_2}$ so that the $j^{\text{th}}$ element of $D$ is given by $D_j = \frac{\left(\frac{1}{n}\sum_{i=1}^{n}X_{ij}X_{i\cdot}^{\intercal}u^{*}-x_{*,j}\right)}{\|u^{*}\|_2}$ where $x_{*,j}$ denotes the $j^{\text{th}}$ component of $x_*$. Due to sub-gaussianity of the design $\{X_{i\cdot}\}_{i=1}^{n}$, $D_j$ is a sub-exponential random variable. Since $\mathbb{E}(D_j) = 0$ for all $1\leq j\leq p$, we apply Corollary $5.17$ in \cite{subgaussian} and establish,
    \begin{equation}
    \begin{aligned}
    \mathbb{P}\left(|D_j|\geq C\sqrt{\frac{\log p}{n}}\right)& \leq p^{-c_0} \quad \text{for some}\quad C,c_0>0 \quad \forall j\\
    \implies \mathbb{P}\left(\|D\|_{\infty}\geq C\sqrt{\frac{\log p}{n}}\right)& \leq p^{1-c_0}
    \end{aligned}
    \label{eq: high prob b1}
    \end{equation}
    
 Let $\widetilde{D} := \frac{x_*^{\intercal}\widehat{\Sigma}u^{*}-\|x_*\|_2^{2}}{\|u^{*}\|_2^{2}}$. Note that $\widetilde{D}$ is centered and  sub-exponential random variable. Hence, by Corollary $5.17$ in \cite{subgaussian},
    \begin{equation}
    \mathbb{P}\left(|\widetilde{D}|\geq C_1\sqrt{\frac{\log p}{n}}\right) \leq p^{-c_0} \quad \text{for some}\quad C_1,c_1>0.
    \label{eq: high prob b2}
    \end{equation}
    
 By Condition {\rm (A1)}, we have $\|\Sigma^{-1}\|\leq C_2$ where $C_2>0$ is a constant. This implies $\|u^{*}\|_2=\|\Sigma^{-1}x_*\|_2 \leq C_2 \|x_*\|_2$. By sub-gaussianity of $X$, using Proposition $5.10$ in \cite{subgaussian}, 
    \begin{equation}
    \|Xu^{*}\|_{\infty} = \max_{1\leq i \leq n}|X_{i\cdot}^{\intercal}u^{*}| \leq \|x_*\|_2\tau_n
    \label{eq: high prob b3}
    \end{equation}
    holds with probability of at least $n^{(1-c_1)}$ where $c_1>0$ is a constant.
  Combining (\ref{eq: high prob b1}), (\ref{eq: high prob b2}) and (\ref{eq: high prob b3}) we establish (\ref{eq: high prob b}).
\color{black}

\subsection{Proof of Lemma \ref{lem: dominating variance}} 
We want to establish
\begin{equation}
\frac{1}{n}\sum_{i=1}^{n}\left(\frac{\exp\left(X_{i\cdot}^{\intercal}{\beta}\right)}{\left(1+\exp\left(X_{i\cdot}^{\intercal}{\beta}\right)\right)^2}\right)^{-1}\widehat{u}^{\intercal}X_{i\cdot}\epsilon_i \rightarrow N(0,{\rm V})
\label{eq: central limit}
\end{equation}
Define 
\begin{equation}
W_i=\frac{1}{\sqrt{\rm V}}{\left(\frac{\exp\left(X_{i\cdot}^{\intercal}{\beta}\right)}{\left(1+\exp\left(X_{i\cdot}^{\intercal}{\beta}\right)\right)^2}\right)^{-1}\widehat{u}^{\intercal}X_{i\cdot}\epsilon_i}
\end{equation}
Conditioning on $X$, then $\{W_i\}_{1\leq i\leq n}$ are independent random variables with $\E(W_i\mid X_i)=0$ and $\sum_{i=1}^{n}{\rm Var}(W_i\mid X_i)=n^{2}$.
To establish \eqref{eq: central limit}, it is sufficient to check the Lindeberg's condition, that is, for any constant $\bar{\epsilon}>0$,
\begin{equation}
\lim_{n\rightarrow\infty}\frac{1}{n^{2}}\sum_{i=1}^{n}\E\left(W_i^2 \mathbf{1}_{\left\{\left|W_i\right|\geq \bar{\epsilon} \sqrt{n}\right\}}\right)=0.
\label{eq: lin condition}
\end{equation}
Note that 
\begin{equation}
\max_{1\leq i\leq n}\left|\frac{1}{\sqrt{\rm V}}{\left(\frac{\exp\left(X_{i\cdot}^{\intercal}{\beta}\right)}{\left(1+\exp\left(X_{i\cdot}^{\intercal}{\beta}\right)\right)^2}\right)^{-1}\widehat{u}^{\intercal}X_{i\cdot}\epsilon_i}\right|\leq 2 \frac{\max_{1\leq i\leq n}\left|\widehat{u}^{\intercal}X_{i\cdot}\right|}{\sqrt{V}}\leq \frac{2\tau_{n}\|\xnew\|_2}{\sqrt{V}}\leq \bar{\epsilon}\sqrt{n}
\label{eq: verification of lin condition}
\end{equation}
where the first inequality follows from the fact that $\left(\frac{\exp\left(X_{i\cdot}^{\intercal}{\beta}\right)}{\left(1+\exp\left(X_{i\cdot}^{\intercal}{\beta}\right)\right)^2}\right)^{-1}\epsilon_i\leq \frac{2}{c_{\min}}$, the second inequality follows from $\left|\widehat{u}^{\intercal}X_{i\cdot}\right|\leq \tau_{n}\|\xnew\|_2$ and the last inequality follows from \eqref{eq: dominating variance} and the condition $\tau_{n}\ll \sqrt{n}$. Then \eqref{eq: lin condition} follows from \eqref{eq: verification of lin condition} and by Lindeberg's central limit theorem, we establish \eqref{eq: central limit}.\\

\subsection{Proof of Proposition \ref{prop: lasso convergence}}
For $t\in(0,1)$, by the definition of $\widehat{\beta}$, we have 
\begin{equation}
\ell(\widehat{\beta})+\lambda \|\widehat{\beta}\|_1\leq \ell(\widehat{\beta}+t(\beta-\widehat{\beta}))+\lambda \|\widehat{\beta}+t(\beta-\widehat{\beta})\|_1\leq \ell(\widehat{\beta}+t(\beta-\widehat{\beta}))+ (1-t) \lambda\|\widehat{\beta}\|_1+t\lambda\|\beta\|_1
\end{equation}
where $\ell(\beta)=\frac{1}{n} \sum_{i=1}^{n}\left(\log\left(1+\exp\left(X_{i\cdot}^{\intercal}\beta\right)\right)-y_i\cdot\left(X_{i\cdot}^{\intercal}\beta\right)\right).$
Then we have 
\begin{equation}
\frac{\ell(\widehat{\beta})-\ell(\widehat{\beta}+t(\beta-\widehat{\beta}))}{t}+\lambda \|\widehat{\beta}\|_1\leq \lambda\|\beta\|_1 \quad \text{for any}\; t\in (0,1) 
\end{equation}
and taking the limit $t\rightarrow 0$, we have 
\begin{equation}
\frac{1}{n}\sum_{i=1}^{n}\left(\frac{\exp(X_{i\cdot}^{\intercal}\widehat{\beta})}{1+\exp(X_{i\cdot}^{\intercal}\widehat{\beta})}-y_i\right)X_{i\cdot}^{\intercal}(\widehat{\beta}-\beta)+\lambda \|\widehat{\beta}\|_1\leq \lambda\|\beta\|_1
\label{eq: basic inequality}
\end{equation}
Note that 
\begin{equation}
\begin{aligned}
&\left(\frac{\exp(X_{i\cdot}^{\intercal}\widehat{\beta})}{1+\exp(X_{i\cdot}^{\intercal}\widehat{\beta})}-y_i\right)X_{i\cdot}^{\intercal}(\widehat{\beta}-\beta)=\left(-\epsilon_i+\left(\frac{\exp(X_{i\cdot}^{\intercal}\widehat{\beta})}{1+\exp(X_{i\cdot}^{\intercal}\widehat{\beta})}-\frac{\exp\left(X_{i\cdot}^{\intercal}{\beta}\right)}{1+\exp\left(X_{i\cdot}^{\intercal}{\beta}\right)}\right)\right)X_{i\cdot}^{\intercal}(\widehat{\beta}-\beta)\\
&= - \epsilon_iX_{i\cdot}^{\intercal}(\widehat{\beta}-\beta)+ \int_{0}^{1} \frac{\exp\left(X_{i\cdot}^{\intercal}{\beta}+tX_{i\cdot}^{\intercal}(\widehat{\beta}-{\beta})\right)}{\left[1+\exp\left(X_{i\cdot}^{\intercal}{\beta}+tX_{i\cdot}^{\intercal}(\widehat{\beta}-{\beta})\right)\right]^2} \left(X_{i\cdot}^{\intercal}(\widehat{\beta}-{\beta})\right)^2dt
\end{aligned}
\label{eq: lower bound 1}
\end{equation}
By \eqref{eq: inequality 1}, we have 
\begin{equation}
\begin{aligned}
&\frac{\exp\left(X_{i\cdot}^{\intercal}{\beta}+tX_{i\cdot}^{\intercal}(\widehat{\beta}-{\beta})\right)}{\left[1+\exp\left(X_{i\cdot}^{\intercal}{\beta}+tX_{i\cdot}^{\intercal}(\widehat{\beta}-{\beta})\right)\right]^2}\geq \frac{\exp\left(X_{i\cdot}^{\intercal}{\beta}\right)}{\left[1+\exp\left(X_{i\cdot}^{\intercal}{\beta}\right)\right]^2}\exp\left(-t\left|X_{i\cdot}^{\intercal}(\widehat{\beta}-{\beta})\right|\right)\\
&\geq \frac{\exp\left(X_{i\cdot}^{\intercal}{\beta}\right)}{\left[1+\exp\left(X_{i\cdot}^{\intercal}{\beta}\right)\right]^2}\exp\left(-t\max_{1\leq i\leq n}\left|X_{i\cdot}^{\intercal}(\widehat{\beta}-{\beta})\right|\right)
\end{aligned}
\label{eq: lower bound 2}
\end{equation}
Combined with \eqref{eq: lower bound 1}, we have 
\begin{equation}
\begin{aligned}
&\int_{0}^{1} \frac{\exp\left(X_{i\cdot}^{\intercal}{\beta}+tX_{i\cdot}^{\intercal}(\widehat{\beta}-{\beta})\right)}{\left[1+\exp\left(X_{i\cdot}^{\intercal}{\beta}+tX_{i\cdot}^{\intercal}(\widehat{\beta}-{\beta})\right)\right]^2} \left(X_{i\cdot}^{\intercal}(\widehat{\beta}-{\beta})\right)^2dt\\
&\geq \frac{\exp\left(X_{i\cdot}^{\intercal}{\beta}\right)}{\left[1+\exp\left(X_{i\cdot}^{\intercal}{\beta}\right)\right]^2}\left(X_{i\cdot}^{\intercal}(\widehat{\beta}-{\beta})\right)^2 \int_{0}^{1}\exp\left(-t\max_{1\leq i\leq n}\left|X_{i\cdot}^{\intercal}(\widehat{\beta}-{\beta})\right|\right)d t\\
&= \frac{\exp\left(X_{i\cdot}^{\intercal}{\beta}\right)}{\left[1+\exp\left(X_{i\cdot}^{\intercal}{\beta}\right)\right]^2}\left(X_{i\cdot}^{\intercal}(\widehat{\beta}-{\beta})\right)^2 \frac{1-\exp\left(-\max_{1\leq i\leq n}\left|X_{i\cdot}^{\intercal}(\widehat{\beta}-{\beta})\right|\right)}{\max_{1\leq i\leq n}\left|X_{i\cdot}^{\intercal}(\widehat{\beta}-{\beta})\right|}
\end{aligned}
\end{equation}
Together with \eqref{eq: basic inequality}, we have 
\begin{equation}
\begin{aligned}
 &\frac{1-\exp\left(-\max_{1\leq i\leq n}\left|X_{i\cdot}^{\intercal}(\widehat{\beta}-{\beta})\right|\right)}{\max_{1\leq i\leq n}\left|X_{i\cdot}^{\intercal}(\widehat{\beta}-{\beta})\right|} \left(\frac{1}{n}\sum_{i=1}^{n}\frac{\exp\left(X_{i\cdot}^{\intercal}{\beta}\right)}{\left[1+\exp\left(X_{i\cdot}^{\intercal}{\beta}\right)\right]^2}\left(X_{i\cdot}^{\intercal}(\widehat{\beta}-{\beta})\right)^2\right)+\lambda \|\widehat{\beta}\|_1 \\
 &\leq \lambda \|\beta\|_1 + \frac{1}{n}\sum_{i=1}^{n}\epsilon_iX_{i\cdot}^{\intercal}(\widehat{\beta}-\beta)\leq \lambda \|\beta\|_1+\lambda_0\|\widehat{\beta}-\beta\|_1.
 \end{aligned}
 \end{equation}
 By the fact that $ \|\widehat{\beta}\|_1=\|\widehat{\beta}_{S}\|_1+\|\widehat{\beta}_{S^{c}}-\beta_{S^{c}}\|_1$ and $\|\beta\|_1-\|\widehat{\beta}_{S}\|_1\leq \|\widehat{\beta}_{S}-\beta_{S}\|_1$, then we have 
 \begin{equation}
 \begin{aligned}
& \frac{1-\exp\left(-\max_{1\leq i\leq n}\left|X_{i\cdot}^{\intercal}(\widehat{\beta}-{\beta})\right|\right)}{\max_{1\leq i\leq n}\left|X_{i\cdot}^{\intercal}(\widehat{\beta}-{\beta})\right|} \left(\frac{1}{n}\sum_{i=1}^{n}\frac{\exp\left(X_{i\cdot}^{\intercal}{\beta}\right)}{\left[1+\exp\left(X_{i\cdot}^{\intercal}{\beta}\right)\right]^2}\left(X_{i\cdot}^{\intercal}(\widehat{\beta}-{\beta})\right)^2\right)\\
&+\delta_0\lambda_0\|\widehat{\beta}_{S^{c}}-\beta_{S^{c}}\|_1\leq \left(2+\delta_0\right)\lambda_0\|\widehat{\beta}_{S}-\beta_{S}\|_1
 \end{aligned}
 \end{equation}
Then we deduce \eqref{eq: est property} and
\begin{equation}
\begin{aligned}
& \frac{1-\exp\left(-\max_{1\leq i\leq n}\left|X_{i\cdot}^{\intercal}(\widehat{\beta}-{\beta})\right|\right)}{\max_{1\leq i\leq n}\left|X_{i\cdot}^{\intercal}(\widehat{\beta}-{\beta})\right|} \left(\frac{1}{n}\sum_{i=1}^{n}\frac{\exp\left(X_{i\cdot}^{\intercal}{\beta}\right)}{\left[1+\exp\left(X_{i\cdot}^{\intercal}{\beta}\right)\right]^2}\left(X_{i\cdot}^{\intercal}(\widehat{\beta}-{\beta})\right)^2\right)\\
 &\leq \left(2+\delta_0\right)\lambda_0\|\widehat{\beta}_{S}-\beta_{S}\|_1.
 \end{aligned}
 \label{eq: basic inequality a}
\end{equation}

\begin{Lemma} On the event $\mathcal{A}_2\cap \mathcal{A}_3$, then 
\begin{equation}
\frac{1}{n}\sum_{i=1}^{n}\frac{\exp\left(X_{i\cdot}^{\intercal}{\beta}\right)}{\left[1+\exp\left(X_{i\cdot}^{\intercal}{\beta}\right)\right]^2}\left(X_{i\cdot}^{\intercal}(\widehat{\beta}-{\beta})\right)^2\geq c\lambda_{\min}\left(\Sigma\right)\|\widehat{\beta}-{\beta}\|_2^2
\end{equation}
\end{Lemma}
\noindent Then \eqref{eq: basic inequality a} is further simplified as 
\begin{equation}
 \frac{1-\exp\left(-\max_{1\leq i\leq n}\left|X_{i\cdot}^{\intercal}(\widehat{\beta}-{\beta})\right|\right)}{\max_{1\leq i\leq n}\left|X_{i\cdot}^{\intercal}(\widehat{\beta}-{\beta})\right|} c\lambda_{\min}\left(\Sigma\right)\|\widehat{\beta}-{\beta}\|_2^2\leq  \left(2+\delta_0\right)\lambda_0\|\widehat{\beta}_{S}-\beta_{S}\|_1
  \label{eq: basic inequality c}
\end{equation}
{\bf Case 1:} Assume that 
\begin{equation}
\max_{1\leq i\leq n}\left|X_{i\cdot}^{\intercal}(\widehat{\beta}-{\beta})\right|\leq c_1 \quad \text{for some}\; c_1>0 
\label{eq: assisting assumption}
\end{equation} then we have 
\begin{equation}
\begin{aligned}
&\frac{1-\exp\left(-\max_{1\leq i\leq n}\left|X_{i\cdot}^{\intercal}(\widehat{\beta}-{\beta})\right|\right)}{\max_{1\leq i\leq n}\left|X_{i\cdot}^{\intercal}(\widehat{\beta}-{\beta})\right|}=\int_{0}^{1}\exp\left(-t\max_{1\leq i\leq n}\left|X_{i\cdot}^{\intercal}(\widehat{\beta}-{\beta})\right|\right)d t\\
&\geq \int_{0}^{1}\exp\left(-t c_1\right)dt=\frac{1-\exp\left(-c_1\right)}{c_1}
\end{aligned}
\end{equation}
Define $c_2=\frac{c\lambda_{\min}\left(\Sigma\right)}{2+\delta_0} \frac{1-\exp\left(-c_1\right)}{c_1}$, then we have
\begin{equation}
c_2 \|\widehat{\beta}-{\beta}\|_2^2\leq \lambda_0\|\widehat{\beta}_{S}-\beta_{S}\|_1\leq \sqrt{k}\lambda_0 \|\widehat{\beta}_{S}-\beta_{S}\|_2
\end{equation}
and hence 
\begin{equation}
\|\widehat{\beta}-\beta\|_2\lesssim \frac{1}{\lambda_{\min}}\sqrt{k}\lambda_0 \quad \text{and} \quad \|\widehat{\beta}-\beta\|_1\leq {k}\lambda_0
\label{eq: error bound under assumption}
\end{equation}
{\bf Case 2:}  Assume that \eqref{eq: assisting assumption} does not hold, then 
\begin{equation}
\frac{1-\exp\left(-\max_{1\leq i\leq n}\left|X_{i\cdot}^{\intercal}(\widehat{\beta}-{\beta})\right|\right)}{\max_{1\leq i\leq n}\left|X_{i\cdot}^{\intercal}(\widehat{\beta}-{\beta})\right|}\geq \frac{1-\exp(-c_1)}{\max_{1\leq i\leq n}\left|X_{i\cdot}^{\intercal}(\widehat{\beta}-{\beta})\right|}
\end{equation}
Together with \eqref{eq: basic inequality c}, we have
\begin{equation}
c_2 c_1\|\widehat{\beta}-{\beta}\|_2^2\leq \lambda_0\|\widehat{\beta}_{S}-\beta_{S}\|_1\max_{1\leq i\leq n}\left|X_{i\cdot}^{\intercal}(\widehat{\beta}-{\beta})\right|
\label{eq: bound 1}
\end{equation}
By $\max_{1\leq i\leq n}\left|X_{i\cdot}^{\intercal}(\widehat{\beta}-{\beta})\right|\leq  \max\left|X_{ij}\right|\|\widehat{\beta}-\beta\|_1$ and \eqref{eq: est property}, we further have
\begin{equation}
\begin{aligned}
\lambda_0\|\widehat{\beta}_{S}-\beta_{S}\|_1\max_{1\leq i\leq n}\left|X_{i\cdot}^{\intercal}(\widehat{\beta}-{\beta})\right|&\leq \frac{2+2\delta_0}{\delta_0}  \max\left|X_{ij}\right| \lambda_0\|\widehat{\beta}_{S}-\beta_{S}\|_1^2\\
&\leq  \frac{2+2\delta_0}{\delta_0} \max\left|X_{ij}\right| k \lambda_0\|\widehat{\beta}_{S}-\beta_{S}\|_2^2,
\end{aligned}
\label{eq: bound 2}
\end{equation}
where the last inequality follows from Cauchy inequality. Combining \eqref{eq: bound 1} and \eqref{eq: bound 2}, we have shown that if \eqref{eq: assisting assumption} does not hold, then \begin{equation}
\max\left|X_{ij}\right| \frac{2+2\delta_0}{\delta_0} k \lambda_0\geq c_2 c_1,
\end{equation}
Since this contradicts the assumption that $\max\left|X_{ij}\right| k\lambda_0< \frac{c_2 c_1\delta_0}{2+2\delta_0}$, we establish \eqref{eq: error bound under assumption} and hence  \eqref{eq: est property}.

\subsection{Proof of Lemma \ref{lem: taylor expansion of logit}}
\label{sec: logit bound}
We first introduce the following version of Taylor expansion.
\begin{Lemma}  If $f''(x)$ is continuous on an open interval $\mathcal{I}$ that contains $a$, and $x\in \mathcal{I}$ , then
\begin{equation}
f(x)-f(a)=f'(a)(x-a)+\int_{0}^{1} (1-t)(x-a)^2f''(a+t(x-a)) dt
\end{equation}
\label{lem: Taylor expansion}
\end{Lemma}
By applying Lemma \ref{lem: Taylor expansion}, we have
\begin{equation*}
h(x)-h(a)=h'(a)(x-a)+\int_{0}^{1} (1-t)(x-a)^2h''(a+t(x-a)) dt
\end{equation*}
Divide both sides by $\left(h'(a)\right)^{-1}$, we establish \eqref{eq: Taylor expansion}.
The inequality \eqref{eq: inequality 1} follows from 
\begin{equation*}
\frac{h'(x)}{h'(a)}=\exp(x-a)\frac{[1+\exp(a)]^2}{[1+\exp(x)]^2}\leq \exp(x-a) \exp(2(a-x)_{+})=\exp(|x-a|)
\end{equation*}
and 
\begin{equation*}
\frac{h'(a)}{h'(x)}\leq \exp(|x-a|)
\end{equation*}
The control of \eqref{eq: taylor error bound} follows from the following inequality,
\begin{equation*}
\begin{aligned}
\frac{h''(a+t(x-a))}{h'(a)}&=\frac{2\exp(2a+2t(x-a))}{(1+\exp(a+t(x-a)))^3}\cdot \frac{(1+\exp(a))^2}{\exp(a)}\\
&\leq 2\exp(t(x-a))\cdot\frac{(1+\exp(a))^2}{(1+\exp(a+t(x-a)))^2}\leq 2\exp(t|x-a|)
\end{aligned}
\end{equation*}
\subsection{Proof of Lemma \ref{lem: symmetrization}}
\label{sec: proof sym}
We start with the conditional expectation $\E_{y\mid X} \Phi\left(\sup_{t\in \mathcal{T}}\left|\sum_{i=1}^{n} g_i(t_i) \epsilon_i\right|\right)$ and note that
$$\E_{y\mid X} \Phi\left(\sup_{t\in \mathcal{T}}\left|\sum_{i=1}^{n} g_i(t_i) \epsilon_i\right|\right)=\E_{y\mid X} \Phi\left(\sup_{t\in \mathcal{T}}\left|\sum_{i=1}^{n} g_i(t_i) \epsilon_i-\E_{y'\mid X}\sum_{i=1}^{n} g_i(t_i) \epsilon'_i\right|\right).$$
Since $\sup_{t\in \mathcal{T}}\left|\sum_{i=1}^{n} g_i(t_i) \epsilon_i-\E_{y'\mid X}\sum_{i=1}^{n} g_i(t_i) \epsilon'_i\right|\leq \E_{y'\mid X} \sup_{t\in \mathcal{T}}\left|\sum_{i=1}^{n} g_i(t_i) (\epsilon_i-\epsilon'_i)\right|$ and $\Phi$ is a non-decreasing function, we have 
$$\Phi\left(\sup_{t\in \mathcal{T}}\left|\sum_{i=1}^{n} g_i(t_i) \epsilon_i-\E_{y'\mid X}\sum_{i=1}^{n} g_i(t_i) \epsilon'_i\right|\right)\leq \Phi\left(\E_{y'\mid X} \sup_{t\in \mathcal{T}}\left|\sum_{i=1}^{n} g_i(t_i) (\epsilon_i-\epsilon'_i)\right|\right).$$
Since $\Phi$ is a convex function, we have 
$$
\E_{y\mid X}\Phi\left(\E_{y'\mid X} \sup_{t\in \mathcal{T}}\left|\sum_{i=1}^{n} g_i(t_i) (\epsilon_i-\epsilon'_i)\right|\right)\leq \E_{(y,y')\mid X} \Phi\left(\sup_{t\in \mathcal{T}}\left|\sum_{i=1}^{n} g_i(t_i) (\epsilon_i-\epsilon'_i)\right|\right).$$
Integration of both sides of the above inequality leads to \eqref{eq: symmetrization}.

\subsection{Proof of Lemma \ref{lem: contraction}}
\label{sec: proof contraction}
The proof follows from that of Theorem 2.2 in \cite{koltchinskii2011oracle} and some modification is necessary to extend the results to the general random variables $\xi_1, \xi_2,\cdots, \xi_{n}$ which are independent and follow the probability distribution \eqref{eq: distribution}.

We start with proving the following inequality for a function $A: \mathcal{T}\rightarrow \R,$
\begin{equation}
\E \Phi\left(\sup_{t\in \mathcal{T}}[A(t)+\sum_{i=1}^{n}\phi_i(t_i)\xi_i]\right)\leq \E\Phi\left(\sup_{t\in \mathcal{T}}[A(t)+\sum_{i=1}^{n} t_i\xi_i]\right),
\label{eq: inter contraction}
\end{equation}
We first prove the special case $n=1$, which is reduced to be the following inequality,
\begin{equation}
\E \Phi\left(\sup_{t\in \mathcal{T}}[t_1+\phi(t_2)\xi_0]\right)\leq \E \Phi\left(\sup_{t\in \mathcal{T}}[t_1+t_2\xi_0]\right),
\label{eq: n=1}
\end{equation}
where $\mathcal{T}\subset \R^2$ and $\PP(\xi_0=1)=\PP(\xi_0=-1)\in[0,\frac{1}{2}]$ and $\PP(\xi_0=0)=1-2\PP(\xi=1).$ It suffices to verify \eqref{eq: n=1} by establishing the following inequality,
\begin{equation*}
\begin{aligned}
&\PP(\xi_0=1)\Phi\left(\sup_{t\in \mathcal{T}}[t_1+\phi(t_2)]\right)+\PP(\xi_0=-1)\Phi\left(\sup_{t\in \mathcal{T}}[t_1-\phi(t_2)]\right)+\PP(\xi_0=0)\Phi\left(\sup_{t\in \mathcal{T}}[t_1]\right)\\
&\leq \PP(\xi_0=1)\Phi\left(\sup_{t\in \mathcal{T}}[t_1+t_2]\right)+\PP(\xi_0=-1)\Phi\left(\sup_{t\in \mathcal{T}}[t_1-t_2]\right)+\PP(\xi_0=0)\Phi\left(\sup_{t\in \mathcal{T}}[t_1]\right)
\end{aligned}
\end{equation*}
This is equivalent to show 
\begin{equation}
\Phi\left(\sup_{t\in \mathcal{T}}[t_1+\phi(t_2)]\right)+\Phi\left(\sup_{t\in \mathcal{T}}[t_1-\phi(t_2)]\right)\leq \Phi\left(\sup_{t\in \mathcal{T}}[t_1+t_2]\right)+\Phi\left(\sup_{t\in \mathcal{T}}[t_1-t_2]\right)
\label{eq: key inequality}
\end{equation}
The above inequality follows from the same line of proof as that in \cite{koltchinskii2011oracle}. It remains to prove the lemma by applying induction and \eqref{eq: n=1}, that is,
\begin{equation*}
\begin{aligned}
&\E_{ ({\xi}_1,\cdots, \xi_{n}) \mid X} \Phi\left(\sup_{t\in \mathcal{T}}[A(t)+\sum_{i=1}^{n}\phi_i(t_i)\xi_i]\right)= \E_{ ({\xi}_1,\cdots, \xi_{n-1}) \mid X} \E_{ \xi_{n} \mid X} \Phi\left(\sup_{t\in \mathcal{T}}[A(t)+\sum_{i=1}^{n}\phi_i(t_i)\xi_i]\right)\\
&\leq \E_{ ({\xi}_1,\cdots, \xi_{n-1}) \mid X} \E_{ \xi_{n} \mid X} \Phi\left(\sup_{t\in \mathcal{T}}[A(t)+\sum_{i=1}^{n-1}\phi_i(t_i)\xi_i+t_{n}\xi_{n}]\right)\\
&=\E_{ \xi_{n} \mid X}\E_{ ({\xi}_1,\cdots, \xi_{n-1}) \mid X}  \Phi\left(\sup_{t\in \mathcal{T}}[A(t)+\sum_{i=1}^{n-1}\phi_i(t_i)\xi_i+t_{n}\xi_{n}]\right)
\end{aligned}
\end{equation*}
Continuing the above equation, we establish $\E_{ ({\xi}_1,\cdots, \xi_{n}) \mid X} \Phi\left(\sup_{t\in \mathcal{T}}[A(t)+\sum_{i=1}^{n}\phi_i(t_i)\xi_i]\right)\leq \E_{ ({\xi}_1,\cdots, \xi_{n}) \mid X} \Phi\left(\sup_{t\in \mathcal{T}}[A(t)+\sum_{i=1}^{n} t_i\xi_i]\right)$. Integration with respect to $X$ leads to \eqref{eq: inter contraction}. In the following, we will apply \eqref{eq: inter contraction} to establish \eqref{eq: contraction}. Note that 
\begin{equation*}
\begin{aligned}
&\E \Phi\left(\frac{1}{2}\sup_{t\in \mathcal{T}}\left|\sum_{i=1}^{n}\phi_i(t_i)\xi_i\right|\right)=\E \Phi\left(\frac{1}{2}\left(\sup_{t\in \mathcal{T}}\sum_{i=1}^{n}\phi_i(t_i)\xi_i\right)_{+}+\frac{1}{2}\left(\sup_{t\in \mathcal{T}}\sum_{i=1}^{n}\phi_i(t_i)(-\xi_i)\right)_{+}\right)\\
&\leq\frac{1}{2}\left[\E \Phi\left(\left(\sup_{t\in \mathcal{T}}\sum_{i=1}^{n}\phi_i(t_i)\xi_i\right)_{+}\right)+\E \Phi\left(\left(\sup_{t\in \mathcal{T}}\sum_{i=1}^{n}\phi_i(t_i)(-\xi_i)\right)_{+}\right)\right]
\end{aligned}
\end{equation*}
By applying \eqref{eq: inter contraction} to the function $u\rightarrow \Phi(u_{+})$, which is convex and non-decreasing, we have $\E \Phi\left(\left(\sup_{t\in \mathcal{T}}\sum_{i=1}^{n}\phi_i(t_i)\xi_i\right)_{+}\right)\leq \E\Phi\left(\sup_{t\in \mathcal{T}}\sum_{i=1}^{n}t_i\xi_i\right)\leq \E\Phi\left(\sup_{t\in \mathcal{T}}\left|\sum_{i=1}^{n} t_i \xi_i\right|\right)$. Then we establish \eqref{eq: contraction}.

\section{Additional Simulation Studies}

\subsection{Scale of \texorpdfstring{$\lambda_n$}{lambda}}
\label{sec: mu and lambda}
For our implemented algorithm, we identify the tuning parameter $\lambda_n$ using the following steps. Recall $H=\left[b, \mathbb{I}_{p \times p}\right]$ and   $b=\frac{1}{\left\|\xnew\right\|_{2}}\xnew.$ For $t=0$, we initialise $\lambda^{0} = \sqrt{2.01\cdot \log p/n}$ and calculate $$\widehat{v}=\arg \min _{v \in \mathbb{R}^{p+1}} \frac{1}{4} v^{\intercal} H^{\intercal} \widehat{\Sigma} H v+b^{\intercal} H v+\lambda^{t}\|v\|_{1}.$$
\begin{enumerate} 
\item If $\|\widehat{v}\|_{2}<\infty$, then, for $t \geq 0,$ we set $ \lambda^{t+1} = \lambda^{t}/1.5$
and calculate $$\widehat{v}=\arg \min _{v \in \mathbb{R}^{p+1}} \frac{1}{4} v^{\intercal} H^{\intercal} \widehat{\Sigma} H v+b^{\intercal} H v+\lambda^{t+1}\|v\|_{1}.$$
Repeat until $\widehat{v}$ cannot be solved or $t =5.$
\item If $\|\widehat{v}\|_{2}=\infty$, then, for $t\geq 0,$ we set $\lambda^{t+1} = \lambda^{t}\cdot 1.5$
and calculate $$\widehat{v}=\arg \min _{v \in \mathbb{R}^{p+1}} \frac{1}{4} v^{\intercal} H^{\intercal} \widehat{\Sigma} H v+b^{\intercal} H v+\lambda^{t+1}\|v\|_{1}.$$
Repeat until $\widehat{v}$ can be solved.
\end{enumerate}
By the above algorithm, we choose the smallest $\lambda_n>0$ such that the dual problem has a finite minimum value.
Through the above algorithm, $\lambda_n$, starting from $\sqrt{2.01\cdot \log p/n}$, can be at most reduced to $\frac{\sqrt{2.01\cdot \log p/n}}{(1.5)^{6}}$. 
In theory, we can also increase $\lambda$ but our observation is that the decreasing of $\lambda$ happens for almost all settings if we start with $\sqrt{2.01\cdot \log p/n}$. We provide results of the numerical experiment for our simulation setting $(\rm S1)$ in Table \ref{tab: mu}. The table shows $\lambda_n \approx 0.4 \sqrt{{\log p}/{n}}$ in this specific simulation setting.

\begin{table}[htp!]
\centering
\scalebox{0.72}{
\begin{tabular}{|rrr|}
\hline
\multicolumn{3}{|c|}{{\bf Setting $(\rm S1)$, Loading 1 with $q = 1$}} \\
\hline
$n$ & $\lambda_n$ & $\sqrt{{\log p}/{n}}$ \\ 
\hline
200& 0.07 & 0.18 \\ 
400& 0.05 & 0.12 \\ 
600& 0.04 & 0.10 \\
\hline
\end{tabular}
}
\caption{Report of $\lambda_n$ and $\sqrt{{\log p}/{n}}$ for the Setting $(\rm S1)$.}
\label{tab: mu}
\end{table}
    
\subsection{Constraint (\ref{eq: constraint 3})}
\label{sec: tau}

We now investigate whether our constructed projection direction $\widehat{u}$ in \eqref{eq: implementation} satisfies the constraint (\ref{eq: constraint 3}) through computing the ratio $\|X\widehat{u}\|_{\infty}/\|x_*\|_2.$ Note that as long as $\|X\widehat{u}\|_{\infty}/\|x_*\|_2\leq C\sqrt{\log n}$ for some positive constant $C>0,$ then the constraint (\ref{eq: constraint 3}) is satisfied, which is sufficient for us to establish the central limit theorem.

\begin{figure}[htp!]
\centering
\includegraphics[scale=0.45]{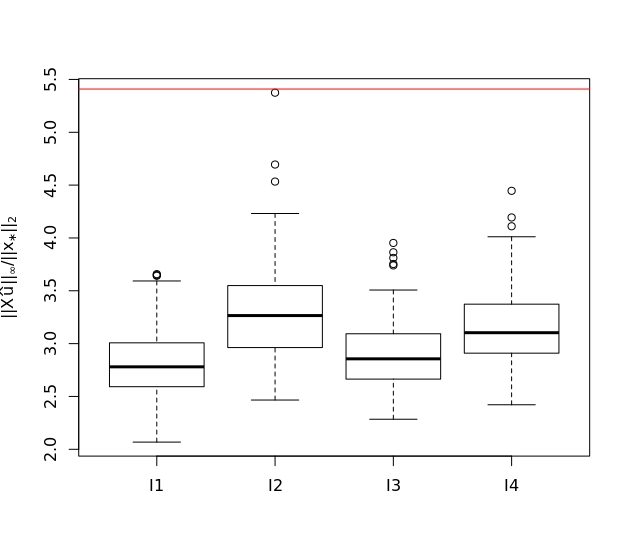}
\vspace{-10mm}
\caption{Boxplot showing the distribution of $\|X\widehat{u}\|_{\infty}/\|x_*\|_2$ for $n = 200$, summarized over 500 simulations. Setting indices ${\rm I1}$ and ${\rm I2}$ denote setting $(\rm S1)$ with $r = 1$ and $r = \frac{1}{25}$ respectively while setting indices ${\rm I3}$ and ${\rm I4}$ denote setting $(\rm S2) (\rm decay = 1)$ with $r = 1$ and $r = \frac{1}{25}$ respectively. The red line is corresponding to $y = 2.35\cdot\sqrt{\log n}$}
\label{fig:Boxplot 1}
\end{figure}

\begin{figure}[htp!]
\centering
\includegraphics[scale=0.45]{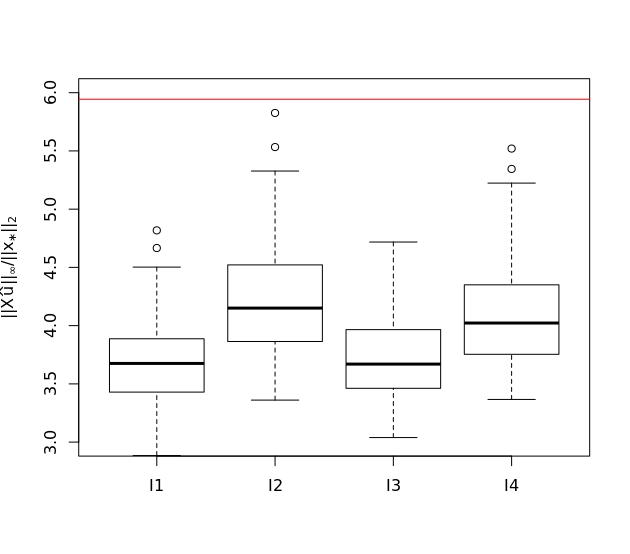}
\vspace{-10mm}
\caption{Boxplot showing the distribution of $\|X\widehat{u}\|_{\infty}/\|x_*\|_2$ for $n = 600$, summarized over 500 simulations. Setting indices ${\rm I1}$ and ${\rm I2}$ denote setting $(\rm S1)$ with $r = 1$ and $r = \frac{1}{25}$ respectively while setting indices ${\rm I3}$ and ${\rm I4}$ denote setting $(\rm S2) (\rm decay = 1)$ with $r = 1$ and $r = \frac{1}{25}$ respectively.  The red line is corresponding to $y = 2.35\cdot\sqrt{\log n}$}
\label{fig:Boxplot 2}
\end{figure}

For $n=200,$ the boxplot \ref{fig:Boxplot 1} verifies Constraint (\ref{eq: constraint 3}) for Settings $(\rm S1)$ and $(\rm S2) (\rm decay = 1)$ and {Loading 1} with $q = 1$ and $r\in \{1,1/25\}.$ The boxplot \ref{fig:Boxplot 2} summarizes the same for $n=600.$ These boxplots show that $\|X\widehat{u}\|_{\infty}/\|x_*\|_2$ is bounded above by $2.35\cdot\sqrt{\log n}$. They demonstrate that the constraint (\ref{eq: constraint 3}) is satisfied by our constructed $\widehat{u}.$ 

\subsection{Comparison of Proposed Method with Post Selection Method}
\label{sec: naive}
We now consider a challenging setting for post-selection and compare the post-selection method with the proposed LiVE method.
\begin{enumerate}
    \item[(S9)] $\beta_1 = 0$; $\beta_{j}={(j-1)}/{20}$ for $2 \leq j \leq 11$ but  $j \neq 9, 10$; $\beta_j = 0.01$ for $j = 9, 10$ and $\beta_j = 0$ for $12 \leq j \leq 501$. 
\end{enumerate}   
\noindent The loading $\xnew$ is generated as follows :\\
\textbf{ Loading 3:} We set $x_{{\rm basis},1}=1$ and generate $x_{{\rm basis},-1}\in \R^{500}$ following $N(0,\Sigma)$ with $\widetilde{\Sigma} = \{0.5^{1+|j-l|}\}_{1\leq j,l\leq 500}$ and generate $\xnew$ as
$$
x_{*,j}=\begin{cases}
x_{{\rm basis},j} &  \text{for} \; 1\leq j\leq 11 \quad ; \quad  j \neq 9,10\\
10 &  \text{for} \; j=9,10 \\
\frac{1}{25} \cdot x_{{\rm basis},j} &  \text{for}\; 12\leq j \leq 501\\
\end{cases}
$$

\noindent We construct the new $\beta$ and $\xnew$ as we believe this is a challenging setting for post selection. The insignificant regression coefficients $\beta_{9}, \beta_{10}$ make the corresponding covariates $X_{\cdot,9}$ and $X_{\cdot,10}$ unlikely to be selected by Lasso in the first step. However, with  enlarged entries $x_{*,9}, x_{*,10}$, the corresponding covariates comprise a major part of the magnitude of the case probability $h(x_{*}^{\intercal}\beta)$, thereby leading to a large omitted variable bias when these relevant covariates are not selected by Lasso. We have observed in Table \ref{tab: bias in naive} that the post selection estimator has a large omitted variable bias and also produces a under-covered confidence interval.

\begin{table}[htp!]
\centering
\scalebox{0.8}{
\begin{tabular}{|rrr|r|rrrrrr|rrrrrr|}
  \hline
\multicolumn{16}{|c|}{{\bf Setting $(\rm S9)$, Loading 3}} \\
  \hline
\multicolumn{3}{|c|}{}&&\multicolumn{6}{c|}{LiVE} &\multicolumn{6}{c|}{Post Selection} \\
   \hline
$\|x_{\rm new}\|_{2}$ &{\rm r}& Prob &$n$& Cov & ERR & Len & RMSE & Bias & SE & Cov & ERR & Len & RMSE & Bias & SE \\ 
  \hline
\multirow{3}{*}{14.19} & \multirow{3}{*}{$\frac{1}{25}$} & \multirow{3}{*}{0.578} &200& 0.91 & 0.11 & 0.87 & 0.36 & 0.05 & 0.35 & 0.54 & 0.29 & 0.29 & 0.24 & 0.05 & 0.23 \\ 
  & & &400& 0.95 & 0.08 & 0.85 & 0.29 & 0.01 & 0.29 & 0.73 & 0.25 & 0.32 & 0.21 & 0.07 & 0.20 \\ 
  & & &600& 0.96 & 0.10 & 0.81 & 0.27 & 0.01 & 0.27 & 0.75 & 0.30 & 0.30 & 0.21 & 0.07 & 0.19 \\ 
 \hline
\end{tabular}
}
\caption{\textbf{Comparison of the proposed method and the post selection method.} ``r" and ``Prob" represent the shrinkage parameter and Case Probability respectively. The columns indexed with ``Cov" and ``Len" represent the empirical coverage and length of the constructed CIs respectively; the column indexed with ``ERR" represents the empirical rejection rate of the testing procedure; the columns indexed with ``RMSE", ``Bias" and ``SE" represent the RMSE, bias and standard error, respectively. The columns under ``LiVE" and ``Post Selection" correspond to the proposed estimator and post model selection estimator respectively.}
\label{tab: bias in naive}
\end{table}

\subsection{Exactly Sparse with Intercept}
\label{sec: exact with intercept}

Here we explore the performance of the inference procedures in presence of an intercept. We generate $\beta$ as in $(\rm S1)$ and instead of having null intercepts we consider two values for $\beta_{1}$, $\beta_{1}=-1$ and $\beta_{1}=1$ leading to two different target case probabilities $0.501$ and $0.881$ respectively. We investigate the finite sample performance of the inference methods for {Loading 1} with $q = 1$.

\noindent We report the simulation results based on 500 replications in Tables \ref{tab: with intercept 2i} and \ref{tab: with intercept 2ii}. Table \ref{tab: with intercept 2i} shows the proposed inference procedure continue to produce valid confidence intervals and the confidence intervals have shorter lengths for a larger sample size or a smaller $\ell_2$ norm $\|\xnew\|_{2}$. In comparison, \texttt{hdi} undercovers in general while the over-coverage issue of \texttt{WLDP} is still persistent. For $\beta_{1}=-1$, the case probability represents an alternative in the indistinguishable region and hence the testing procedures do not have power in general while for $\beta_{1}=1$, the case probability is well above $0.5$ and corresponds to an alternative to the null hypothesis, thereby the ERR, an empirical measure of power, increases with a larger sample size for all the testing procedures except for the one based on \texttt{WLDP}. It should be mentioned here that the comparison of our proposed method with \texttt{hdi} and \texttt{WLDP} in the setting with intercepts is unfair since \texttt{hdi} and \texttt{WLDP} are not designed to handle case probability and their output does not handle inference for the intercept. However, in practical applications, the intercept is an important term in capturing the case probabilities in logistic model. 

\begin{table}[htp!]
\centering
\scalebox{0.8}{
\begin{tabular}{|rrr|r|rrr|rrr|rrr|rrr|}
  \hline
\multicolumn{16}{|c|}{{\bf Setting $(\rm S1)$ with $\beta_1 = -1$, Loading 1 with $q = 1$}} \\
\hline
\multicolumn{3}{|c|}{}&&\multicolumn{3}{c|}{LiVE} &\multicolumn{3}{c|}{Post Selection} &\multicolumn{3}{c|}{\texttt{hdi}}&\multicolumn{3}{c|}{\texttt{WLDP}} \\
   \hline
$\|\xnew\|_{2}$ &{\rm r} & Prob &$n$& Cov & ERR& Len & Cov & ERR & Len & Cov & ERR & Len & Cov & ERR & Len\\ 
  \hline
\multirow{3}{*}{16.1} & \multirow{3}{*}{1} & \multirow{3}{*}{0.501} &200& 0.97 & 0.01 & 0.93 & 0.65 & 0.22 & 0.49 & 0.91 & 0.11 & 0.97 & 1.00 & 0.00 & 1.00 \\ 
  & & &400& 0.96 & 0.02 & 0.85 & 0.61 & 0.20 & 0.41 & 0.90 & 0.13 & 0.89 & 1.00 & 0.00 & 1.00 \\ 
  & & &600& 0.96 & 0.03 & 0.80 & 0.71 & 0.21 & 0.38 & 0.91 & 0.12 & 0.83 & 1.00 & 0.00 & 1.00 \\ 
 \hline
\multirow{3}{*}{1.09} & \multirow{3}{*}{$\frac{1}{25}$} & \multirow{3}{*}{0.501} &200& 0.93 & 0.02 & 0.42  & 0.76 & 0.10 & 0.41 & 0.23 & 0.84 & 0.34 & 0.94 & 0.25 & 0.61 \\ 
  & & &400& 0.97 & 0.01 & 0.30 & 0.88 & 0.09 & 0.33 & 0.15 & 0.98 & 0.21 & 0.90 & 0.31 & 0.54 \\ 
  & & &600& 0.98 & 0.02 & 0.22 & 0.94 & 0.02 & 0.26 & 0.07 & 0.99 & 0.21 & 0.88 & 0.34 & 0.51  \\ 
 \hline
  \hline
  \multicolumn{16}{|c|}{{\bf Setting $(\rm S1)$ with $\beta_1 = 1$, Loading 1 with $q = 1$}} \\
  \hline
\multicolumn{3}{|c|}{}&&\multicolumn{3}{c|}{LiVE} &\multicolumn{3}{c|}{Post Selection} &\multicolumn{3}{c|}{\texttt{hdi}}&\multicolumn{3}{c|}{\texttt{WLDP}} \\
   \hline
$\|\xnew\|_{2}$ &{\rm r} & Prob &$n$& Cov & ERR& Len & Cov & ERR & Len & Cov & ERR & Len & Cov & ERR & Len\\ 
  \hline
\multirow{3}{*}{16.1} & \multirow{3}{*}{1} & \multirow{3}{*}{0.881} &200& 0.97 & 0.07 & 0.91 & 0.65 & 0.80 & 0.28 & 0.91 & 0.07 & 0.98 & 1.00 & 0.00 & 1.00 \\ 
  & & &400& 0.97 & 0.14 & 0.79 & 0.61 & 0.84 & 0.24 & 0.93 & 0.07 & 0.91 & 1.00 & 0.00 & 1.00 \\ 
  & & &600& 0.96 & 0.18 & 0.74 & 0.71 & 0.88 & 0.20 & 0.91 & 0.06 & 0.86 & 1.00 & 0.00 & 1.00 \\ 
 \hline
\multirow{3}{*}{1.09} & \multirow{3}{*}{$\frac{1}{25}$} & \multirow{3}{*}{0.881} &200& 0.93 & 0.97 & 0.24  & 0.65 & 0.99 & 0.16 & 0.51 & 0.79 & 0.43 & 1.00 & 0.17 & 0.63 \\ 
  & & &400& 0.94 & 0.98 & 0.15 & 0.71 & 1.00 & 0.11 & 0.39 & 0.94 & 0.24 & 1.00 & 0.30 & 0.55 \\ 
  & & &600& 0.91 & 1.00 & 0.12 & 0.85 & 1.00 & 0.11 & 0.21 & 0.96 & 0.20 & 1.00 & 0.32 & 0.52  \\ 
 \hline
\end{tabular}
}
\caption{\textbf{Exactly sparse regression with intercept.} ``r" and ``Prob" represent the shrinkage parameter and Case Probability respectively. The columns indexed with ``Cov" and ``Len" represent the empirical coverage and length of the constructed CIs respectively; the column indexed with ``ERR" represents the empirical rejection rate of the testing procedure. The columns under ``LiVE" ,``Post Selection", ``\texttt{hdi}" and ``\texttt{WLDP}" correspond to the proposed estimator, the post model selection estimator, the plug-in debiased estimator using \texttt{hdi} and \texttt{WLDP} respectively.}
\label{tab: with intercept 2i}
\end{table}

\begin{table}[htp!]
\centering
\scalebox{0.7}{
\begin{tabular}{|rrr|r|rrr|rrr|rrr|rrr|rrr|}
  \hline
\multicolumn{19}{|c|}{{\bf Setting $(\rm S1)$ with $\beta_1 = -1$, Loading 1 with $q = 1$}} \\
\hline
\multicolumn{3}{|c|}{}&&\multicolumn{3}{c|}{LiVE}&\multicolumn{3}{c|}{Post Selection} &\multicolumn{3}{c|}{\texttt{hdi}}&\multicolumn{3}{c|}{\texttt{WLDP}}&\multicolumn{3}{c|}{Lasso} \\
   \hline
$\|\xnew\|_{2}$ &{\rm r}& Prob & $n$& RMSE & Bias & SE & RMSE & Bias & SE & RMSE & Bias & SE & RMSE & Bias & SE & RMSE & Bias & SE\\ 
  \hline
\multirow{3}{*}{16.1} & \multirow{3}{*}{1} & \multirow{3}{*}{0.501} &200& 0.35 & 0.01 & 0.35 & 0.31 & 0.01 & 0.31 & 0.40 & 0.15 & 0.38 & 0.41 & 0.19 & 0.36 & 0.14 & -0.09 & 0.12 \\ 
  & & &400& 0.32 & -0.01 & 0.32 & 0.27 & -0.01 & 0.27 & 0.37 & 0.11 & 0.35 & 0.38 & 0.15 & 0.35 & 0.11 & -0.07 & 0.09\\ 
  & & &600& 0.26 & 0.02 & 0.26 & 0.20 & 0.02 & 0.20 & 0.31 & 0.16 & 0.27 & 0.32 & 0.10 & 0.25 & 0.08 & -0.05 & 0.07\\ 
 \hline
\multirow{3}{*}{1.09} & \multirow{3}{*}{$\frac{1}{25}$} & \multirow{3}{*}{0.501} &200& 0.12 & -0.04 & 0.11 & 0.21 & -0.04 & 0.21 & 0.27 & 0.26 & 0.08 & 0.31 & 0.31 & 0.07 & 0.13 & -0.10 & 0.08\\ 
  & & &400& 0.08 & -0.02 & 0.08 & 0.12 & -0.02 & 0.12 & 0.23 & 0.23 & 0.05 & 0.30 & 0.30 & 0.04 & 0.09 & -0.07 & 0.06\\ 
  & & &600& 0.05 & -0.02 & 0.05 & 0.08 & 0.02 & 0.07 & 0.22 & 0.22 & 0.04 & 0.29 & 0.29 & 0.04 & 0.08 & 0.06 & 0.05 \\ 
 \hline
   \hline
  \multicolumn{19}{|c|}{{\bf Setting $(\rm S1)$ with $\beta_1 = 1$, Loading 1 with $q = 1$}} \\
  \hline
\multicolumn{3}{|c|}{}&&\multicolumn{3}{c|}{LiVE}&\multicolumn{3}{c|}{Post Selection} &\multicolumn{3}{c|}{\texttt{hdi}}&\multicolumn{3}{c|}{\texttt{WLDP}}&\multicolumn{3}{c|}{Lasso} \\
   \hline
$\|\xnew\|_{2}$ &{\rm r} & Prob &$n$& RMSE & Bias & SE & RMSE & Bias & SE & RMSE & Bias & SE & RMSE & Bias & SE & RMSE & Bias & SE\\ 
  \hline
\multirow{3}{*}{16.1} & \multirow{3}{*}{1} & \multirow{3}{*}{0.881} &200& 0.33 & -0.15 & 0.30 & 0.18 & -0.02 & 0.18 & 0.48 & -0.28 & 0.38 & 0.45 & -0.25 & 0.38 & 0.13 & -0.10 & 0.08\\ 
  & & &400& 0.22 & -0.07 & 0.21 & 0.17 & -0.01 & 0.17 & 0.41 & -0.20 & 0.33 & 0.35 & -0.15 & 0.31 & 0.09 & -0.07 & 0.05\\ 
  & & &600& 0.23 & -0.06 & 0.21 & 0.10 & -0.01 & 0.10 & 0.38 & -0.20 & 0.30 & 0.36 & -0.20 & 0.30 &  0.07 & -0.06 & 0.04\\ 
 \hline
\multirow{3}{*}{1.09} & \multirow{3}{*}{$\frac{1}{25}$} & \multirow{3}{*}{0.881} &200& 0.06 & -0.02 & 0.06 & 0.08 & 0.03 & 0.07 & 0.16 & -0.14 & 0.08 & 0.12 & -0.09 & 0.07 & 0.11 & -0.10 & 0.06\\ 
  & & &400& 0.04 & -0.01 & 0.04 & 0.06 & 0.04 & 0.05 & 0.16 & -0.15 & 0.06 & 0.10 & -0.09 & 0.05 & 0.08 & -0.06 & 0.04\\ 
  & & &600& 0.03 & -0.01 & 0.03 & 0.03 & 0.02 & 0.03 & 0.16 & -0.16 & 0.05 & 0.11 & -0.10 & 0.04 & 0.06 & -0.05 & 0.03\\ 
 \hline
\end{tabular}}
\caption{\textbf{Exactly sparse regression with intercept}. ``r" and ``Prob" represent the shrinkage parameter and Case Probability respectively. The columns indexed with ``RMSE", ``Bias" and ``SE" represent the RMSE, bias and standard error, respectively. The columns under ``LiVE", ``Post Selection", ``\texttt{hdi}", ``\texttt{WLDP}" and ``Lasso" correspond to the proposed estimator, the post model selection estimator, the plug-in \texttt{hdi}, the plug-in \texttt{WLDP} and the Plug-in Lasso estimator respectively.}
\label{tab: with intercept 2ii}
\end{table}
    
\subsection{Additional Simulation Results for Section \ref{sec: varying n and norm}}
\label{sec: varying n and norm supp}

We consider the exactly sparse regression setup $(\rm S1)$ and report the inference results for Loading 2 with $q = 1$ in Table \ref{tab: Setting 2 i}. The CIs constructed by LiVE and \texttt{hdi} have coverage over different scenarios while \texttt{WLDP} and the post-selection suffer from the issue of over-coverage and under-coverage respectively. The case probability being less than $0.5$ ($0.293$) the proposed LiVE method, \texttt{hdi} and \texttt{WLDP} have type I error controlled across all sample sizes while post selection does not have it controlled for the sample size at $n=200$.

In Table \ref{tab: Setting 1,2 ii}, we compare the proposed  estimator, the post selection estimator, the plug-in \texttt{hdi}, \texttt{WLDP} and Lasso estimator in terms of Root Mean Squared Error (RMSE), bias and standard error with respect to regression setting $(\rm S1)$. Through comparing the proposed and plug-in Lasso estimators, we observe that the bias component is reduced at the expense of increasing the variance. Although the bias component is reduced, the total RMSE is not necessarily decreasing after correcting the bias, since the increased variance can lead to a larger RMSE in total. The increase in variance is proportional to the loading norm $\|\xnew\|_2$; specifically, if the loading norm is large, we may suffer from a larger total RMSE after bias-correction; if the loading norm is relatively small,  the variance only increases slightly and the total RMSE decreases due to the reduction of the bias. This matches with the theoretical results presented in Theorem \ref{thm: limiting distribution}.

\begin{table}[htp!]
\centering
\scalebox{0.75}{
\begin{tabular}{|rrr|r|rrrr|rrrr|rrrr|rrrr|}
\hline
\multicolumn{20}{|c|}{{\bf Setting $(S1)$, Loading 2 with $q = 1$}} \\
\hline
 \multicolumn{3}{|c|}{}&&\multicolumn{4}{c|}{LiVE} &\multicolumn{4}{c|}{Post Selection} &\multicolumn{4}{c|}{\texttt{hdi}}&\multicolumn{4}{c|}{\texttt{WLDP}} \\
\hline
$\|\xnew\|_{2}$ &{\rm r}& Prob & $n$& Cov & ERR& Len & t & Cov & ERR & Len & t & Cov & ERR & Len & t & Cov & ERR & Len & t \\
 \hline
 \multirow{3}{*}{16.6} & \multirow{3}{*}{1} & \multirow{3}{*}{0.293} &200& 0.94 & 0.02 & 0.95 & 4 & 0.66 & 0.20 & 0.62 & 1 & 0.92 & 0.04 & 0.93 & 370 & 0.98 & 0.00 & 0.97 & 32 \\
 & & &400& 0.95 & 0.01 & 0.91 & 13 & 0.82 & 0.07 & 0.66 & 2 & 0.94 & 0.02 & 0.92 & 743 & 1.00 & 0.00 & 0.98 & 56 \\ 
 & & &600& 0.96 & 0.03 & 0.90 & 22 & 0.79 & 0.03 & 0.61 & 5 & 0.95 & 0.03 & 0.89 & 3132 & 1.00 & 0.00 & 0.98 & 115 \\ 
 \hline
 \multirow{3}{*}{5.38} & \multirow{3}{*}{$\frac{1}{25}$} & \multirow{3}{*}{0.293} &200& 0.93 & 0.02 & 0.81 & 4 & 0.66 & 0.17 & 0.62 & 1 & 0.93 & 0.01 & 0.78 & 369 & 1.00 & 0.00 & 0.72 & 32\\ 
 & & &400& 0.96 & 0.00 & 0.67 & 13 & 0.83 & 0.03 & 0.67 & 2 & 0.95 & 0.00 & 0.68 & 742 & 0.99 & 0.00 & 0.65 & 56\\ 
 & & &600& 0.95 & 0.01 & 0.61 & 21 & 0.82 & 0.03 & 0.59 & 5 & 0.94 & 0.01 & 0.58 & 3131 & 0.98 & 0.00 & 0.60 & 115 \\ 
 \hline
\end{tabular}
}
\caption{\textbf{Varying $n$ and $\|x_*\|_2$.} ``r" and ``Prob" represent the shrinkage parameter and Case Probability respectively. The columns indexed with ``Cov" and ``Len" represent the empirical coverage and length of the CIs; the column indexed with ``ERR" represents the empirical rejection rate of the test; ``t" represents the averaged computation time (in seconds). The columns under ``LiVE" ,``Post Selection", ``\texttt{hdi}" and ``\texttt{WLDP}" correspond to the proposed estimator, the post selection estimator, the plug-in debiased estimator using \texttt{hdi} and \texttt{WLDP}, respectively.}
\label{tab: Setting 2 i}
\end{table}

\begin{table}
\centering
\scalebox{0.7}{
\begin{tabular}{|rrr|r|rrr|rrr|rrr|rrr|rrr|}
  \hline
\multicolumn{19}{|c|}{{\bf Setting $(\rm S1)$, Loading 1 with $q = 1$}} \\
\hline
\multicolumn{3}{|c|}{{}}&&\multicolumn{3}{c|}{LiVE} &\multicolumn{3}{c|}{Post Selection}&\multicolumn{3}{c|}{\texttt{hdi}}&\multicolumn{3}{c|}{\texttt{WLDP}} &\multicolumn{3}{c|}{Lasso} \\
   \hline
$\|\xnew\|_{2}$ & {\rm r} & Prob &$n$& RMSE & Bias & SE & RMSE & Bias & SE & RMSE & Bias & SE & RMSE & Bias & SE & RMSE & Bias & SE\\ 
  \hline
\multirow{3}{*}{16.1} & \multirow{3}{*}{1} & \multirow{3}{*}{0.732} &200& 0.33 & -0.10 & 0.32 & 0.26 & -0.02 & 0.26 & 0.38 & -0.11 & 0.37 & 0.37 & -0.04 & 0.37 & 0.14 & -0.11 & 0.09 \\ 
  & & &400& 0.26 & -0.04 & 0.26 & 0.21 & -0.02 & 0.21 & 0.31 & -0.06 & 0.31 & 0.31 & 0.01 & 0.31 & 0.11 & -0.08 & 0.07 \\ 
  & & &600& 0.24 & -0.05 & 0.24 & 0.20 & -0.02 & 0.20 & 0.30 & -0.08 & 0.29 & 0.30 & -0.03 & 0.30 & 0.08 & -0.06 & 0.06\\ 
 \hline
\multirow{3}{*}{1.09} & \multirow{3}{*}{$\frac{1}{25}$} & \multirow{3}{*}{0.732} &200& 0.10 & -0.03 & 0.09 & 0.14 & 0.04 & 0.14 & 0.07 & 0.02 & 0.07 & 0.11 & 0.10 & 0.06 & 0.13 & -0.11 & 0.07\\ 
  & & &400& 0.07 & -0.02 & 0.07 & 0.09 & 0.03 & 0.09 & 0.06 & 0.02 & 0.06 & 0.09 & 0.08 & 0.05 & 0.10 & -0.08 & 0.06\\ 
  & & &600& 0.06 & -0.02 & 0.06 & 0.07 & 0.02 & 0.07 & 0.05 & 0.01 & 0.05 & 0.08 & 0.07 & 0.04 & 0.08 & -0.06 & 0.05\\ 
 \hline
  \hline
  \multicolumn{19}{|c|}{{\bf Setting $(\rm S1)$, Loading 2 with $q = 1$}}  \\
   \hline
   \multicolumn{3}{|c|}{}&&\multicolumn{3}{c|}{LiVE} &\multicolumn{3}{c|}{Post Selection} &\multicolumn{3}{c|}{\texttt{hdi}}&\multicolumn{3}{c|}{\texttt{WLDP}}&\multicolumn{3}{c|}{Lasso} \\
   \hline
$\|\xnew\|_{2}$ & {\rm r} & Prob &$n$& RMSE & Bias & SE & RMSE & Bias & SE & RMSE & Bias & SE & RMSE & Bias & SE & RMSE & Bias & SE\\
  \hline
\multirow{3}{*}{16.6} & \multirow{3}{*}{1} & \multirow{3}{*}{0.293} &200& 0.45 & 0.16 & 0.42 & 0.43 & 0.19 & 0.40 & 0.44 & 0.17 & 0.40 & 0.35 & 0.12 & 0.32 & 0.29 & 0.20 & 0.22\\
  & & &400& 0.38 & 0.10 & 0.37 & 0.33 & 0.08 & 0.32 & 0.39 & 0.10 & 0.37 & 0.29 & 0.08 & 0.28 & 0.23 & 0.13 & 0.19 \\ 
  & & &600& 0.36 & 0.04 & 0.35 & 0.25 & 0.03 & 0.25 & 0.34 & 0.05 & 0.34 & 0.22 & 0.01 & 0.22 & 0.22 & 0.13 & 0.19 \\ 
 \hline
\multirow{3}{*}{5.38} & \multirow{3}{*}{$\frac{1}{25}$} & \multirow{3}{*}{0.293} &200& 0.29 & 0.05 & 0.29 & 0.39 & 0.14 & 0.37 & 0.28 & 0.07 & 0.27 & 0.22 & -0.02 & 0.22 & 0.27 & 0.17 & 0.21\\ 
  & & &400& 0.21 & 0.01 & 0.21 & 0.30 & 0.06 & 0.29 & 0.21 & 0.03 & 0.21 & 0.20 & -0.03 & 0.20 & 0.22 & 0.13 & 0.18\\ 
  & & &600& 0.22 & 0.03 & 0.22 & 0.24 & 0.01 & 0.24 & 0.20 & 0.03 & 0.20 & 0.17 & 0.01 & 0.17 & 0.21 & 0.11 & 0.18\\ 
 \hline  
\end{tabular}
}
\caption{\textbf{Varying $n$ and $\|x_*\|_2$.} ``r" and ``Prob" represent the shrinkage parameter and Case Probability respectively. The columns indexed with ``RMSE", ``Bias" and ``SE" represent the RMSE, bias and standard error, respectively. The columns under ``LiVE", ``Post Selection", ``\texttt{hdi}", ``\texttt{WLDP}" and ``Lasso" correspond to the proposed estimator, the post model selection estimator, the plug-in \texttt{hdi}, the plug-in \texttt{WLDP} and the plug-in Lasso estimator respectively. }
\label{tab: Setting 1,2 ii}
\end{table}

The inference results in the approximately sparse regression setup $(\rm S2)$ with $\text{\rm decay} = 2$ are summarized in Table \ref{tab: decaying coef i decay2}. The main observations are similar to that for ${\rm decay} = 1$. However for $\text{\rm decay}= 2,$ the testing procedures based on the proposed LiVE, \texttt{hdi} and \texttt{WLDP} have type I error controlled for both $r=1$ and $r=1/25$ while the post selection method suffers from an inflated Type I error for the setting $r=1$.

Here we also report the estimation accuracy results for the approximately sparse regression setup $(\rm S2)$ with $\text{\rm decay} \in \left\{1,2\right\}$. Table \ref{tab: decaying coef ii} summarizes the estimation accuracy results for Loading 1 and the results are similar to the exactly sparse setting in Table \ref{tab: Setting 1,2 ii}. Table \ref{tab: decaying coef ii} shows again the plug-in Lasso estimator cannot be used for confidence interval construction owing to its large bias. 

\begin{table}[htp!]
\centering
\scalebox{0.72}{
\begin{tabular}{|rrr|r|rrrr|rrrr|rrrr|rrrr|}
  \hline
    \multicolumn{20}{|c|}{{\bf Setting $(\rm S2)$ with ${\rm decay}= 2$, Loading 1 with $q = 1$}} \\
  \hline
\multicolumn{3}{|c|}{}&&\multicolumn{4}{c|}{LiVE} &\multicolumn{4}{c|}{Post Selection} &\multicolumn{4}{c|}{\texttt{hdi}}&\multicolumn{4}{c|}{\texttt{WLDP}} \\
   \hline
$\|\xnew\|_{2}$ &{\rm r}& Prob &$n$& Cov & ERR& Len & t & Cov & ERR & Len & t & Cov & ERR & Len & t & Cov & ERR & Len & t\\ 
 \hline
\multirow{3}{*}{16.1} & \multirow{3}{*}{1} & \multirow{3}{*}{0.488} &200& 0.95 & 0.04 & 0.91 & 5 & 0.66 & 0.14 & 0.37 & 1 & 0.94 & 0.03 & 0.92 & 370 & 1.00 & 0.00 & 1.00 & 34 \\ 
  & & &400& 0.96 & 0.03 & 0.86 & 14 & 0.60 & 0.18 & 0.29 & 2 & 0.95 & 0.04 & 0.88 & 751 & 1.00 & 0.00 & 1.00 & 56 \\ 
  & & &600& 0.96 & 0.03 & 0.78 & 23 & 0.67 & 0.15 & 0.27 & 6 & 0.93 & 0.03 & 0.85 & 3212 & 1.00 & 0.00 & 1.00 & 118 \\ 
 \hline
\multirow{3}{*}{1.09} & \multirow{3}{*}{$\frac{1}{25}$} & \multirow{3}{*}{0.481} &200& 0.96 & 0.03 & 0.35 & 5 & 0.87 & 0.05 & 0.22 & 1 & 0.95 & 0.03 & 0.38 & 371 & 1.00 & 0.00 & 0.75 & 34\\ 
  & & &400& 0.93 & 0.04 & 0.27 & 14 & 0.83 & 0.06 & 0.15 & 2 & 0.93 & 0.03 & 0.27 & 751 & 1.00 & 0.00 & 0.68 & 54\\ 
  & & &600& 0.96 & 0.02 & 0.22 & 22 & 0.73 & 0.02 & 0.12 & 5 & 0.94 & 0.02 & 0.23 & 3211 & 1.00 & 0.00 & 0.65 & 118 \\ 
 \hline
\end{tabular}
}
\caption{\textbf{Varying $n$ and $\|x_*\|_2$.} ``r" and``Prob" represent the shrinkage parameter and Case Probability respectively. The columns indexed with ``Cov" and ``Len" represent the empirical coverage and length of the constructed CIs respectively; the column indexed with ``ERR" represents the empirical rejection rate of the testing procedure; ``t" represents the averaged computation time (in seconds). The columns under ``LiVE" ,``Post Selection", ``\texttt{hdi}" and ``\texttt{WLDP}" correspond to the proposed estimator, the post selection estimator, the plug-in debiased estimator using \texttt{hdi} and \texttt{WLDP}  respectively.}
\label{tab: decaying coef i decay2}
\end{table}

\begin{table}[htp!]
\centering
\scalebox{0.7}{
\begin{tabular}{|rrr|r|rrr|rrr|rrr|rrr|rrr|}
  \hline
\multicolumn{19}{|c|}{{\bf Setting $(\rm S2)$ with ${\rm decay}=1$, Loading 1 with $q = 1$}} \\
  \hline
\multicolumn{3}{|c|}{}&&\multicolumn{3}{c|}{LiVE}&\multicolumn{3}{c|}{Post Selection} &\multicolumn{3}{c|}{\texttt{hdi}}&\multicolumn{3}{c|}{\texttt{WLDP}}&\multicolumn{3}{c|}{Lasso} \\
   \hline
$\|\xnew\|_{2}$ &{\rm r}& Prob &$n$& RMSE & Bias & SE & RMSE & Bias & SE & RMSE & Bias & SE & RMSE & Bias & SE & RMSE & Bias & SE\\ 
 \hline
\multirow{3}{*}{16.1} & \multirow{3}{*}{1} & \multirow{3}{*}{0.645} &200& 0.37 & -0.03 & 0.37 & 0.31 & -0.09 & 0.31 & 0.38 & -0.04 & 0.38 & 0.38 & -0.02 & 0.38 & 0.18 & -0.14 & 0.11\\ 
  & & &400& 0.29 & -0.05 & 0.28 & 0.28 & -0.10 & 0.27 & 0.32 & -0.06 & 0.31 & 0.32 & 0.02 & 0.32 & 0.16 & -0.13 & 0.09\\ 
  & & &600& 0.24 & -0.03 & 0.24 & 0.23 & -0.05 & 0.23 & 0.28 & -0.02 & 0.28 & 0.29 & 0.07 & 0.29 &  0.15 & -0.13 & 0.08\\ 
 \hline
\multirow{3}{*}{1.09} & \multirow{3}{*}{$\frac{1}{25}$} & \multirow{3}{*}{0.523} & 200 & 0.10 & -0.01 & 0.10 & 0.18 & -0.03 & 0.18 & 0.10 & 0.02 & 0.10 & 0.12 & 0.05 & 0.11 & 0.07 & -0.03 & 0.06\\ 
  & & &400& 0.08 & -0.01 & 0.08 & 0.13 & -0.01 & 0.13 & 0.08 & 0.01 & 0.08 & 0.09 & 0.04 & 0.09 & 0.05 & -0.03 & 0.04\\ 
  & & &600& 0.06 & -0.02 & 0.06 & 0.09 & -0.03 & 0.08 & 0.05 & 0.01 & 0.05 & 0.05 & 0.05 & 0.05 & 0.05 & -0.03 & 0.04\\ 
 \hline
 \hline
 \multicolumn{19}{|c|}{{\bf Setting $(\rm S2)$ with ${\rm decay}=2$, Loading 1 with $q = 1$}} \\
  \hline
\multicolumn{3}{|c|}{}&&\multicolumn{3}{c|}{LiVE}&\multicolumn{3}{c|}{Post Selection} &\multicolumn{3}{c|}{\texttt{hdi}}&\multicolumn{3}{c|}{\texttt{WLDP}}&\multicolumn{3}{c|}{Lasso} \\
   \hline
$\|\xnew\|_{2}$ &{\rm r}& Prob &$n$& RMSE & Bias & SE & RMSE & Bias & SE & RMSE & Bias & SE & RMSE & Bias & SE & RMSE & Bias & SE\\ 
 \hline
\multirow{3}{*}{16.1} & \multirow{3}{*}{1} & \multirow{3}{*}{0.488} &200& 0.36 & 0.02 & 0.36 & 0.25 & -0.02 & 0.25 & 0.38 & 0.01 & 0.38 & 0.41 & 0.02 & 0.41 & 0.08 & 0.00 & 0.08\\ 
  & & &400& 0.31 & 0.01 & 0.31 & 0.19 & -0.01 & 0.19 & 0.32 & -0.01 & 0.32 & 0.38 & 0.00 & 0.38 & 0.06 & 0.00 & 0.06\\ 
  & & &600& 0.26 & -0.03 & 0.26 & 0.15 & 0.00 & 0.15 & 0.32 & -0.02 & 0.32 & 0.26 & -0.03 & 0.26 &  0.04 & 0.00 & 0.04\\ 
 \hline
\multirow{3}{*}{1.09} & \multirow{3}{*}{$\frac{1}{25}$} & \multirow{3}{*}{0.481} & 200 & 0.09 & 0.01 & 0.09 & 0.10 & -0.01 & 0.10 & 0.10 & 0.01 & 0.10 & 0.11 & 0.01 & 0.11 & 0.04 & 0.01 & 0.04\\ 
  & & &400& 0.07 & 0.01 & 0.07 & 0.05 & 0.00 & 0.05 & 0.08 & -0.01 & 0.08 & 0.09 & -0.01 & 0.09 & 0.03 & 0.01 & 0.03\\ 
  & & &600& 0.06 & 0.01 & 0.06 & 0.05 & -0.02 & 0.05 & 0.06 & -0.01 & 0.06 & 0.06 & -0.01 & 0.06 & 0.03 & 0.01 & 0.03\\ 
 \hline
\end{tabular}}
\caption{\textbf{Varying $n$ and $\|x_*\|_2$.} ``r" and ``Prob" represent the shrinkage parameter and Case Probability respectively. The columns indexed with ``RMSE", ``Bias" and ``SE" represent the RMSE, bias and standard error, respectively. The columns under ``LiVE", ``Post Selection", ``\texttt{hdi}", ``\texttt{WLDP}" and ``Lasso" correspond to the proposed estimator, the post model selection estimator, the plug-in \texttt{hdi}, the plug-in \texttt{WLDP} and the Plug-in Lasso estimator respectively.}
\label{tab: decaying coef ii}
\end{table}

\subsection{Additional Simulation Results for Section \ref{sec: larger p, larger beta}}
\label{sec: larger p, larger beta supp}

The inference results for the exactly sparse regression setup $(\rm S5)$ with respect to Loading 2 with $q = 1/2$ are reported in Table \ref{tab: ES p1000 Loading 2}. 

\begin{table}[htp!]
    \centering
    \scalebox{0.72}{
    \begin{tabular}{|rrrr|r|rrrrrr|}
    \hline
    \multicolumn{11}{|c|}{{\bf Setting $(\rm S5)$, Loading 2 with $q = 1/2$}} \\
    \hline
    $p$ &$\|x_*\|_{2}$ &{\rm r}& Prob &$n$& Cov & ERR & Len & RMSE & Bias & SE \\ 
    \hline
    \multirow{3}{*}{1001} &\multirow{3}{*}{3.35} & \multirow{3}{*}{$\frac{1}{5}$} & \multirow{3}{*}{0.278} &400& 0.97 & 0.00 & 0.56 & 0.14 & -0.03 & 0.14 \\ 
    & & & &600& 0.97 & 0.00 & 0.53 & 0.13 & -0.03 & 0.13 \\ 
    & & & &1000& 0.97 & 0.00 & 0.47 & 0.12 & -0.02 & 0.12 \\
    \hline
    \multirow{3}{*}{2001} &\multirow{3}{*}{4.89} & \multirow{3}{*}{$\frac{1}{5}$} & \multirow{3}{*}{0.508} &400&
    0.95 & 0.00 & 0.78 & 0.26 & -0.14 & 0.22 \\
    & & & &600& 0.93 & 0.01 & 0.73 & 0.24 & -0.12 & 0.20 \\ 
    & & & &1000& 0.95 & 0.00 & 0.67 & 0.19 & -0.10 & 0.17 \\
    \hline
    \multirow{2}{*}{5001} &\multirow{2}{*}{7.10} & \multirow{2}{*}{$\frac{1}{5}$} & \multirow{2}{*}{0.363} &400& 0.98 & 0.02 & 0.88 & 0.29 & 0.08 & 0.28 \\
    & & & &600& 0.99 & 0.00 & 0.85 & 0.23 & 0.07 & 0.21 \\ 
    & & & &1000& 0.97 & 0.00 & 0.80 & 0.20 & 0.02 & 0.20 \\
    \hline
    \end{tabular}
    }
    \caption{\textbf{Inference properties of LiVE with increasing $p$ and coefficient magnitudes.} ``r" and``Prob" represent the shrinkage parameter and Case Probability respectively. The columns indexed with ``Cov" and ``Len" represent the empirical coverage and length of the constructed CIs respectively; the column indexed with ``ERR" represents the empirical rejection rate of the testing procedure; The columns indexed with ``RMSE", ``Bias" and ``SE" represent the RMSE, bias and standard error, respectively.}
    \label{tab: ES p1000 Loading 2}
    \end{table}
    
\noindent We summarize the results for the approximately sparse regression setup $(\rm S6)$ with ${\rm decay} = 1 $ and  $\rm decay = 2$ in Tables \ref{tab: AS decay 1 p large} and \ref{tab: AS decay 2 p large} respectively. Tables \ref{tab: ES p1000 Loading 2}, \ref{tab: AS decay 1 p large} and \ref{tab: AS decay 2 p large} further support the validity of the constructed confidence intervals. The proposed test has type I error controlled when the case probabilities correspond to the null hypothesis. However for $p = 2001$ the testing procedure does not have power since the case probabilities ($0.508, 0.530, 0.533$) correspond to alternatives in the indistinguishable region.

\begin{table}[htp!]
    \centering
    \scalebox{0.72}{
    \begin{tabular}{|rrrr|r|rrrrrr|}
    \hline
    \multicolumn{11}{|c|}{{\bf Setting $(\rm S6)$ with $\rm decay = 1$, Loading 1 with $q = 1/2$}} \\
    \hline
    $p$ &$\|x_*\|_{2}$ &{\rm r}& Prob &$n$& Cov & ERR & Len & RMSE & Bias & SE \\ 
    \hline
    \multirow{3}{*}{1001} &\multirow{3}{*}{3.21} & \multirow{3}{*}{$\frac{1}{5}$} & \multirow{3}{*}{0.252} &400& 0.93 & 0.00 & 0.40 & 0.11 & 0.05 & 0.10 \\ 
    & & & &600& 0.95 & 0.00 & 0.37 & 0.10 & 0.04 & 0.09 \\
    & & & &1000& 0.95 & 0.00 & 0.28 & 0.07 & 0.03 & 0.07 \\
    \hline
    \multirow{3}{*}{2001} &\multirow{3}{*}{4.60} & \multirow{3}{*}{$\frac{1}{5}$} & \multirow{3}{*}{0.412} &400& 0.95 & 0.03 & 0.56 & 0.16 & 0.07 & 0.14 \\
    & & & &600& 0.97 & 0.01 & 0.49 & 0.14 & 0.06 & 0.13 \\ 
    & & & &1000& 0.96 & 0.01 & 0.45 & 0.11 & 0.04 & 0.11 \\
    \hline
    \multirow{2}{*}{5001} &\multirow{2}{*}{7.07} & \multirow{2}{*}{$\frac{1}{5}$} & \multirow{2}{*}{0.373} &400& 0.99 & 0.00 & 0.71 & 0.19 & 0.03 & 0.19 \\
    & & & &600& 0.98 & 0.00 & 0.63 & 0.19 & 0.03 & 0.19 \\ 
    & & & &1000& 0.97 & 0.00 & 0.55 & 0.13 & 0.00 & 0.13 \\
    \hline
    \hline
    \multicolumn{11}{|c|}{{\bf Setting $(\rm S6)$ with $\rm decay = 1$, Loading 2 with $q = 1/2$}} \\
    \hline
    $p$ &$\|x_*\|_{2}$ &{\rm r}& Prob &$n$& Cov & ERR & Len & RMSE & Bias & SE \\ 
    \hline
    \multirow{3}{*}{1001} &\multirow{3}{*}{3.35} & \multirow{3}{*}{$\frac{1}{5}$} & \multirow{3}{*}{0.302} &400& 0.95 & 0.00 & 0.55 & 0.15 & -0.06 & 0.14 \\ 
    & & & &600& 0.96 & 0.00 & 0.53 & 0.14 & -0.04 & 0.13 \\ 
    & & & &1000& 0.96 & 0.00 & 0.47 & 0.13 & -0.03 & 0.12 \\
    \hline
    \multirow{3}{*}{2001} &\multirow{3}{*}{4.89} & \multirow{3}{*}{$\frac{1}{5}$} & \multirow{3}{*}{0.530} &400& 0.97 & 0.02 & 0.80 & 0.25 & -0.10 & 0.23 \\
    & & & &600& 0.96 & 0.01 & 0.74 & 0.22 & -0.08 & 0.20 \\ 
    & & & &1000& 0.97 & 0.01 & 0.67 & 0.19 & -0.07 & 0.17 \\
    \hline
    \multirow{2}{*}{5001} &\multirow{2}{*}{7.10} & \multirow{2}{*}{$\frac{1}{5}$} & \multirow{2}{*}{0.414} &400& 0.99 & 0.01 & 0.89 & 0.28 & 0.08 & 0.27 \\
    & & & &600& 0.97 & 0.00 & 0.86 & 0.25 & 0.04 & 0.25 \\ 
    & & & &1000& 0.98 & 0.00 & 0.80 & 0.20 & 0.01 & 0.20 \\
    \hline
    \end{tabular}
    }
    \caption{\textbf{Inference properties of LiVE with increasing $p$ and coefficient magnitudes.} ``r" and``Prob" represent the shrinkage parameter and Case Probability respectively. The columns indexed with ``Cov" and ``Len" represent the empirical coverage and length of the constructed CIs respectively; the column indexed with ``ERR" represents the empirical rejection rate of the testing procedure; The columns indexed with ``RMSE", ``Bias" and ``SE" represent the RMSE, bias and standard error, respectively.}
    \label{tab: AS decay 1 p large}
\end{table}

\begin{table}[htp!]
    \centering
    \scalebox{0.72}{
    \begin{tabular}{|rrrr|r|rrrrrr|}
    \hline
    \multicolumn{11}{|c|}{{\bf Setting $(\rm S6)$ with $\rm decay = 2$, Loading 1 with $q = 1/2$}} \\
    \hline
    $p$ &$\|x_*\|_{2}$ &{\rm r}& Prob &$n$& Cov & ERR & Len & RMSE & Bias & SE \\ 
    \hline
    \multirow{3}{*}{1001} &\multirow{3}{*}{3.21} & \multirow{3}{*}{$\frac{1}{5}$} & \multirow{3}{*}{0.257} &400& 0.95 & 0.00 & 0.40 & 0.11 & 0.05 & 0.10 \\ 
    & & & &600& 0.95 & 0.00 & 0.37 & 0.10 & 0.04 & 0.09 \\ 
    & & & &1000& 0.96 & 0.00 & 0.28 & 0.07 & 0.02 & 0.07 \\
    \hline
    \multirow{3}{*}{2001} &\multirow{3}{*}{4.60} & \multirow{3}{*}{$\frac{1}{5}$} & \multirow{3}{*}{0.365} &400& 0.97 & 0.01 & 0.55 & 0.15 & 0.05 & 0.14 \\
    & & & &600& 0.95 & 0.01 & 0.48 & 0.13 & 0.03 & 0.13 \\ 
    & & & &1000& 0.96 & 0.00 & 0.43 & 0.10 & 0.02 & 0.10 \\
    \hline
    \multirow{2}{*}{5001} &\multirow{2}{*}{7.07} & \multirow{2}{*}{$\frac{1}{5}$} & \multirow{2}{*}{0.396} &400& 0.96 & 0.01 & 0.71 & 0.20 & 0.03 & 0.20 \\
    & & & &600& 0.96 & 0.00 & 0.66 & 0.16 & 0.02 & 0.16 \\ 
    & & & &1000& 0.95 & 0.00 & 0.55 & 0.14 & 0.00 & 0.14 \\
    \hline
    \hline
    \multicolumn{11}{|c|}{{\bf Setting $(\rm S6)$ with $\rm decay = 2$, Loading 2 with $q = 1/2$}} \\
    \hline
    $p$ &$\|x_*\|_{2}$ &{\rm r}& Prob &$n$& Cov & ERR & Len & RMSE & Bias & SE \\ 
    \hline
    \multirow{3}{*}{1001} &\multirow{3}{*}{3.35} & \multirow{3}{*}{$\frac{1}{5}$} & \multirow{3}{*}{0.322} &400& 0.97 & 0.00 & 0.58 & 0.15 & -0.05 & 0.14 \\ 
    & & & &600& 0.96 & 0.00 & 0.55 & 0.14 & -0.04 & 0.14 \\ 
    & & & &1000& 0.97 & 0.00 & 0.49 & 0.13 & -0.03 & 0.12 \\
    \hline
    \multirow{3}{*}{2001} &\multirow{3}{*}{4.89} & \multirow{3}{*}{$\frac{1}{5}$} & \multirow{3}{*}{0.533} &400& 0.98 & 0.02 & 0.80 & 0.24 & -0.06 & 0.23 \\
    & & & &600& 0.96 & 0.02 & 0.75 & 0.20 & -0.04 & 0.20 \\ 
    & & & &1000& 0.97 & 0.03 & 0.67 & 0.19 & -0.04 & 0.18 \\
    \hline
    \multirow{2}{*}{5001} &\multirow{2}{*}{7.10} & \multirow{2}{*}{$\frac{1}{5}$} & \multirow{2}{*}{0.431} &400& 0.98 & 0.03 & 0.89 & 0.32 & 0.11 & 0.30 \\
    & & & &600& 0.99 & 0.06 & 0.81 & 0.28 & 0.06 & 0.27 \\ 
    & & & &1000& 0.97 & 0.00 & 0.80 & 0.24 & 0.06 & 0.23 \\
    \hline
    \end{tabular}
    }
    \caption{\textbf{Inference properties of LiVE with increasing $p$ and coefficient magnitudes.} ``r" and``Prob" represent the shrinkage parameter and Case Probability respectively. The columns indexed with ``Cov" and ``Len" represent the empirical coverage and length of the constructed CIs respectively; the column indexed with ``ERR" represents the empirical rejection rate of the testing procedure; The columns indexed with ``RMSE", ``Bias" and ``SE" represent the RMSE, bias and standard error, respectively.}
    \label{tab: AS decay 2 p large}
\end{table}

\subsection{Additional Simulation Results for Section \ref{sec: A2}}

\begin{figure}[htp!]
\centering
\includegraphics[scale=0.75]{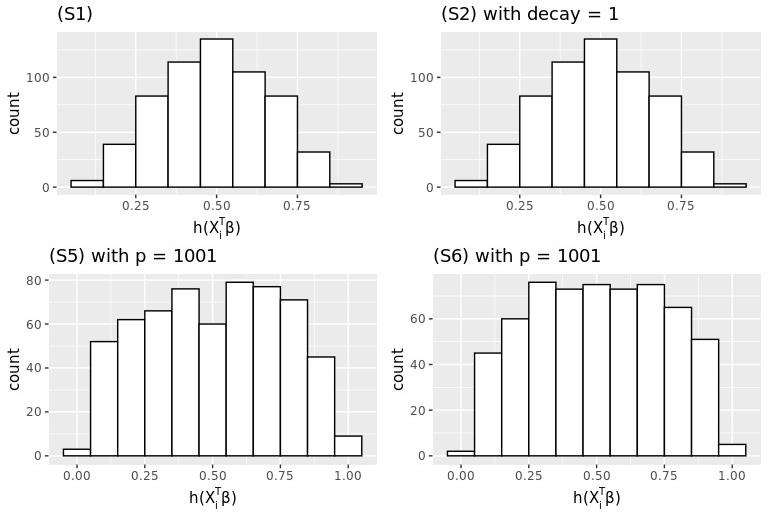}
\caption{Histogram showing the distribution of $\left\{h(X_{i \cdot}^{\intercal}\beta)\right\}_{i=1}^{n}$ for sample $1$ with respect to  regression settings $(\rm S1)$ (top left), $(\rm S2)$ with $\rm decay = 1$ (top right), $(\rm S5)$ with $p = 1001$ (bottom left) and $(\rm S6)$ with $\rm decay = 1$ and $p = 1001$ (bottom right).  Here sample size $n = 600$} 
\label{fig:histogram s1s2s5s6}
\end{figure}

We plot the histograms of conditional case probabilities for settings $(\rm S1)$, $(\rm S2)$, $(\rm S5)$ and $(\rm S6)$ in the figure \ref{fig:histogram s1s2s5s6}. The inverted U-shape of the histograms indicates that assumption $(\rm A2)$ is plausible or weakly violated. Consequently the LiVE method performs well with respect to CI construction and hypothesis testing as indicated earlier in sections \ref{sec: varying n and norm} and \ref{sec: larger p, larger beta}.

\section{Additional Real Data Analysis}
\begin{figure}[htp!]
\centering
\includegraphics[scale=0.4]{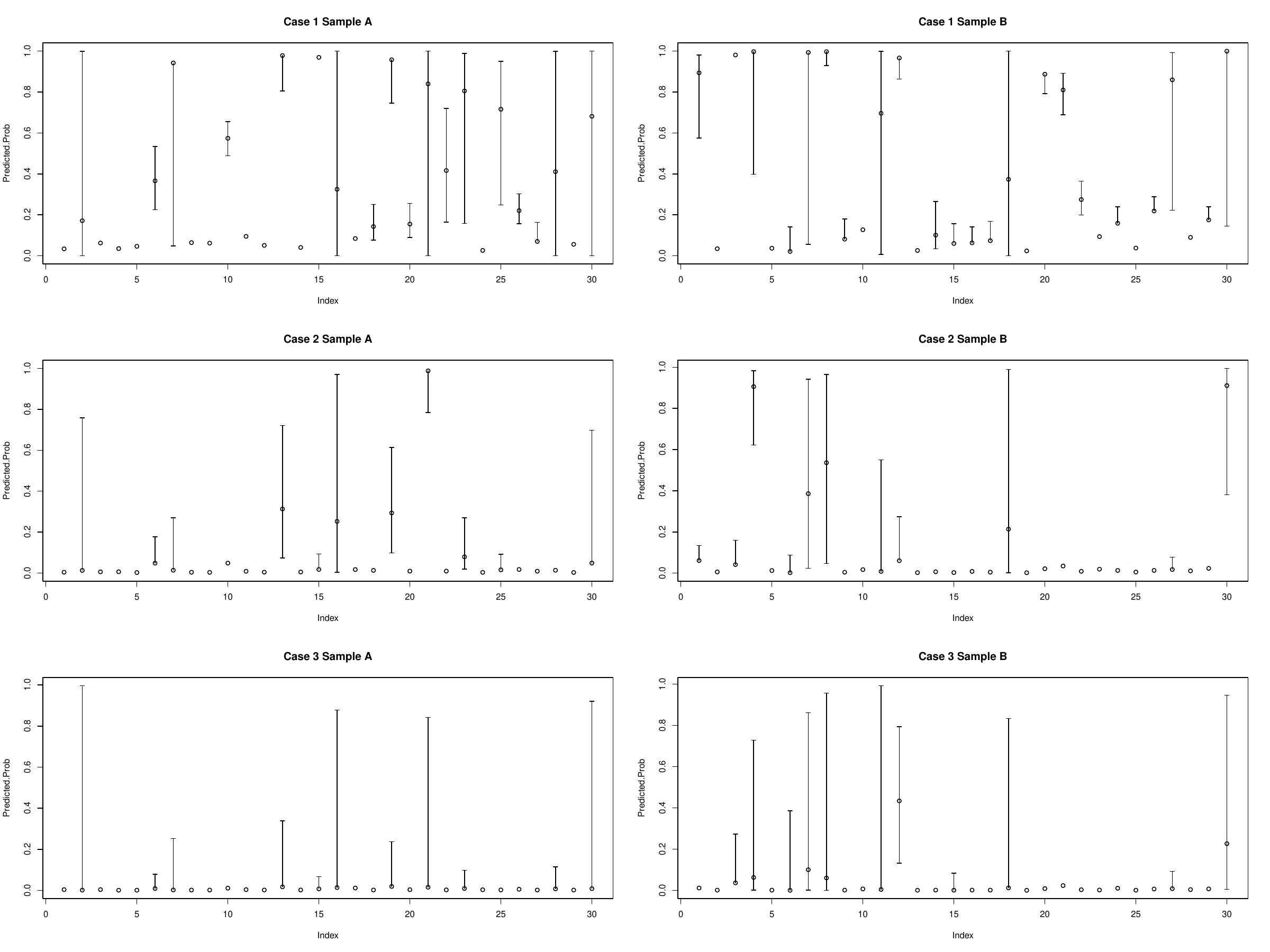}
\caption{Confidence interval construction for the random subsamples}
\label{fig: CI comp}
\end{figure}

Figure \ref{fig: CI comp} presents confidence intervals constructed using our method for the predicted probabilities shown in all six panels in Figure \ref{fig: pred}, corresponding to prediction of hypertension, resistant hypertension and high blood pressure with unexplained low blood potassium across two random subsamples.

\end{document}